\documentclass[aps,prd,twocolumn,10pt,longbibliography,superscriptaddress,amssymb,amsmath,nofootinbib]{revtex4-2}

\usepackage[utf8]{inputenc}
\usepackage{lmodern}
\usepackage[T1]{fontenc}
\usepackage{graphicx} 
\usepackage{hyperref}
\usepackage[dvipsnames]{xcolor}
\usepackage{mathtools}
\usepackage{mathrsfs}
\usepackage{booktabs}
\usepackage{orcidlink}
\usepackage{stmaryrd}
\usepackage{bm}
\usepackage[capitalise]{cleveref}

\newcommand{\beq}{\begin{equation}}
\newcommand{\eeq}{\end{equation}}
\newcommand{\e}{\varepsilon}
\newcommand{\secref}[1]{Sec.~\ref{sec:#1}}
\newcommand{\gexact}{{\sf g}}
\newcommand{\hexact}{{\sf h}}

\newcommand{\nn}{\nonumber}

\renewcommand{\SS}{\mathrm{SS}}
\newcommand{\SR}{\mathrm{SR}}
\newcommand{\R}{\mathrm{R}}
\newcommand{\ms}{\mathrm{ms}}
\newcommand{\brE}{\breve{E}}
\newcommand{\brG}{\breve{G}}

\newcommand{\eff}{\mathrm{eff}}

\newcommand{\barh}{\bar{h}}
\newcommand{\jump}[1]{\llbracket #1 \rrbracket}
\newcommand{\rminus}{r_{-}}
\newcommand{\rp}{r_{p}}
\newcommand{\rplus}{r_{+}}
\newcommand{\sigmaplus}{\sigma_{+}}
\newcommand{\sigmap}{\sigma_{p}}
\newcommand{\sigmaminus}{\sigma_{-}}

\font\ec=ecrm0800 at 10pt
\def\thorn{\hbox{\ec\char'336}}

\def\thornp{\hbox{\ec\char'336}'}

\def\mb{{\bar{m}}}

\DeclareFontFamily{OT1}{pzc}{}
\DeclareFontShape{OT1}{pzc}{m}{it}{<-> s * [1.10] pzcmi7t}{}
\DeclareMathAlphabet{\mathpzc}{OT1}{pzc}{m}{it}

\newcommand{\s}{\mathpzc{s}}
\renewcommand{\b}{\mathpzc{b}}
\renewcommand{\l}{\mathpzc{l}}
\newcommand{\m}{\mathpzc{m}}
\renewcommand{\i}{\mathpzc{i}}

\renewcommand{\S}{{\mathcal S}}
\newcommand{\E}{{\mathcal E}}
\newcommand{\T}{{\mathcal T}}
\newcommand{\Lie}{{\mathcal L}}
\renewcommand{\O}{{\mathcal O}}
\newcommand{\calP}{\mathcal{P}}
\newcommand{\calR}{\mathcal{R}}

\def\beq{\begin{equation}}
\def\eeq{\end{equation}}
\def\beqs{\begin{subequations}}
\def\eeqs{\end{subequations}}
\def\bal{\begin{align}}
\def\eal{\end{align}}
\def\no{\nonumber}

\renewcommand{\Re}{\operatorname{Re}}

\newcommand{\scri}{$\mathscr{I}^+$}
\newcommand{\n}{{\tilde{n}}}

\definecolor{colour1}{HTML}{0571b0} 
\definecolor{colour2}{HTML}{92c5de} 
\definecolor{colour3}{HTML}{f4a582} 
\definecolor{colour4}{HTML}{ca0020} 
\definecolor{colour5}{HTML}{fe4a49} 

\hypersetup{colorlinks=true, linkcolor=colour1, citecolor=colour1,
filecolor=colour1, urlcolor=colour1}

\graphicspath{{./}{Figures/}}

\newcommand{\soton}{\affiliation{School of Mathematical Sciences and STAG Research Centre, University of Southampton, Southampton, United Kingdom, SO17 1BJ}}
\newcommand{\ucd}{\affiliation{School of Mathematics \& Statistics, University College Dublin, Belfield, Dublin 4, Ireland, D04 V1W8}}

\begin{document}

\title{Second-order Teukolsky calculations for nonspinning, quasicircular binaries}

\author{Benjamin Leather\,\orcidlink{0000-0001-6186-7271}}
\soton
\author{Andrew Spiers\,\orcidlink{0000-0003-0222-7578}}
\soton
\affiliation{School of Mathematical Sciences \& School of Physics and Astronomy,
University of Nottingham, University Park, Nottingham, NG7 2RD, UK}
\ucd
\author{Adam Pound\,\orcidlink{0000-0001-9446-0638}}
\soton
\author{Samuel D.\ Upton\,\orcidlink{0000-0003-2965-7674}}
\soton
\author{Barry Wardell\,\orcidlink{0000-0001-6176-9006}}
\ucd
\author{Leanne Durkan\,\orcidlink{0000-0001-8593-5793}}
\ucd
\affiliation{Center for Gravitational Physics, The University of Texas at Austin, Austin, Texas 78712, USA}
\author{Niels Warburton\,\orcidlink{0000-0003-0914-8645}}
\ucd

\date{\today}

\begin{abstract}
    Currently, the only second-order gravitational self-force calculations have been based on directly solving the perturbative Einstein equations in the Lorenz gauge. That method relied on the complete separability of the Einstein equations in a Schwarzschild background. In this paper, we present a new scheme based on the second-order Teukolsky equation. Crucially, this method promises to extend (reasonably straightforwardly) to the more realistic case of a Kerr background. Here we implement the scheme in the simplest setting of quasicircular orbits around a Schwarzschild black hole. In addition to working with the Teukolsky equation, our scheme incorporates several other advances over previous second-order self-force calculations: compactified hyperboloidal slicing, transformation to a Bondi--Sachs gauge, and a combination of spectral and variation-of-parameters methods. We also use these tools to re-examine the infrared divergences that arise in second-order Lorenz-gauge calculations, showing they are less pronounced in the Teukolsky case and completely eliminated in the Bondi-Sachs gauge. We conclude by calculating the asymptotic energy fluxes and benchmarking them against previous Lorenz-gauge calculations.
\end{abstract}

\maketitle

\tableofcontents

\section{Introduction}
Among the sources that the 
Laser Interferometer Space Antenna (LISA) is expected to observe, extreme-mass-ratio-inspirals (EMRIs) occupy a privileged position. 
A stellar-mass compact object of mass $m$ spiralling into a massive black hole of mass $M$, with mass ratio $\e := m / M \sim 10^{-7} - 10^{-4}$, executes of order $1/\e$ orbits in band, each one tracing the strong-field geometry of the central object.
The scientific return is correspondingly rich: measurements of the primary's mass and spin to 4 or 5 digits, tests of the Kerr hypothesis at a precision no other source class approaches, and a census of the massive black hole population in a mass range that electromagnetic observations only reach indirectly~\cite{lisa,LISA:2024hlh}.
That return, however, rests on stringent requirements on EMRI modeling.
Accurately extracting the parameters of an EMRI from the LISA data stream demands waveform templates that remain phase accurate to a fraction of a radian over the $\sim 10^{5}$ radians of accumulated orbital phase~\cite{Burke:2023lno}.
No numerical relativity simulation can span that many orbits~\cite{LISAConsortiumWaveformWorkingGroup:2023arg}, and no post-Newtonian expansion is sufficiently accurate in such a strong-field regime~\cite{Honet:2025gge}.
The disparity of scales that defeats those methods is, however, precisely what makes the problem perturbative: the small body moves on the background of the larger central body, and gravitational self-force theory emerges from the expansion of the Einstein field equations (EFEs) about that limit~\cite{Barack:2018yvs,Pound:2021qin}.

The structure of the expansion dictates how far it should be carried. In a multiscale~\cite{Hinderer:2008dm,Pound:2021qin,Mathews:2025nyb} treatment of the inspiral,
the orbital phase takes a form such that the leading adiabatic (0PA) phase is of order $1/\e$, and the first post-adiabatic (1PA) correction enters at order unity. 
The adiabatic term requires only the time-averaged dissipative piece of the first-order self-force, which first-order perturbation theory supplies.
The 1PA term is where the requirement becomes substantially more challenging: it requires the full first-order self-force, the corrections to the small body's structure, and the dissipative second-order self-force.
A 0PA model nevertheless remains valuable for signal searches and preliminary parameter estimation.
However, because the omitted 1PA phase is formally of order unity, such a model will generically accumulate an order-unity dephasing over an inspiral.
Parameter inference across the full LISA observation therefore requires the 1PA contribution.
Second-order self-force is thus a necessary ingredient of a complete 1PA waveform model, rather than merely a higher-order correction to first-order self-force.

The requirement has now been met in the simplest setting. 
Over the past several years, the second-order metric perturbation has been computed for quasicircular orbits in Schwarzschild spacetime by directly solving the second-order EFEs in the Lorenz gauge, yielding the binding energy~\cite{Pound:2019lzj,Bonetto:2021exn}, the energy flux~\cite{Warburton:2021kwk}, and complete 1PA waveforms~\cite{Wardell:2021fyy, Mathews:2025txc}. 
Those results carry two messages. The first is that the machinery works: the delicate infrastructure of punctures~\cite{Pound:2014xva,Miller:2023ers, Upton:2025bja}, effective sources~\cite{Miller:2023ers,Upton:2025bja}, and multiscale~\cite{Miller:2020bft,Miller:2023ers, Upton:2025bja} (slow-evolution) terms assembles into results that agree with post-Newtonian theory where it is valid~\cite{Warburton:2024xnr} and with numerical relativity where comparison is possible~\cite{Warburton:2021kwk,Albertini:2022rfe,Mathews:2025txc}.
The second is that the agreement with numerical relativity persists to mass ratios as close to unity as $q := \e^{-1} = 10$ and remains respectable even closer to equal masses~\cite{Mathews:2025txc}. 
This does not by itself establish a complete intermediate-mass-ratio waveform model: at such mass ratios second post-adiabatic (2PA) contributions may also be quantitatively important~\cite{Albertini:2022rfe,Mathews:InPrep}.
It does, however, demonstrate the unexpectedly broad utility of the second-order approximation and suggests it can provide both an important ingredient and a stringent benchmark for modelling the strong-field inspiral of intermediate-mass-ratio binaries. 

What these results cannot do is describe an astrophysical EMRI, for the simple reason that astrophysical massive black holes spin. The second-order Schwarzschild results have been extended to linear order in the primary's spin (as well as the secondary's)~\cite{Mathews:2025txc}, but linear-in-spin corrections are insufficient for astrophysical EMRIs, as massive black holes are expected not only to spin but to spin rapidly~\cite{reynolds2021observational, lisa}.
The nonlinear spin of the primary enters the phase at adiabatic order and cannot be treated perturbatively for an EMRI: the background must be Kerr.

The key simplification of the Lorenz-gauge EFEs in Schwarzschild does not hold in Kerr spacetime. 
The Schwarzschild calculations leaned, at every stage, on the complete separability of the linearised EFEs on a Schwarzschild background: the decomposition into tensor spherical harmonics reduces the field equations to decoupled 1+1-dimensional systems, one per $(\l, \m)$-mode, and the entire numerical apparatus was built on that reduction.
The Kerr background offers no direct analogue of this decomposition. 
What Kerr does retain is Teukolsky's discovery~\cite{teuk1972,teuk1973}: the linear perturbations of the Weyl scalars $\psi_0$ and $\psi_4$ obey a master equation that separates fully, and those scalars carry the complete radiative flux content at first order. Moreover, most of the metric perturbation can also be reconstructed from the Weyl scalars~\cite{chrzanowski1975vector, cohen1975space, kegeles1979constructive, Wald:1978vm, Green:2019nam,Loutrel:2020wbw, Hollands:2024iqp, Li:2026rkf}; the majority of work in the first-order Kerr self-force programme is built on such metric reconstruction~\cite{Shah:2012gu,Pound:2013faa,vandeMeent:2015lxa,Bini:2016dvs,Merlin:2016boc, Kavanagh:2016idg,vandeMeent:2017bcc, Bini:2018ylh, Toomani:2021jlo, Bourg:2024vre, Nasipak:2025tby}.

The structure of the Teukolsky formalism survives extension to second order.
Given any linearised EFE, with or without a source, one can always construct an associated Teukolsky equation for a corresponding Weyl scalar (or multiple associated Teukolsky equations~\cite{Spiers:2023cip}). Applying this approach to the second-order EFE yields a second-order Teukolsky equation~\cite{Campanelli:1998jv, Green:2019nam, Spiers:2023cip}, a separable second-order differential equation whose homogeneous solutions, Green's functions, and asymptotics are known. Crucially, in an asymptotically regular gauge, the second-order Weyl scalar contains the complete gravitational wave fluxes of the second-order perturbation~\cite{Spiers:2026yqx}. From these fluxes, the second-order 1PA contribution can be constructed using flux balance laws~\cite{Wardell:2021fyy,Trestini:2026tky}, up to the evolution of the Carter constant in the case of inclined orbits. Extracting the 1PA evolution of the Carter constant from the second-order Weyl scalars is currently an active avenue of research.

The consequence of this packaging is that every difficulty in the problem migrates into the source of the second-order Teukolsky equation. 
Much of this paper is devoted to constructing the source while confronting these difficulties. 
The most significant challenge is the strong singularity in the quadratic source, which we handle, as in the Lorenz-gauge calculations, with a puncture scheme: the singular field is subtracted analytically and the Teukolsky equation is solved for a regular residual field driven by an effective source. More insidiously, the multiscale expansion generically produces infrared divergences in retarded integrals at second order~\cite{Pound:2015wva, Cunningham:2024dog}.
Applying the physically correct boundary conditions involves hereditary integrals over the binary's past.

We confront the boundary-condition issues using two ingredients that were previously absent at second order: compactified hyperboloidal slicing and a preliminary first-order transformation to a Bondi-Sachs gauge. Compactified hyperboloidal time slices asymptote to advanced time at the future horizon and retarded time at future null infinity, so that the computational domain includes both boundaries and, when the fields are regular there, the field equations themselves enforce the boundary conditions~\cite{Zenginoglu:2007jw, Zenginoglu:2011jz, PanossoMacedo:2018hab, PanossoMacedo:2019npm, PanossoMacedo:2022fdi, PanossoMacedo:2024nkw, PanossoMacedo:2024pox, Leather:2024mls}. The Bondi-Sachs gauge ensures that these slices, designed to asymptote to outgoing null cones in the background spacetime, remain asymptotically null in the perturbed spacetime~\cite{Bondi:1962px,Sachs:1962wk,Madler:2016xju,Flanagan:2015pxa,Compere:2019gft,Spiers:2026yqx}. While all previous second-order calculations employed hyperboloidal slicing, they avoided compactification due to their use of variation of parameters~\cite{Miller:2023ers}. And while second-order results have all relied on transformations to a Bondi-Sachs gauge to extract quantities at future null infinity~\cite{Cunningham:2024dog}, they only did so after carrying out all first- and second-order calculations in the Lorenz gauge. We show that compactification does not impact the second-order infrared problems, but it facilitates use of numerical methods that avoid some infrared problems associated with variation of parameters, and it naturally combines with the Bondi-Sachs gauge conditions to allow regularity conditions to do all the work of boundary conditions. Moreover, we recover Ref.~\cite{Spiers:2026yqx}'s result that the second-order Teukolsky equation, even in the Lorenz gauge, is free from the infrared divergences associated with hereditary integrals that enter the EFE.

Our numerical strategy assigns distinct roles to hyperboloidal compactification, spectral collocation, and variation of parameters.
Hyperboloidal compactification maps the horizon and future null infinity to finite coordinate locations and allows regularity of the field equation to impose the physical boundary conditions.
A multidomain spectral discretisation then exploits the smoothness of the field and source on each open subdomain, yielding rapid convergence without extrapolation from a finite radius.
Inside the worldtube, the effective source is of limited ($C^1$~\cite{Upton:2025bja}) differentiability at the worldline, and the operator that converts an EFE source into a Teukolsky source acts on it with two further derivatives.
Constructing the source directly from tabulated mode data would consequently require numerical radial differentiation of these components, amplifying interpolation and discretisation errors precisely where the data are least regular.
To circumvent the potentially challenging numerically differentiated source, we implement a variation-of-parameters scheme in a neighbourhood of the particle.
The solution is written as an integral of the effective source against homogeneous solutions, and integration by parts transfers the offending derivatives from the source onto the homogeneous solutions, which are smooth.
The derivatives are performed on spectrally resolved homogeneous solutions, while the finite differentiability of the source is handled analytically rather than numerically.
The resulting combination of hyperboloidal compactification, Bondi-Sachs asymptotics, multidomain spectral collocation, and an integration-by-parts variation-of-parameters construction provides a natural framework for second-order Teukolsky calculations.

In this paper we implement the scheme in the simplest setting: a nonspinning, quasicircular binary, treated as a particle of mass $m$ on a quasicircular inspiral on a Schwarzschild background.
The restriction is a deliberate strategy rather than a limitation of the method.
The Schwarzschild setting is the one in which our calculation can be checked against an independent calculation, and we use that to our advantage: our first-order inputs and effective sources are converted from the Lorenz-gauge infrastructure of Refs.~\cite{Warburton:2021kwk, Wardell:2021fyy}, and our final energy fluxes are benchmarked mode-by-mode against published Lorenz-gauge results.
Every ingredient of this scheme---the Teukolsky equation, the hyperboloidal compactification, the Bondi-Sachs transformation, the spectral variation-of-parameters solver---is formulated so as to carry over to a Kerr background, where the spin-weighted spherical harmonics deform to spheroidal ones. While nontrivial mode mixing in Kerr poses a distinct challenge, it can be addressed using the source decomposition method of Ref.~\cite{Spiers:2024src}.

The significance of this scheme extends beyond the 1PA waveforms that motivate it. 
The second-order Teukolsky source constructed here is the same object that drives quadratic quasinormal mode couplings in ringdown, where nonlinearities of the black hole spectrum have recently become measurable targets~\cite{London:2014cma, Cheung:2022rbm,   Mitman:2022qdl, Redondo-Yuste:2023seq, Baibhav:2023clw, Lagos:2024ekd, Yi:2024elj, Ma:2024qcv, Khera:2024bjs, Bourg:2025lpd, Bourg:2024jme, Bucciotti:2024zyp, Bucciotti:2024jrv}; the infrared analysis bears directly on gravitational-wave memory~\cite{Cunningham:2024dog,Spiers:2026yqx}; and the efficiency of the compactified spectral method opens the door to the dense coverage of parameter space that LISA data analysis will ultimately require.

The remainder of this paper is organised as follows. 
In \secref{multiscale Teukolsky} we review the multiscale formulation of the second-order Teukolsky equation. 
Section~\ref{sec:compactification} introduces the compactified hyperboloidal method, analyses the infrared divergences that survive compactification, and describes the advantages of the Bondi-Sachs gauge.
Section~\ref{sec:source} presents the construction of the source terms: the nonlinear coupling term, the slow-evolution term, and the punctures. 
Section~\ref{sec:spectral_method} presents the spectral method, and \secref{mixed method} develops the variation-of-parameters construction. 
We present our results in \secref{results}, benchmarking the second-order energy fluxes against previous Lorenz-gauge calculations. In Sec.~\ref{sec:conclusion} we conclude by discussing the prospects for  applying our scheme in Kerr spacetime and how it might dovetail with emerging Lorenz-gauge methods in Kerr~\cite{Dolan:2021ijg,Wardell:2024yoi,Hollands:2024iqp,Vu:2026ypc,Osburn:2026uct}. Appendices~\ref{sec:Lorenz conversion}--\ref{sec:Lorenz-Bondi consistency} describe various aspects of how our calculations and analyses relate to our previous, Lorenz-gauge ones.

We use a mostly positive metric signature $(-+++)$ and geometric units with $G=c=1$. Greek letters are used for spacetime indices. Lowercase Latin letters from the beginning of the alphabet ($a,b,c$) are used for tetrad indices. Lowercase Latin letters from the middle of the alphabet ($i,j,k$) are used both for spatial indices and for discretization indices, though never with both meanings in the same expression. Calligraphic lowercase letters ($\s,\l,\m$) are used for spherical-harmonic and Fourier mode indices. Uppercase Latin letters ($A,B,C$) are used as indices on tuples of slowly evolving parameters.


\section{Second-order Teukolsky equation in a multiscale expansion}
\label{sec:multiscale Teukolsky}

Our calculations are specialized to a particle of mass $m$ on a quasicircular inspiral around a slowly spinning black hole of mass $M_{\rm BH}=M+\e \delta M$ and spin $S_{\rm BH}=\e \delta S$, where $\delta M$ and $\delta S$ evolve due to absorption of gravitational radiation through the black hole's horizon. The background spacetime, on which we calculate perturbations, is then a Schwarzschild metric of constant mass~$M$.

In this section, we review Ref.~\cite{Miller:2023ers}'s multiscale formulation of the Einstein and Teukolsky equations for such a binary. Our review of the Teukolsky formalism largely follows the conventions of Ref.~\cite{Spiers:2023mor}.

\subsection{Binary evolution and spacetime foliation}

In the multiscale expansion, the binary's evolution is divided into ``fast variables'', which describe periodic behavior on the short time scale $\sim M$, and ``slow variables'', which describe the binary's evolution due to dissipation on the long time scale $\sim M/\e$. For a quasicircular inspiral, the only fast variable is the particle's orbital phase $\phi_p$, while there are three slow variables: the orbital frequency $\Omega\coloneqq d\phi_p/dt$ and the perturbations $\delta M_A = (\delta M, \delta S)$ to the primary black hole's mass and spin. Altogether, the binary's governing equations then take the form
\begin{align}
    \dot\phi_p &= \Omega,\label{phidot}\\
    \dot\Omega &= \e F_{\Omega}^{(0)}(\Omega) + \e^2 F_\Omega^{(1)}({\cal J}_I) + \O(\e^3),\label{Omegadot}\\
    \dot{\delta M_A} &= \e F_A^{(1)}(\Omega) + \O(\e^2),\label{Mdot}
\end{align}
where a dot denotes $d/dt$ and we have defined ${\cal J}_I=(\Omega,\delta M_A)$. $F_A^{(1)}$ represents the standard adiabatic (0PA) flux of energy and angular momentum into the black hole, $F_{\Omega}^{(0)}$ represents the standard 0PA orbital evolution, and $F_\Omega^{(1)}$ is the 1PA correction first computed in Ref.~\cite{Wardell:2021fyy}.\footnote{To date, calculations of $F_\Omega^{(1)}$ have been based on applications of an energy balance law, and they have omitted numerically small contributions to that balance law: memory distortion~\cite{Cunningham:2024dog}, subleading horizon fluxes, and Schott terms in the binding energy~\cite{Trestini:2025nzr, Grant:inprep}; see Ref.~\cite{Trestini:2026tky} for a recent discussion. Here we again do not compute those missing contributions. Instead, we focus on the asymptotic energy flux first computed in Ref.~\cite{Warburton:2021kwk}.} Here, numerical labels denote the post-adiabatic order at which the quantity enters.

In the above expressions, $t$ denotes Schwarzschild time. The particle's spatial trajectory $x^i_p=(r_p,\theta_p,\phi_p)$ in Schwarzschild coordinates is then written as a function of $(\phi_p,\Omega,\e)$ and expanded for small $\e$:
\begin{equation}
    x^i_p(\phi_p,\Omega,\e) = (r_p(\Omega,\e),\pi/2,\phi_p),
\end{equation}
where 
\begin{equation}
r_p = r_{(0)}(\Omega) + \e r_{(1)}(\Omega) + \O(\e^2).    
\end{equation}
Here $r_{(0)}(\Omega)=(M/\Omega^2)^{1/3}$ is the standard geodesic relationship, and $r_{(1)}(\Omega)$ is the correction due to the radial self-force. We refer to Ref.~\cite{Miller:2020bft} for explicit expressions and a detailed derivation of the 1PA orbital evolution.

Now, the core idea in the multiscale expansion is that all time dependence in the spacetime metric is encoded in the time dependence of the binary's ``mechanical'' variables $(\phi_p,{\cal J}_I)$. 
To make such a formulation well behaved toward future null infinity (\scri) and the future horizon ($\mathscr{H}^+$), we foliate the spacetime into slices of constant hyperboloidal time $s$. This time is related to Schwarzschild time by 
\beq\label{eq:s def}
s \coloneqq  t - k(r^*),
\eeq
where $k$ is referred to as a height function and $r^*$ is the tortoise coordinate. The height function is chosen to ensure $s$ asymptotes to advanced time $v$ at the future horizon and to retarded time $u$ at future null infinity.

\begin{figure}[t!]
	\centering
	\includegraphics[width=\columnwidth]{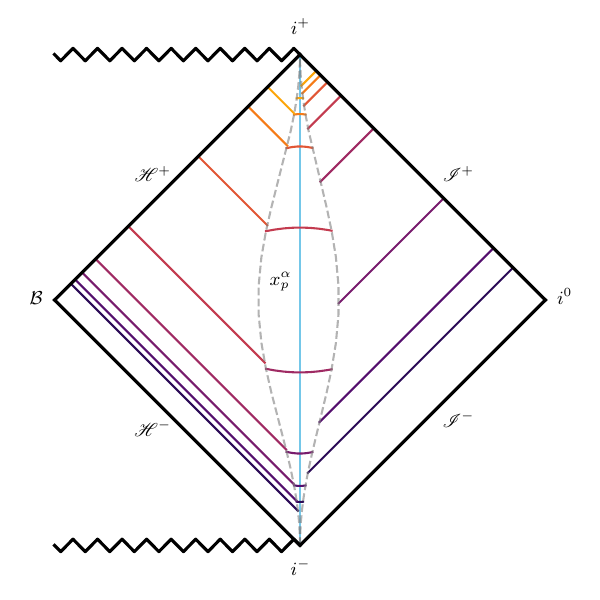}
	\caption{Penrose diagram of the Schwarzschild exterior foliated with (discontinuous) $v$-$t$-$u$ slices.
	The solid, colored curves depict hyperboloidal time surfaces $s=\text{constant}$, with curves of the same color corresponding to the same value of~$s$.
	These slices foliate the full exterior spacetime and extend to future null infinity (\scri) and the future horizon ($\mathscr{H}^+$), which means that both \scri and $\mathscr{H}^+$ can be included in the computational domain.
    To avoid compressing the worldtube into the corner at $i^0$, we have centered the compactification on the particle's
    worldline $x_p^{\alpha}$ (blue vertical line at $r=\rp$), which runs
    straight from $i^{-}$ to $i^{+}$.
    The dashed curves are the worldtube boundaries, 
    where the slicing
    switches from $v$ to $t$ and from $t$ to $u$, and where the discontinuity in each
    $s=\text{constant}$ surface is apparent.}
\label{fig:slicing}
\end{figure}

Such slicing eliminates oscillations that occur at large $|r^*|$ on slices of constant $t$, avoiding the resulting divergences that arise toward the boundaries in $t$ slicing at second order~\cite{Pound:2015wva,Miller:2020bft}. We assume that $s$ reduces to $t$ in a neighborhood of the particle's worldline, but our restrictions on the choice of $s$ in this section are otherwise mild: surfaces of constant $s$ need not be smooth, and they need not be spacelike everywhere in the black hole's exterior (as we specifically allow null segments, making our slicing only ``quasi-hyperboloidal''). However, in later sections, we highlight limitations of some traditional choices. Figure~\ref{fig:slicing} illustrates an example of the particular type of slicing used in previous second-order calculations, referred to as ``$v$-$t$-$u$'' slicing, which we again adopt for our numerical implementation here. With this choice, $s=v$ in a region extending from the horizon to near the particle, $s=t$ in a region containing the particle, and $s=u$ in a region extending to \scri. Note that, as shown in the figure, $s=\text{constant}$ surfaces are discontinuous with this choice of slicing (contrary to our misleading description in Ref.~\cite{Miller:2023ers}, where this slicing was also used). The recent work~\cite{Vu:2026ypc} utilized a modified, equally useful but more intuitive version of this slicing that enforces continuity.

Given a choice of $s$, we promote $(\phi_p,{\cal J}_I)$ to functions on the whole spacetime by defining them to be constant on each slice of constant $s$.

\subsection{Expansion of the Einstein equation}
\label{sec:EFE_expansion}

The spacetime metric $\gexact_{\mu\nu}$ is fundamentally a function of $(s,x^i)$. In the multiscale expansion, we replace the dependence on $s$ with a dependence on $(\phi_p,{\cal J}_I)$, promoting the metric to be a function on the Cartesian product of space (with coordinates $x^i$) and the binary's mechanical phase space with coordinates $(\phi_p,{\cal J}_I)$. We then expand the metric for small $\e$ at fixed $(\phi_p,{\cal J}_I,x^i)$:
\begin{multline}
\gexact_{\mu\nu} = g_{\mu\nu}(x^i) +\e h^{(1)}_{\mu\nu}(\phi_p,{\cal J}_I,x^i) \\+ \e^2 h^{(2)}_{\mu\nu}(\phi_p,{\cal J}_I,x^i) + \O(\e^3),\label{g tt expansion}
\end{multline}
where $g_{\alpha\beta}$ is the Schwarzschild background metric of mass $M$ (which we use to raise and lower indices), and $x^i$ are any set of spatial coordinates on the slices of constant $s$. Since all functions of $\phi_p$ are periodic, we also expand the metric perturbations in discrete Fourier series:
\beq
h^{(n)}_{\alpha\beta} = \sum_{\m=-\infty}^\infty h^{(n)}_{\alpha\beta,\m}({\cal J}_I,x^i)e^{-i\m\phi_p},\label{Fourier}
\eeq
using calligraphic font for mode indices. We refer to Ref.~\cite{Lewis:2025ydo} for a first-principles derivation of this form of the metric, starting from the more general ``self-consistent'' formulation of self-force theory~\cite{Pound:2009sm}.

When substituting the above metric into the EFEs, we apply time derivatives\footnote{In the case of mechanical variables, our choice of slicing is such that $t$ and $s$ derivatives are equivalent so we do not distinguish between them.} using the chain rule $\partial_s = \dot\phi_p(s) \partial_{\phi_p}+\dot{\cal J}(s)\partial_{\cal J_I}$. Appealing to the evolution equations~\eqref{phidot}--\eqref{Mdot}, we can then expand the background covariant derivative, $\nabla_\alpha$, as
\beq\label{nabla}
\nabla_\alpha = \nabla^{(0)}_\alpha + \e s_\alpha \vec{\partial}_{\cal V} + \O(\e^2),
\eeq
where $s_\alpha \coloneqq  \partial_\alpha s$. The zeroth-order term, $\nabla^{(0)}_\alpha$, is simply the background derivative with the parameters ${\cal J}_I$ treated as constants. When acting on a Fourier series such as~\eqref{Fourier},
\beq\label{eq:nabla0 modes}
\nabla^{(0)}_\alpha = e^i_\alpha\frac{\partial}{\partial x^i} -i\omega_\m s_\alpha  + \text{Christoffel terms},
\eeq
where $e^i_\alpha \coloneqq  \frac{\partial x^i}{\partial x^\alpha}$, $\omega_\m \coloneqq  \m\Omega$, and the Christoffel terms are unaffected by the multiscale expansion. 
Subleading terms in Eq.~\eqref{nabla} arise from the evolution of ${\cal J}_I$ in Eqs.~\eqref{Omegadot}--\eqref{Mdot}. To describe the effect of that evolution, in  Eq.~\eqref{nabla} we introduce ${\cal V}_I\coloneqq(F^{(0)}_\Omega,F^{(1)}_A)$ as the leading-order velocity through parameter space and 
\beq
\vec{\partial}_{\cal V}\coloneqq {\cal V}_I\frac{\partial}{\partial J_I} = F_\Omega^{(0)}\frac{\partial}{\partial\Omega} + F_A^{(1)}\frac{\partial}{\partial\delta M_A}
\eeq
as a directional derivative in the parameter space. For the calculation of fluxes, we only require the oscillatory, $\m\neq0$ modes in $h^{(2)}_{\alpha\beta}$. Consequently, in this paper we can omit $F^{(1)}_A$ from ${\cal V}_I$ because $\delta M_A$ only contributes to the $\m=0$ modes of $h^{(1)}_{\alpha\beta}$ (since a perturbation of the mass and spin represents a perturbation toward a slowly evolving Kerr solution, which must be quasistationary).

On the left-hand side of the EFE, we use these expansions of the metric and of the background derivative to expand the Einstein tensor as
\begin{multline}
    \hspace{-10pt}G_{\alpha\beta}[\gexact] = G_{\alpha\beta}[g] + \e G^{(1,0)}_{\alpha\beta}[h^{(1)}] + \e^2 \Bigl(G^{(1,0)}_{\alpha\beta}[h^{(2)}] \\
    + G^{(2,0)}_{\alpha\beta}[h^{(1)},h^{(1)}]+G^{(1,1)}_{\alpha\beta}[h^{(1)}]\Bigr) + \O(\e^3).\label{G expansion}
\end{multline}
The operator $G^{(1,0)}_{\mu\nu}$ is the linearized Einstein tensor with ${\cal J}_I$ treated as a constant, which we write as
\begin{align}\label{eq:LinearEinstein}
G^{(1,0)}_{\alpha\beta}[h] &=  -\tfrac{1}{2}E^{(0)}_{\alpha\beta}[\bar h]+\overline{\nabla^{(0)}_{(\alpha}Z^{(0)}_{\beta)}}[\bar h]
\end{align}
for generic $h_{\alpha\beta}$. Here an overline denotes trace reversal, as in $\bar h_{\mu\nu}\coloneqq h_{\mu\nu}-\tfrac{1}{2}g_{\mu\nu}g^{\alpha\beta}h_{\alpha\beta}$, and the Lorenz-gauge linearized Einstein operator
\beq
E^{(0)}_{\alpha\beta}[\bar h]\coloneqq\Box^{(0)} \bar h_{\alpha\beta}+2R_\alpha{}^\mu{}_\beta{}^\nu \bar h_{\mu\nu},
\eeq
where $\Box\coloneqq g^{\mu\nu}\nabla_{\!\mu}\nabla_{\!\nu}$ is the d'Alembertian and $R_{\alpha\mu\beta\nu}$ is the Riemann tensor of the background. The divergence
\beq
 Z_\alpha[\bar h] \coloneqq  g^{\beta\gamma}\nabla_\gamma\bar h_{\alpha\beta}
\eeq
vanishes in the Lorenz gauge. Zero superscripts on differential operators, as in $\Box^{(0)}$, indicate the replacement $\nabla\to\nabla^{(0)}$.

The second-order term in Eq.~\eqref{G expansion} involves three pieces. The first two are the usual linearized Einstein tensor associated with $h^{(2)}_{\alpha\beta}$ and the second-order (quadratic) Einstein tensor constructed from $h^{(1)}_{\alpha\beta}$. See Eq.~(4) of Ref.~\cite{Pound:2021qin} for the explicit $G^{(2,0)}_{\alpha\beta}$, with the replacement $\nabla\to\nabla^{(0)}$; we will not require that explicit expression here. The final operator in Eq.~\eqref{G expansion}, $G^{(1,1)}_{\mu\nu}$, is given by the terms in the linearised Einstein tensor that are linear in the velocity ${\cal V}_I$. Explicitly, 
\beq\label{G11}
G^{(1,1)}_{\mu\nu}[h] = -\frac{1}{2}E^{(1)}_{\mu\nu}[\bar h] + \overline{\nabla^{(0)}_{(\mu}Z^{(1)}_{\nu)}}[\bar h] + \overline{\nabla^{(1)}_{(\mu}Z^{(0)}_{\nu)}}[\bar h],
\eeq
with
\begin{align}
E^{(1)}_{\mu\nu}[\bar h] &= -i\m\, s_\alpha s^\alpha F_\Omega^{(0)}\bar h_{\mu\nu} + 2 s^\alpha\nabla^{(0)}_\alpha\!\bigl(\vec{\partial}_{\cal V}\bar h_{\mu\nu}\bigr) \nonumber\\
&\quad +\bigl(\nabla^{(0)}_\alpha s^\alpha\bigr)\vec{\partial}_{\cal V} \bar h_{\mu\nu},\label{eq:E1}\\
Z^{(1)}_\mu[\bar h] &= s^\alpha \vec{\partial}_{\cal V}\bar h_{\mu\alpha},\label{eq:Z1}\\
\nabla^{(1)}_{\mu}Z^{(0)}_{\nu}[\bar h] &= s_\mu  Z^{(0)}_\nu[\vec{\partial}_{\cal V}\bar h] - i\m s_\mu s^\beta F^{(0)}_\Omega \bar h_{\nu\beta},\label{eq:D1Z0}
\end{align}
when acting on Fourier modes, where we recall $F^{(0)}_\Omega$ is the 0PA term in $\dot\Omega$.

On the right-hand side of the EFE, as mentioned in the Introduction, the secondary is represented as a point singularity, with a form that is derived rigorously from matched asymptotic expansions~\cite{Pound:2015tma,Barack:2018yvs,Upton:2021oxf}. This singularity can be incorporated into the field equations in either of two equivalent ways: using a puncture scheme or ascribing the particle a point-mass stress-energy tensor known as the Detweiler stress-energy~\cite{Detweiler:2011tt,Upton:2021oxf}, 
\beq\label{eq:Detweiler T}
T_{\alpha\beta} = m \int \tilde u_{\alpha} \tilde u_\beta \frac{\delta^4(x^\mu-x^\mu_p(\tilde\tau))}{\sqrt{-\tilde g}}d\tilde\tau.
\eeq
Here  
\beq
\tilde g_{\alpha\beta} = g_{\alpha\beta} + \e h^{(1){\rm R}}_{\alpha\beta} + \e^2 h^{(2){\rm R}}_{\alpha\beta} +\O(\e^3)
\eeq
is a certain regular effective metric that (i) is a smooth vacuum solution along the particle's worldline $x^\alpha_p$ and (ii) determines the self-force on the particle, via the equation of motion
\begin{multline}\label{eq:eom}
    u^\beta\nabla_{\!\beta} u^\alpha = -\frac{1}{2}(g^{\alpha}{}_\beta+u^\alpha u_\beta)(g^{\beta\gamma}-h^{{\rm R}\beta\gamma})\\
    \times (2\nabla_{\!\rho}h^{\rm R}_{\sigma\gamma}-\nabla_{\!\gamma} h^{\rm R}_{\rho\sigma})u^\rho u^\sigma + \O(\e^3).
\end{multline}
Here $h^{\rm R}_{\alpha\beta} = \e h^{(1)\rm R}_{\alpha\beta} + \e^2 h^{(2)\rm R}_{\alpha\beta} + \O(\e^3)$. In Eq.~\eqref{eq:Detweiler T} we have written the stress-energy tensor in terms of effective-metric quantities:  $\tilde g$ is the determinant of that metric, $\tilde\tau$ is the proper time measured in it, and $\tilde u_\alpha\coloneqq \tilde g_{\alpha\beta}dx^\beta_p/d\tilde\tau$ is the particle's four-velocity normalized in it. In Eq.~\eqref{eq:eom}, in order to introduce the self-force on the right-hand side, we have instead written the covariant acceleration in the background metric: $\nabla_\alpha$ is compatible with $g_{\alpha\beta}$, and $u^\alpha \coloneqq dx^\alpha/d\tau$ is normalized in it. If written in terms of $\tilde g_{\alpha\beta}$, Eq.~\eqref{eq:eom} simply reduces to the geodesic equation in the effective metric~\cite{Pound:2015fma}.

Given the multiscale expansions of the particle's trajectory and of the metric, we can expand the stress-energy tensor in the form
\beq
T_{\mu\nu} = \e T^{(1)}_{\mu\nu}(\phi_p,\Omega,x^i) + \e^2 T^{(2)}_{\mu\nu}(\phi_p,{\cal J}_I,x^i) + \O(\e^3)\label{T tt expansion},
\eeq
each term of which is immediately expandable in a discrete Fourier series:
\begin{equation}
    T^{(n)}_{\mu\nu}(\phi_p,\Omega,x^i) = \sum_{\m=-\infty}^\infty T^{(n)}_{\mu\nu,\m}(\Omega,x^i)e^{-i\m\phi_p}.
\end{equation}
We will not require the explicit coefficients in these expansions.

Combining the above results, we write the EFE $G_{\mu\nu}[\gexact]=8\pi T_{\mu\nu}$ as a hierarchical sequence of equations. After the zeroth-order equation $G_{\mu\nu}[g] = 0$, the first two perturbative equations are
\begin{align}
 G^{(1,0)}_{\mu\nu}[h^{(1)}] &= 8\pi T^{(1)}_{\mu\nu},\label{tt EFE1}\\
 G^{(1,0)}_{\mu\nu}[h^{(2)}] &= 8\pi T^{(2)}_{\mu\nu} - G^{(2,0)}_{\mu\nu}[h^{(1)},h^{(1)}]\nonumber\\
 &\quad -G^{(1,1)}_{\mu\nu}[h^{(1)}].\label{tt EFE2}
\end{align}
The strong singularity at the particle makes $G^{(2,0)}_{\mu\nu}$ non-integrable there, but the right-hand side of Eq.~\eqref{tt EFE2} can nevertheless be made well defined by adopting the canonical distributional definition of $G^{(2,0)}_{\mu\nu}$ from Ref.~\cite{Upton:2021oxf} (see also Ref.~\cite{Musaeus:inprep}). 

Equation~\eqref{tt EFE2} has never been solved directly in practice. Instead, the field equations have been solved in the puncture scheme mentioned above. In this approach, one splits $h^{(n)}_{\alpha\beta}$ into two pieces,
\begin{equation}
    h^{(n)}_{\alpha\beta} = h^{(n)\cal P}_{\alpha\beta} + h^{(n)\cal R}_{\alpha\beta}.
\end{equation}
The puncture fields $h^{(n)\cal P}_{\alpha\beta}$, derived from matched expansions, capture the singularity at the particle. They are given in covariant form in Ref.~\cite{Pound:2014xva} and expanded in multiscale form in Ref.~\cite{Upton:2025bja}. The residual fields $h^{(n)\cal R}_{\alpha\beta}$ agree with the regular fields $h^{(n)\rm R}_{\alpha\beta}$ at the particle, in the sense that
\begin{align}
    h^{(n)\cal R}_{\alpha\beta}(x_p) &= h^{(n)\rm R}_{\alpha\beta}(x_p), \\
    \partial_\mu h^{(n)\cal R}_{\alpha\beta}(x_p) &= \partial_\mu h^{(n)\rm R}_{\alpha\beta}(x_p),
\end{align}
such that the self-force on the right-hand side of Eq.~\eqref{eq:eom} can be computed from the residual fields. The puncture fields are constructed as expansions in powers of distance from the particle, and the order to which those expansions are carried determines the residual fields' degree of regularity. In this paper we employ the same punctures as in Ref.~\cite{Upton:2025bja}, for which $h^{(1)\cal R}_{\alpha\beta}$ is $C^2$ at the particle, and $h^{(2)\cal R}_{\alpha\beta}$ is $C^1$; see Ref.~\cite{Upton:2025bja} for details.

Moving the puncture to the right-hand side of the field equations yields equations for the residual fields,\footnote{\label{fn:Teff}In many places in the self-force literature, the stress-energy tensors $T^{(n)}_{\mu\nu}$ do not appear on the right-hand sides of these equations, and $T^{(2)}_{\mu\nu}$ in particular has never been used in practice. This is because the effective sources are more primitively defined and calculated without any knowledge of $T_{\mu\nu}$: one begins with the vacuum field equations away from the worldline, moves the puncture fields to the right-hand side, and then promotes the combined effective source, as an integrable function, to the whole domain including $\gamma$. See the Introduction of Ref.~\cite{Upton:2021oxf}.}
\begin{align}
 G^{(1,0)}_{\mu\nu}[h^{(1)\cal R}] &= 8\pi T^{(1)}_{\mu\nu} - G^{(1,0)}_{\mu\nu}[h^{(1)\cal P}],\label{tt EFE1R}\\
 G^{(1,0)}_{\mu\nu}[h^{(2)\cal R}] &= 8\pi T^{(2)}_{\mu\nu} - G^{(2,0)}_{\mu\nu}[h^{(1)},h^{(1)}] - G^{(1,1)}_{\mu\nu}[h^{(1)}] \nonumber\\
 &\quad- G^{(1,0)}_{\mu\nu}[h^{(2)\cal P}].\label{tt EFE2R}
\end{align}
We write these together as
\begin{equation}\label{Tneff def}
    G^{(1,0)}_{\mu\nu}[h^{(n)\cal R}] = 8\pi T^{(n)\rm eff}_{\mu\nu}.
\end{equation}
With the punctures we use in this paper, the effective sources $T^{(1)\rm eff}_{\mu\nu}$ and $T^{(2)\rm eff}_{\mu\nu}$ are $C^0$ and $C^{-1}$ (diverging as one over distance from the particle), respectively. We again refer to Ref.~\cite{Upton:2025bja} for details. 

In practice, one solves the effective-source equation~\eqref{tt EFE2R} inside a finite region around the particle (``the worldtube'') and the physical equation~\eqref{tt EFE2} outside the worldtube. However, we note that existing second-order calculations in the Lorenz gauge~\cite{Pound:2019lzj,Miller:2020bft,Warburton:2021kwk,Wardell:2021fyy,Miller:2023ers,Upton:2025bja} solved a slightly different set of equations. Our calculations make direct use of the Lorenz-gauge effective source from those references, meaning we require the relationship between the two sets of equations. We describe that relationship in Appendix~\ref{sec:Lorenz conversion}.

For conciseness, we will use $8\pi T^{(2)\rm eff}_{\mu\nu}$ to denote the right-hand side of the EFE even outside the worldtube, in the regions where both the puncture and $T^{(2)}_{\mu\nu}$ vanish. The entire physical second-order source in those regions, $-(G^{(2,0)}_{\mu\nu}[h^{(1)},h^{(1)}] +G^{(1,1)}_{\mu\nu}[h^{(1)}])$, is then considered an ``effective stress-energy tensor''.

\subsection{Second-order Teukolsky equation}
\label{sec:second-order-teukolsky}

Starting from the second-order EFE~\eqref{tt EFE2R}, we can immediately construct a second-order Teukolsky equation, whose solution encodes the second-order gravitational-wave content (and therefore the second-order energy flux)~\cite{Green:2019nam,Spiers:2023cip,Spiers:2023mor}. This follows from the fact that, given a linearized EFE 
\beq\label{eq:G10=T}
G^{(1,0)}_{\alpha\beta}[h]=8\pi T_{\alpha\beta}, 
\eeq
for any perturbation $h_{\alpha\beta}$ and source $T_{\alpha\beta}$, we can always construct an associated spin-weight $-2$ Teukolsky equation 
\beq\label{eq:Opsi4=S}
\hat\O_4[\psi_{4L}]=S_4, 
\eeq
where $\hat\O_4$ is a linear (hyperbolic) operator recalled below. Here $\psi_{4L}$ is a linearized Weyl scalar constructed from $h_{\alpha\beta}$ through application of another linear operator, $\hat\T_4$:
\beq
\psi_{4L}[h]=\hat\T_4[h]. 
\eeq
Likewise, the source $S_4$ is constructed from the source $T_{\alpha\beta}$ through application of a linear operator $\hat\S_4$:
\beq
S_4 = 8\pi\hat{\cal S}_4[T].
\eeq
The fact that Eq.~\eqref{eq:G10=T} implies Eq.~\eqref{eq:Opsi4=S} is an immediate consequence of Wald's operator identity $\hat\S_4\hat\E=\hat\O_4{\cal T}_4$ ~\cite{Wald:1978vm,Pound:2021qin}, where $\hat\E$ is shorthand for the Einstein operator $G^{(1,0)}_{\alpha\beta}$.\footnote{It might not be obvious that this standard identity remains valid when we omit $\dot{\cal J}_I$ terms in the operators, as we do for all operators in this section. However, omitting $\dot{\cal J}_I$ terms simply puts the operator identity in its frequency-domain form.}

The linear operators $\hat\O_4$, $\hat\T_4$, and $\hat\S_4$ are most succinctly written in terms of Geroch-Held-Penrose (GHP) quantities~\cite{Geroch:1973am,Pound:2021qin}, which are defined in terms of a Newman-Penrose null tetrad $\{l^\alpha,n^\alpha,m^\alpha,\bar{m}^\alpha\}$. These tetrad legs satisfy the orthonormality conditions
\begin{equation}
l^\alpha n_\alpha=-1, \quad m^\alpha\bar{m}_\alpha=1,
\end{equation}
with all other combinations of contracted tetrad legs vanishing. The GHP expressions for the operators are given in Ref.~\cite{Pound:2021qin}, for example, but we will not need them here; we only explicitly require the mode-decomposed versions, which we provide in Sec.~\ref{sec:sph_harm} below. In the current section we only stress that we always use leading-order multiscale versions of the operators, making the replacement $\nabla_\mu\to\nabla^{(0)}_\mu$.  

Using these operators, we write the Teukolsky equation associated with Eq.~\eqref{tt EFE2} as
\begin{align}
\hat\O_4[\psi^{(2)}_{4L}] &= \hat\S_4\!\left[8\pi T^{(2)}_{\mu\nu} - G^{(2,0)}_{\mu\nu}[h^{(1)},h^{(1)}]-G^{(1,1)}_{\mu\nu}[h^{(1)}]\right]\nonumber\\
&\coloneqq S^{(2)}_4,\label{Opsi}
\end{align}
where $\psi^{(2)}_{4L}\coloneqq \hat\T_4[h^{(2)}]$. Analogously, the Teukolsky equation associated with Eq.~\eqref{tt EFE2R} is
\beq
\hat\O_4\bigl[\psi^{(2)\cal R}_{4L}\bigr] = \hat\S_4\bigl[8\pi T^{(2)\rm eff}_{\mu\nu}\bigr] \coloneqq S^{(2)}_{4,\rm eff},\label{OpsiRv1}
\eeq
where $\psi^{(2)\cal R}_{4L}\coloneqq \hat\T_4[h^{(2)\cal R}]$. This last equation can also be obtained by introducing a puncture 
\beq
\psi^{(2)\cal P}_{4L}\coloneqq {\cal T}_4[h^{(2)\cal P}],
\label{eq:worldtube_puncture_definition}
\eeq
which implies
\beq
\hat\O_4\bigl[\psi^{(2)\cal R}_{4L}\bigr] = S_4^{(2)} - \hat\O_4\bigl[\psi^{(2)\cal P}_{4L}\bigr] = S^{(2)}_{4,\rm eff}.\label{OpsiRv2}
\eeq
Consistency between Eqs.~\eqref{OpsiRv1} and \eqref{OpsiRv2} follows from applying the operator identity $\hat\S_4\hat\E=\hat\O_4{\cal T}_4$ to $h^{(2)\cal P}_{\alpha\beta}$.

We note that $\psi^{(2)}_{4L}$ represents only the linear part of the second-order Weyl scalar $\psi^{(2)}_4$. The complete second-order perturbation to the Weyl scalar divides into linear, quadratic, and slow-evolution parts,
\begin{align}\label{eq:psi42deconstructed}
\psi_4^{(2)}
&=\psi_{4L}^{(2)}+\psi_{4Q}^{(2)} + \psi^{(2)}_{4\cal V},
\end{align}
where $\psi_{4Q}^{(2)}$ is quadratic in $h^{(1)}_{\alpha\beta}$ and in perturbations to the tetrad legs, and $\psi^{(2)}_{4\cal V}$ is linear in $\vec\partial_{\cal V}h^{(1)}_{\alpha\beta}$; this last piece is obtained by replacing $\nabla^{(0)}_\alpha$ with $\nabla_\alpha$ in $\hat\T_4$ and then reading off the linear-in-$\vec\partial_{\cal V}$ term in the result. Since Eq.~\eqref{Opsi} only yields the linear piece, it is referred to as the \emph{reduced} second-order Teukolsky equation~\cite{Spiers:2023cip}. Crucially, in an asymptotically regular gauge, at large $r$ the $1/r$ term in $\psi^{(2)}_{4L}$ agrees with the $1/r$ term in $\psi^{(2)}_{4}$, meaning $\psi^{(2)}_{4L}$ contains the complete gravitational-wave content in $\psi^{(2)}_{4}$~\cite{Campanelli:1998jv,Spiers:2023cip}. We refer to Ref.~\cite{Spiers:2023cip} and Appendix~C of Ref.~\cite{Bourg:2025lpd} for further discussion. 

Finally, we stress that unlike the first-order Weyl scalar $\psi^{(1)}_4$, $\psi^{(2)}_{4L}$ is not gauge invariant (nor is $\psi^{(2)}_{4}$). More precisely, while $\psi^{(2)}_{4L}$ is invariant under linear second-order transformations, $h^{(2)}_{\alpha\beta}\to h^{(2)}_{\alpha\beta} + {\cal L}_\xi g_{\alpha\beta}$, it is \emph{not} invariant under first-order transformations $h^{(1)}_{\alpha\beta}\to h^{(1)}_{\alpha\beta} + {\cal L}_\xi g_{\alpha\beta}$. Such first-order transformations have a knock-on second-order effect,
\begin{equation}\label{eq:h2-gauge-transform}
    h^{(2)}_{\alpha\beta}\to h^{(2)}_{\alpha\beta} + {\cal L}^{(0)}_\xi h^{(1)}_{\alpha\beta} + \frac{1}{2}{\cal L}_\xi^{(0)}{\cal L}_\xi^{(0)} g_{\alpha\beta}+\Lie_\xi^{(1)}g_{\alpha\beta},
\end{equation}
that alters $\psi^{(2)}_{4L}$: 
\begin{equation}\label{eq:psi4L2GaugeTransform}
    \psi^{(2)}_{4L}\to \psi^{(2)}_{4L} + \hat\T_4\left[{\cal L}^{(0)}_\xi h^{(1)} + \frac{1}{2}{\cal L}^{(0)}_\xi {\cal L}^{(0)}_\xi g + \Lie_\xi^{(1)}g\right].
\end{equation}
Here, we have introduced the leading and first subleading terms in the multiscale expansion of the Lie derivative, $\Lie_\xi = \Lie_\xi^{(0)} + \varepsilon \Lie_\xi^{(1)}$. For a generic symmetric tensor $A_{\alpha\beta}$ we have the explicit expressions
\begin{subequations}
\label{eq:multiscale-Lie}
\begin{align}
{\cal L}^{(0)}_\xi A_{\alpha\beta} &= \xi^\gamma\partial^{(0)}_\gamma A_{\alpha\beta} + 2\partial^{(0)}_{(\alpha} \xi^\gamma A_{\beta)\gamma},\\
{\cal L}^{(1)}_\xi A_{\alpha\beta} &= s_\gamma\xi^\gamma\vec\partial_{\cal V} A_{\alpha\beta} + 2s_{(\alpha}\vec\partial_{\cal V}  \xi^\gamma A_{\beta)\gamma},
\end{align}
\end{subequations}
where $\partial^{(0)}_\alpha \coloneqq e^i_\alpha \partial_i -i\omega_m s_\alpha$. This will be important in our treatment, which relies on transforming $h^{(1)}_{\alpha\beta}$ to a Bondi-Sachs gauge. We refer to Ref.~\cite{Spiers:2023cip} for a detailed analysis of the properties of $\psi^{(2)}_{4L}$ and $\psi^{(2)}_{4}$.


\subsection{Regularity conditions and choice of tetrad}\label{sec:regularity and tetrad}

In the next sections, a key point will be the regularity of the source at the horizon and infinity. The degree of regularity depends on the choice of tetrad. We use the Carter tetrad~\cite{Pound:2021qin} in most of our computations, but in this section we remain agnostic to the choice of tetrad.

To assess regularity at the horizon, we write the tetrad in advanced Eddington-Finkelstein coordinates $(v,r,\theta,\phi)$~\cite{Spiers:2023mor}:%
\begin{align}
    l^\alpha &= \frac{\gamma}{\sqrt{2 f}}(2,f,0,0),\label{l def}\\
    n^\alpha &= \frac{1}{\gamma \sqrt{2f}}(0,-f,0,0),\label{n def}
\end{align}
where $f=1-2M/r$ and $\gamma$ is freely specified. $\gamma=1$ for the Carter tetrad~\cite{Pound:2021qin}, $\gamma=\sqrt{2/f}$ for the Kinnersley tetrad~\cite{Kinnersley:1969zza}, and $\gamma=\sqrt{f/2}$ for the Hartle-Hawking tetrad~\cite{Hawking:1972hy}. Therefore, for a generic tensor $T_{\alpha\beta}$ that is smooth at the horizon, the tetrad components behave as%
\begingroup\allowdisplaybreaks%
\begin{subequations}\label{eq:near-horizon scaling}%
\begin{align}
    T_{ll} &= \O(\gamma^2 f^{-1}),\\
    T_{ln} &= \O(f^0),\\
    T_{lm} &= \O(\gamma f^{-1/2}),\\
    T_{nn} &= \O(\gamma^{-2} f),\\
    T_{nm} &= \O(\gamma^{-1} f^{1/2}),\\
    T_{mm} &= \O(f^0).
\end{align}
\end{subequations}\endgroup%

Given the form of the operator ${\cal S}_4$, the above scalings imply that for any smooth source $T_{\alpha\beta}$ in a linearized EFE $G^{(1,0)}_{\alpha\beta}[h]=8\pi T_{\alpha\beta}$, the corresponding Teukolsky source behaves as
\begin{align}
     \hat\S_4[T] &= \O(\gamma^{-2}f).
\end{align}
We emphasize that $T_{\alpha\beta}$ here is a generic source term in a linearized EFE, not necessarily a stress-energy tensor; in our case, it will be the right-hand side of the second-order EFE~\eqref{tt EFE2}.

Turning now to behavior near future null infinity, we write the tetrad in retarded coordinates $(u,r,\theta,\phi)$:
\begin{align}
    l^\alpha &= \frac{\gamma}{\sqrt{2 f}}(0,f,0,0),\\
    n^\alpha &= \frac{1}{\gamma \sqrt{2f}}(2,-f,0,0).
\end{align}
We note that outgoing wave solutions behave as 
\begin{align}
    \psi^{(2)}_{4L} &\sim \frac{A^{(2)}_4(u,\theta,\phi)}{r}
\end{align}
in an asymptotically regular gauge~\cite{Spiers:2023cip,Spiers:2026yqx}. For these to be freely propagating waves, they must satisfy the homogeneous Teukolsky equation up to $\hat\O_4[\psi^{(2)}_{4L}]=\O(r^{-4})$
~\cite{Spiers:2023cip}, implying the falloff requirement
\begin{align}
     \hat\S_4[T] &= \O(r^{-4}).
\end{align}
These imply falloff requirements on $T_{\alpha\beta}$, derived in Ref.~\cite{Spiers:2023cip} and discussed in Sec.~\ref{sec:compactification} below. 

\subsection{Expansion in spin-weighted spherical harmonics}
\label{sec:sph_harm}

To take advantage of the background spacetime's spherical symmetry, we expand all functions in a basis of spin-weighted spherical harmonics ${}_\s Y_{\l\m}$~\cite{Newman:1966ub}. Each tetrad component of the $n$th-order metric perturbation~\eqref{Fourier} is expanded as 
\beq
h^{(n)}_{ab} = \sum_{\l\geq |\s|}\sum_{\m =-\l}^\l h^{(n)}_{ab,\l\m }(r,{\cal J}_I)\,{}_\s Y_{\l\m }\,e^{-i\m\phi_p}, \label{eq:hYslm}
\eeq
where Latin indices from the beginning of the alphabet denote tetrad indices, as in $e^\alpha_{a}=(l^\alpha,n^\alpha,m^\alpha,\bar m^\alpha)$, and $\s$ is the spin weight of the tetrad component (where each $m$ component adds 1 to the spin weight and each $\mb$ component reduces it by 1). Analogously, the spin-weight $-2$ Weyl scalar $\psi^{(2)}_{4L}$ is expanded as
\begin{equation}\label{eq:psi24L mode expansion}
    \psi^{(2)}_{4L} = \sum_{\l\geq2}\sum_{\m=-\l}^\l \psi^{(2)}_{4,\l\m}(r,{\cal J}_I)\, {}_{-2}Y_{\l\m}\, e^{-i\m\phi_p}
\end{equation}
and the source as
\begin{align}
    S^{(2)}_4 &= \sum_{\l\geq2}\sum_{\m=-\l}^\l S^{(2)}_{4,\l\m}(r,{\cal J}_I)\, {}_{-2}Y_{\l\m}\, e^{-i\m\phi_p},
\end{align}
with corresponding expansions of the residual field $\psi^{{\cal R}(2)}_{4L}$ and effective source $S^{(2)}_{4,\rm eff}$.

With these expansions, the Teukolsky equation~\eqref{Opsi} is reduced to decoupled radial ordinary differential equations for the $\l\m$ mode coefficients~\cite{Spiers:2023mor}:
\begin{align}\label{eq:Opsi2=S2 multiscale modes}
    \hat\O_{4,\l\m}\,\psi^{(2)}_{4,\l\m} &= S^{(2)}_{4,\l\m},    
\end{align}
with
\begin{equation}\label{eq:GHPO modes}
     \hat\O_{4,\l\m} = \left(\thorn_\m+3\rho\right)\left(\thornp_\m-\rho'\right) +\frac{{}_{-2}\lambda_{\l}}{\ \; 2r^2}+\frac{3M}{r^3}.
\end{equation}
Here $\thorn_\m$ and $\thornp_\m$ are mode-decomposed GHP derivatives, which we can reduce to
\begin{align}
\thorn_\m &\coloneqq \frac{\gamma}{\sqrt{2f}}  \left[f\partial_r - i \omega_\m (1-H)\right]-2\b \epsilon,\label{eq:thorn modes}\\
\thornp_\m &\coloneqq -\frac{1}{\sqrt{2f}\gamma} \left[f\partial_r+i \omega_\m (1+H)\right] + 2\b \epsilon'.\label{eq:thornp modes}
\end{align}
The derivative of the height function appearing here,
\begin{equation}
H\coloneqq dk/dr^*, 
\end{equation}
is defined such that $H=1$ when $s=u$, and $H=-1$ when $s=v$. 
The quantity $\gamma$ in $\thorn_\m$ and $\thorn'_\m$ is the boost parameter introduced in Eq.~\eqref{l def}, and the parameter $\b$ is the boost weight of the object acted upon, where a boost-weighted tensor $\eta$ transforms as $\eta\to \gamma^\b\eta$ under the boost $(l^\alpha, n^\alpha) \to (\gamma l^\alpha,\gamma^{-1}n^\alpha)$; $\thorn$ raises boost weight by 1, while $\thorn'$ reduces it by 1. The spin coefficients appearing in the above expressions are 
\begin{align}
    \rho&=-l^r/r, \qquad \rho'=-n^r/r,\\
    \epsilon &= \frac{1}{2}\partial_r l^r, \qquad \epsilon' = \frac{1}{2}\partial_r n^r,
\end{align}
and the angular eigenvalue is
\beq\label{eq:slambdaell}
{}_\s\lambda_{\l}\coloneqq \l(\l+1)-\s(\s+1).
\eeq

In Eq.~\eqref{eq:Opsi2=S2 multiscale modes}, the field modes $\psi^{(2)}_{4,\l\m}$ and source modes $S^{(2)}_{4,\l\m}$ are related to the metric perturbation modes $h^{(2)}_{ab,\l\m}$ and effective stress-energy tensor modes $T^{(2)\rm eff}_{ab,\l\m}$ using the mode-decomposed versions of the operators $\hat\T$ and $\hat\S$. Adapting those mode-decomposed operators from Ref.~\cite{Spiers:2023mor}, we can write
\begin{align}\label{eq:GHPT modes}
\hat\T_{4,\l\m}[h] &= -\frac{\mu^\l_{\ 2}}{4r^2} h^{\l \m}_{nn} + \frac{\sqrt{{}_{-2}\lambda_\l}}{\sqrt{2}r} (\thorn'_\m - \rho') h^{\l\m}_{n\mb} \no\\
&\quad -\frac{1}{2}(\thorn'_\m-\rho')^2h^{\l\m}_{\mb\mb},
\end{align}
and
\begin{align}\label{eq:GHPS modes}
\hat\S_{4,\l\m}[T] &= -\frac{\mu^\l_{\ 2}}{4r^2} T^{\l\m}_{nn}+\frac{\sqrt{{}_{-2}\lambda_\l}}{\sqrt{2}r}(\thorn'_\m-3\rho')T^{\l\m}_{n\mb}\no\\
		&\quad -\frac{1}{2}(\thorn_\m'-5\rho')(\thorn_\m'-\rho')T^{\l\m}_{\mb\mb},
\end{align}
where we follow the \textsc{PerturbationEquations} package~\cite{PerturbationEquations} by defining
\begin{equation}
    \mu^\l_{\ \s}\coloneqq\sqrt{\frac{(\l+|\s|)!}{(\l-|\s|)!}}
\end{equation}
(denoted $\lambda_{\l,\s}$ in Ref.~\cite{Spiers:2023mor}). Note we freely place mode indices up or down as convenient.

\subsection{Master equation}\label{sec:master equation}

In practice, one often changes the field variable to rewrite Eq.~\eqref{eq:Opsi2=S2 multiscale modes} in the form of the spin-weight-$\s$ Teukolsky master equation: 
\begin{equation}\label{eq:master eqn}
    {}_\s\hat\O_{\l\m}\;{}_{\s}\psi^{(2)}_{\l\m} = {}_{\s}S^{(2)}_{\l\m},
\end{equation}
where
\begin{align}
{}_\s\hat\O_{\l\m} &\coloneqq \Delta\partial_r^2 +2\left[(\s+1)(r-M)+ir^2\omega_\m H\right]\partial_r  \nonumber\\
    &\quad + \frac{\omega_\m}{f} \left[r^2 \omega_\m(1-H^2)-2 i\s M(1-H) + i r^2 H'\right]\nonumber\\
    &\quad + 2 i r\omega_\m\left[\s+(\s+1)H\right]-{}_\s\lambda_\l\,,
\end{align}
with 
\begin{equation}
    \Delta\coloneqq r^2f,
\end{equation}
$H'\coloneqq \partial_{r^*}H$, and the normalization choices of Ref.~\cite{Pound:2021qin}. Equation~\eqref{eq:master eqn}, with $\s=-2$, is obtained from Eq.~\eqref{eq:Opsi2=S2 multiscale modes} via multiplying the equation by $-r^6f\gamma^2$ and converting from $\psi^{(2)}_{4,\l\m}$ to the spin-weight $-2$ master variable ${}_{-2}\psi^{(2)}_{\l\m}$. Unlike $\psi^{(2)}_{4,\l\m}$, the master variable is independent of the choice of tetrad; the two are related by~\cite{Pound:2021qin}
\begin{align}
{}_{-2}\psi^{(2)}_{\l\m} = r^4(\gamma/\gamma_{\rm K})^2\psi^{(2)}_{4,\l\m}, 
\end{align}
where $\gamma_{\rm K}=\sqrt{2/f}$ is the boost parameter in the Kinnersley tetrad. Similarly, the source in Eq.~\eqref{eq:master eqn} is related to the source in Eq.~\eqref{eq:Opsi2=S2 multiscale modes} by
\begin{align}\label{eq:-2Slm}
{}_{-2}S^{(2)}_{\l\m} = - 2r^6(\gamma/\gamma_{\rm K})^2S^{(2)}_{4,\l\m}.
\end{align}
The spin-weight $+2$ case of Eq.~\eqref{eq:master eqn} is obtained via analogous transformations from the Teukolsky equation for the spin-weight $+2$ Weyl scalar $\psi^{(2)}_{0L}$; concretely, ${}_2\psi^{(2)}_{\l\m} = (\gamma_{\rm K}/\gamma)^2\psi^{(2)}_{0,\l\m}$, where $\psi^{(2)}_{0,\l\m}$ are the coefficients in an expansion of $\psi^{(2)}_{0L}$ analogous to Eq.~\eqref{eq:psi24L mode expansion}, and ${}_2 S^{(2)}_{\l\m}=-2r^2(\gamma_{\rm K}/\gamma)^2 S^{(2)}_{0,\l\m}$.

Our numerical calculations in this paper focus exclusively on the spin $-2$ case. However, in order to explore the relevance of spin weight, we will use the generic-$\s$ master equation in our analysis of asymptotics and infrared divergences in the next section.


\section{Compactification, infrared divergences, and the Bondi-Sachs gauge}
\label{sec:compactification}
Rather than directly solving the Teukolsky equation~\eqref{eq:Opsi2=S2 multiscale modes} or \eqref{eq:master eqn}, we follow Refs.~\cite{PanossoMacedo:2022fdi,Leather:2024mls} by compactifying the radial coordinate and factoring out the expected asymptotic behavior of the field variable. This has two main advantages: it allows one to extract the waveforms at \scri and $\mathscr{H}^+$ directly from the field values at the boundaries of the compactified domain, as opposed to extrapolating; and, when combined with hyperboloidal slicing, it can bypass the need to construct large-$r$ and near-horizon expansions as boundary conditions.

That second advantage is sometimes described as boundary conditions not being required at all in this framework. However, the more accurate statement is that regularity at the boundaries suffices as a boundary condition---in the case that a regular solution can be obtained. It is not obvious a priori how this idea applies to second-order self-force calculations, since these calculations typically involve infrared-divergent solutions even when using hyperboloidal slicing~\cite{Pound:2015wva,Cunningham:2024dog,Spiers:2026yqx}. Previous Lorenz-gauge calculations have overcome those divergences using matched asymptotic expansions, matching the multiscale expansion in a ``near zone'' to a self-force-multipolar-post-Minkowskian (SF-MPM) expansion toward~\scri~\cite{Pound:2015wva,Cunningham:2024dog} and to an analogous expansion toward~$\mathscr{H}^+$.

In this section, we study the compactified second-order Teukolsky equation and how ill behavior of the second-order source can spoil the usual approach that relies on regularity. However, we also show that the most pernicious divergences arising in Lorenz-gauge calculations do not arise in the Teukolsky equations, and that no problems arise at the horizon. We then describe how transforming $h^{(1)}_{\mu\nu}$ to the Bondi-Sachs gauge, as proposed in Refs.~\cite{Spiers:2023cip,Spiers:2026yqx}, eliminates the remaining divergences that otherwise do arise in the second-order Teukolsky equation. Following that transformation, one can directly employ the usual regularity-based approach from Refs.~\cite{PanossoMacedo:2022fdi,Leather:2024mls}.

To complement this analysis, in Appendix~\ref{sec:infrared Lorenz} we revisit the infrared divergences found in Lorenz-gauge calculations, showing how they prevent one from following the regularity-based approach: the ``near-zone'' equations fundamentally cannot determine the hereditary effects (memory integrals over the past history) that arise at large distances in the SF-MPM expansion, and it is impossible to obtain correct boundary conditions without these terms.

\subsection{Compactification}

For our compactified radial coordinate we adopt 
\begin{equation}
    \sigma \coloneqq \frac{2M}{r},
    \label{eq:sigma_coordinate}
\end{equation}
which runs from 0 at \scri to 1 at $\mathscr{H}^+$. We next factor out the $r$ dependence of a freely propagating, outgoing (ingoing) wave at \scri ($\mathscr{H}^+$). For the physical variables $\psi_4$ and $\psi_0$, free outgoing waves toward \scri behave as $\sim 1/r$ and $\sim 1/r^5$, respectively, meaning the master variables behave as ${}_\s\psi_{\l\m}\sim r^{-(2\s+1)}\propto \sigma^{2\s+1}$. Free ingoing waves toward $\mathscr{H}^+$ behave as $\psi_4\sim \gamma^{-2}f$ and $\psi_0\sim \gamma^{2}f^{-1}$, as suggested by Eq.~\eqref{eq:near-horizon scaling}; this translates to ${}_\s\psi_{\l\m}\sim f^{-\s}= (1-\sigma)^{-\s}$. Hence, we adopt the ansatz
\begin{equation}
    {}_\s \psi^{(2)}_{\l\m}(\sigma,{\cal J}_I) = (1-\sigma)^{-\s}\sigma^{2\s+1} {}_\s \tilde \psi^{(2)}_{\l\m}(\sigma,{\cal J}_I),
    \label{eq:master_rescaling}
\end{equation}
such that the radial function ${}_\s\tilde\psi^{(2)}_{\l\m}$ goes to a nonzero constant at each boundary if it represents a free wave there. 

At the level of the physical Weyl scalars, the above ansatz corresponds to 
\begin{equation}\label{eq:psi4 to R}
\psi^{(2)}_{4,\l\m}=\frac{4}{r}\frac{(\gamma_{\rm HH}/\gamma)^2}{(2M)^3}{}_{-2}\tilde\psi^{(2)}_{\l\m} 
\end{equation}
and 
\begin{equation}\label{eq:psi0 to R}
\psi^{(2)}_{0,\l\m}=\frac{(\gamma/\gamma_{\rm HH})^2(2M)^{5}}{4r^5}{}_2 \tilde\psi^{(2)}_{\l\m}, 
\end{equation}
where $\gamma_{\rm HH}$ is the boost factor in the Hartle-Hawking tetrad. So, up to unimportant numerical factors, the variable ${}_{\s}\tilde\psi^{(2)}_{\l\m}$ simply represents the spin-$\s$ Weyl scalar in the Hartle-Hawking tetrad, with the dominant falloff at large~$r$ factored out.

For brevity, in the rest of this section we suppress mode indices and adopt the shorthands
\beq\label{eq:scaled variables}
\tilde\psi\coloneqq {}_\s \tilde\psi^{(2)}_{\l\m} \quad \text{and} \quad \tilde S\coloneqq (1-\sigma)^\s\sigma^{-(2\s+1)} {}_\s S^{(2)}_{\l\m}.
\eeq
The equation for the rescaled master variable is then
\beq
(1-\sigma) \sigma^2 \partial^2_\sigma \tilde\psi + {\cal B}\,\partial_\sigma \tilde\psi + {\cal A}\tilde\psi = \tilde S,\label{eq:teukolsky_compactified}
\eeq
where we follow the convention of Ref.~\cite{Miller:2023ers} in naming the coefficients ${\cal A}$ and ${\cal B}$. These coefficients are given by
\begin{align}
{\cal A} &= \frac{\varpi^2 \left(1-H^2\right)}{(1-\sigma) \sigma^2} -{}_\s\lambda_\l - \sigma(\s+1) - i\varpi \partial_\sigma H 
\label{eq:teukolsky_a_coefficient}\\
&\quad + \frac{i\varpi \s\, [2(1-H) -\sigma(3-H)]}{(1-\sigma) \sigma },\\
{\cal B} &= 2\sigma(\s+1) - \sigma^2(\s+3) - 2i \varpi H,
\label{eq:teukolsky_b_coefficient}
\end{align}
with
\begin{equation}
   \varpi \coloneqq 2 M \omega_\m. 
\end{equation}

Now, if both $\tilde\psi$ and $\tilde S$ are smooth at the boundaries ($\sigma=0$ and $\sigma=1$), then the $\partial_\sigma^2$ term in Eq.~\eqref{eq:teukolsky_compactified} vanishes there, such that the field equation reduces to a relationship between $\tilde\psi|_{\sigma=0,1}$ and $\partial_\sigma\tilde\psi|_{\sigma=0,1}$. Although it might appear that ${\cal A}$ is singular at the boundaries, the fact that $H\to\pm1$ there ensures the factors of $1/\sigma$ and $1/(1-\sigma)$ are canceled. Explicitly, Eq.~\eqref{eq:teukolsky_compactified} reduces to
\begin{equation}\label{eq:sigma=0 eqn}
    -\bigl[2i\varpi\partial_\sigma\tilde\psi+({}_\s\lambda_\l+2i\varpi\s)\tilde\psi \bigr]_{\sigma=0} = \tilde S\bigr|_{\sigma=0}
\end{equation}
at \scri and to
\begin{multline}\label{eq:sigma=1 eqn}
    \bigl[(s - 1 + 2 i \varpi ) \partial_\sigma\tilde\psi \\
    - ({}_\s\lambda_\l +\s +1 - 4 i\varpi \s )\tilde\psi\bigr]_{\sigma=1} = \tilde S\bigr|_{\sigma=1}
\end{multline}
at $\mathscr{H}^+$. More generally, smoothness is sufficient but not necessary to arrive at these equations; if $\tilde\psi$ is at least $C^1$ and its second derivative blows up less rapidly than $1/(1-\sigma)$ and $1/\sigma^2$ at the two boundaries, then the above equations follow.

Therefore, if the fields are assumed to be sufficiently regular, the field equation itself enforces a boundary condition at each boundary. In this case, no additional conditions need to be found. This contrasts with the more traditional variation-of-parameters method, where the homogeneous solutions are found by first constructing large-$|r^*|$ expansions as boundary conditions; see Ref.~\cite{PanossoMacedo:2022fdi} and Sec.~\ref{sec:mixed method} for discussion.

In the next section, we analyze the irregularity of the source $\tilde S$ and how its lack of smoothness can spoil this method. 

\subsection{Asymptotic analysis}\label{infrared divergences}

For simplicity, we will analyze the behavior of the compactified field equations in $s=u$ slicing toward \scri and $s=v$ toward $\mathscr{H}^+$. However, we stress that our conclusions carry over immediately to any slicing that asymptotes to these sufficiently rapidly. In $u$ slicing, the coefficients in Eq.~\eqref{eq:teukolsky_compactified} reduce to
\begin{align}
{\cal A}_{[u]} &=  -{}_\s\lambda_\l - \sigma(\s+1)   - \frac{2i\varpi \s\,}{1-\sigma},\label{eq:A[u]}\\
{\cal B}_{[u]} &= 2\sigma(\s+1) - \sigma^2(\s+3) - 2i \varpi;\label{eq:B[u]}
\end{align}
and in $v$ slicing, to 
\begin{align}
{\cal A}_{[v]} &=  -{}_\s\lambda_\l - \sigma(\s+1) + \frac{4i\varpi \s\,}{\sigma},\label{eq:A[v]}\\
{\cal B}_{[v]} &= 2\sigma(\s+1) - \sigma^2(\s+3) + 2i \varpi. \label{eq:B[v]}
\end{align}

In this section, we assume the first-order field is in the Lorenz gauge, before turning to the Bondi-Sachs gauge in the next section. 

\subsubsection{Behavior at $\mathscr{H}^+$}

There are two sources in Eqs.~\eqref{tt EFE2}, and hence in Eq.~\eqref{eq:teukolsky_compactified}, to consider at the horizon: the quadratic source $G^{(2,0)}_{\alpha\beta}$ and the slow-evolution source $G^{(1,1)}_{\alpha\beta}$. So long as we use $v$ slicing in a neighborhood of the horizon, both of these sources are well behaved, and the ``compactification + regularity'' prescription leading to Eq.~\eqref{eq:sigma=1 eqn} applies.

To see this, first note that if $h^{(1)}_{\alpha\beta}$ is smooth at $\mathscr{H}^+$, as it is for us~\cite{Akcay:2010dx}, then $G^{(2,0)}_{\alpha\beta}$ is automatically smooth. $\hat\S_4[G^{(2,0)}_{\alpha\beta}]$ and $\hat\S_0[G^{(2,0)}_{\alpha\beta}]$ are also then automatically smooth at $\mathscr{H}^+$ so long as they are computed in a tetrad that is smooth. Our conversion from $S^{(2)}_{4,\l\m}$ and $S^{(2)}_{0,\l\m}$ in Eq.~\eqref{eq:scaled variables} corresponds to transforming to the Hartle-Hawking tetrad, which is horizon-regular; see Eqs.~\eqref{l def}--\eqref{eq:near-horizon scaling}. Hence, the part of $\tilde S$ constructed from $G^{(2,0)}_{\alpha\beta}$ is smooth at $\sigma=1$.

The same reasoning applies to $G^{(1,1)}_{\alpha\beta}$, as given in Eq.~\eqref{G11}, so long as $s^\alpha$ and $\vec\partial_{\cal V}h^{(1)}_{\alpha\beta}$ are smooth at $\mathscr{H}^+$. For the former, we observe that $s_\alpha$ is indeed smooth in $v$ slicing, with components $s_\alpha = (1,0,0,0)$ and $s^\alpha = (0,1,0,0)$ in ingoing Eddington-Finkelstein coordinates $(v,r,\theta,\phi)$. For $\vec\partial_{\cal V}h^{(1)}_{\alpha\beta}$, we recall that $h^{(1)}_{\alpha\beta}$ has the form $\sum_{m=-\infty}^\infty h^{(1)}_{\alpha\beta,\m}({\cal J}_I,x^i)e^{-i\m\phi_p}$, where the Eddington-Finkelstein components $h^{(1)}_{\alpha\beta,\m}$ are smooth functions of $r$ at $\mathscr{H}^+$. In $v$ slicing, these components can be expanded in a Taylor series in $(1-\sigma)$~\cite{Miller:2020bft}:
\begin{equation}\label{eq:h1 near horizon}
    h^{(1)}_{\alpha\beta,\m} = \sum_{j\geq0} \frac{(-1)^j}{j!}(1-\sigma)^j\partial_\sigma^jh^{(1)}_{\alpha\beta,\m}|_{\sigma=1}.
\end{equation}
Hence, $\vec\partial_{\cal V}h^{(1)}_{\alpha\beta,\m}$ is also smooth at $\sigma=1$, with an expansion
\begin{equation}
    \vec\partial_{\cal V}h^{(1)}_{\alpha\beta,\m} = \sum_{j\geq0} \frac{(-1)^j}{j!}(1-\sigma)^j\vec\partial_{\cal V}\partial_\sigma^jh^{(1)}_{\alpha\beta,\m}|_{\sigma=1}, 
\end{equation}
and we can conclude that $G^{(1,1)}_{\alpha\beta}$ is likewise smooth.

This argument assumes that $s=v$ in an open neighborhood of $\mathscr{H}^+$. If $s$ approaches $v$ too slowly, singularities can arise. To illuminate this, we write $s$, as defined in Eq.~\eqref{eq:s def}, as $s = v - (k+r^*)$. The normal vector is then 
$s_\alpha = (1,-f^{-1}(H+1),0,0)$ in ingoing Eddington-Finkelstein coordinates. This is only smooth at $\sigma=1$ if $H =-1 + A f$ for some smooth $A$. Commonly used hyperboloidal slicings, such as the ``minimal gauge''~\cite{PanossoMacedo:2023qzp} used in Ref.~\cite{PanossoMacedo:2022fdi}, do satisfy this condition.\footnote{Explicitly, $H=(1-2\sigma^2)=-1+2f(1+\sigma)$ in the minimal gauge, and the combination appearing in Eq.~\eqref{eq:dVh1 near horizon generic} is $k+r^*=4M (\sigma^{-1} - \ln \sigma)$.} However, other singularities can arise in $\vec\partial_{\cal V}h^{(1)}_{\alpha\beta,\m}$. 
If we let $h^{[v]}_{\alpha\beta,\m}$ denote the Fourier mode coefficients in $v$ slicing, then coefficients for a generic height function $k(r^*)$ are given by $h^{(1)}_{\alpha\beta,\m}=e^{-i\m\Omega[k(r^*)+r^*]}h^{[v]}_{\alpha\beta,\m}$~\cite{Miller:2020bft}. Equation~\eqref{eq:h1 near horizon} then becomes 
\begin{equation}
    h^{(1)}_{\alpha\beta,\m} = e^{-i\m\Omega[k(r^*)+r^*]}\sum_{j\geq0} \frac{(-1)^j}{j!}(1-\sigma)^j\partial_\sigma^jh^{[v]}_{\alpha\beta,\m}|_{\sigma=1}, 
\end{equation}
implying
\begin{align}\label{eq:dVh1 near horizon generic}
    \vec\partial_{\cal V}h^{(1)}_{\alpha\beta,\m} &= e^{-i\m\Omega[k(r^*)+r^*]}\no\\
    &\quad\times\sum_{j\geq0} \frac{(-1)^j}{j!}(1-\sigma)^j\Bigl\{\vec\partial_{\cal V}\partial_\sigma^jh^{(1)}_{\alpha\beta,\m}|_{\sigma=1}\no\\
    &\quad -i\m(\vec\partial_{\cal V}\Omega)[k(r^*)+r^*]\partial_\sigma^jh^{[v]}_{\alpha\beta,\m}|_{\sigma=1}\Bigr\}.
\end{align}
This is already divergent at the horizon before any additional derivatives are taken in constructing the source, unless $k=-r^*+B$ for some bounded $B$---a condition that is not met in the minimal gauge, for example. Avoiding such ill behavior is one of our motivations for adopting the same $v$-$t$-$u$ slicing used in previous, Lorenz-gauge second-order self-force calculations.

\subsubsection{Behavior at \scri}

The large-$r$ behavior of the Lorenz-gauge source terms was analyzed in Refs.~\cite{Miller:2020bft,Cunningham:2024dog,Spiers:2023cip,Spiers:2026yqx}. Those analyses largely carry over to the present context, but with one significant change: previous analyses never considered $G^{(1,1)}_{\alpha\beta}$ due to the Lorenz-gauge calculations' different formulation of the field equations, discussed in Appendix~\ref{sec:Lorenz conversion}. 

To set the stage, we write the large-$r$ expansion of the Lorenz-gauge tetrad components $h^{(1)}_{ab}$ as
\begin{equation}\label{eq:h1 large r expansion}
    h^{(1)}_{ab} = \frac{h^{\{1\}}_{ab}(\phi_p,{\cal J}_I,\theta,\phi)}{r} + \frac{h^{\{2\}}_{ab}(\phi_p,{\cal J}_I,\theta,\phi)}{r^2} + \O(1/r^3),
\end{equation}
in analogy with Eq.~\eqref{eq:h1 near horizon} and adopting the curly-bracket notation of Ref.~\cite{Spiers:2026yqx}. From Sec.~5.4 of Ref.~\cite{Cunningham:2024dog}, we note the Lorenz-gauge condition implies\footnote{We convert between tetrad and coordinate components in the usual way: $A_a\coloneqq A_\alpha e^\alpha_a$ and $A^a \coloneqq A^\alpha e^a_\alpha$, where $e^a_\alpha\coloneqq g^{ab}g_{\alpha\beta}e^\beta_b$ and $g^{ab}=\left(\begin{smallmatrix}0 & -1 & 0 &0\\ -1 & 0 & 0 & 0\\ 0 & 0 & 0 & 1\\ 0 & 0 & 1 & 0\end{smallmatrix}\right)=g_{ab}$.}
\begin{equation}\label{eq:h1 Lorenz m!=0}
    \bar h^{\{1\}}_{ab}l^b = 0 \quad (\omega_\m\neq0)
\end{equation}
for the $\omega_\m\neq0$ part of $h^{\{1\}}_{ab}$, while the $\omega_m=0$ mode of $h^{\{1\}}_{ab}$ is a pure monopole:
\begin{equation}\label{eq:h1 Lorenz m=0}
    \bar h^{\{1\}}_{ab} = 4M^{(1)}t_a t_b \quad (\omega_\m=0)
\end{equation}
where $M^{(1)}$ is $h^{(1)}_{\mu\nu}$'s contribution to the Bondi mass. Here $t_\alpha = -\frac{1}{\sqrt{2}}(\gamma^{-1}l_\alpha+\gamma n_\alpha)$, with it being understood that only the large-$r$ limit of $\gamma$ is used.

When applying operators to $h^{(1)}_{\alpha\beta}$, we also note that $u$ derivatives do not introduce a power of $1/r$, while all other derivatives and Christoffel symbols do. Therefore, at leading order in $1/r$, we can write the derivative~\eqref{eq:nabla0 modes} as 
\begin{equation}\label{eq:nabla0 large r}
\nabla^{(0)}_\alpha = s_\alpha \Omega\, \partial_{\phi_p} +\O(1/r).     
\end{equation}
Since we use $u$ slicing here, the normal vector $s_\alpha=\partial_\alpha u$ is null, specifically (anti-)aligned with $l_\alpha$: 
\begin{equation}\label{eq:s propto l}
    s^\alpha = - \frac{1}{\gamma}\sqrt{\frac{2}{f}}\;l^\alpha.
\end{equation}
So, for example, the Lorenz-gauge condition, $0=g^{\beta\gamma}\nabla^{(0)}_\gamma \bar h^{(1)}_{\alpha\beta}\approx \Omega s^\beta \partial_{\phi_p}\bar h^{(1)}_{\alpha\beta}$, immediately implies Eq.~\eqref{eq:h1 Lorenz m!=0}.

Using Eqs.~\eqref{eq:h1 large r expansion}--\eqref{eq:nabla0 large r}, one finds that at leading order in $1/r$, the quadratic source $G^{(2,0)}_{\alpha\beta}$ is given by Eq.~(140) of Ref.~\cite{Cunningham:2024dog}. Its tetrad components are
\begin{subequations}\label{eq:G20 asymptotics Lorenz}%
\begin{align}
    G^{(2,0)}_{nn} &= \frac{\Omega^2\Bigl[2\Re \partial_{\phi_p}\bigl(h^{\{1\}}_{mm}\partial_{\phi_p} h^{\{1\}}_{\mb\mb}\bigr)-\bigl|\partial_{\phi_p} h^{\{1\}}_{mm}\bigr|^2\Bigr]}{r^2} \no\\
    &\quad + \O(1/r^3),\\
    G^{(2,0)}_{\mb\mb} &= \frac{4M^{(1)} \Omega^2\partial_{\phi_p}^2 h^{\{1\}}_{\mb\mb}}{r^2} + \O(1/r^3),
\end{align}
\end{subequations}
with all other tetrad components falling off at least as fast as $1/r^{3}$ (except $G^{(2,0)}_{mm}$, which is the conjugate of $G^{(2,0)}_{\mb\mb}$).

The large-$r$ expansion of the other source term, $G^{(1,1)}_{\alpha\beta}$, is readily found from Eq.~\eqref{G11} with Eqs.~\eqref{eq:h1 large r expansion}--\eqref{eq:nabla0 large r}. The first term, $E^{(1)}_{\alpha\beta}$, behaves as in previous Lorenz-gauge analyses: using $s_\alpha s^\alpha=0$ and $\nabla^{(0)}_\alpha s^\alpha = -2/r$, we see the first term vanishes in Eq.~\eqref{eq:E1} and the other two cancel at order $1/r^2$. This implies 
\begin{equation}\label{eq:E1 asymptotics Lorenz}
E^{(1)}_{\alpha\beta}[\bar h^{(1)}] = \O(1/r^3),    
\end{equation}
consistent with previous analyses. However, using Eqs.~\eqref{eq:Z1} and \eqref{eq:h1 Lorenz m!=0}, we find the second term in Eq.~\eqref{G11} decays more slowly:
\begin{multline}
\overline{\nabla^{(0)}_{(\alpha} \vec{\partial}_{\cal V}\bar h^{(1)}_{\beta)\gamma}s^\gamma} = -\frac{i\omega_\m}{2r^2} \bigl(2s_{(\alpha}g_{\beta)}{}^\gamma  - g_{\alpha\beta}s^\gamma \bigr)\vec{\partial}_{\cal V}\bar h^{\{2\}}_{\gamma\delta}s^\delta \\+\O(1/r^3)
\end{multline}
for $\omega_\m\neq0$ modes. Using Eq.~\eqref{eq:h1 Lorenz m=0}, one also finds this term decays as $1/r^2$ for the $\omega_\m=0$ mode, but since that mode of $G^{(1,1)}_{\alpha\beta}$ is monopolar, it does not contribute to the Teukolsky source (which annihilates $\l=0$ and $\l=1$ modes). Finally, the third term in Eq.~\eqref{G11} identically vanishes by virtue of the gauge condition $Z^{(0)}_{\alpha}[\bar h^{(1)}]=0$. Putting these results together and appealing to Eq.~\eqref{eq:s propto l}, we obtain
\begin{subequations}\label{eq:G11 Lorenz large r}
    \begin{align}
        G^{(1,1)}_{nn} &= \frac{2i\omega_\m\vec{\partial}_{\cal V}\bar h^{\{2\}}_{ln}}{\gamma^2r^2} +\O(1/r^3),\\
        G^{(1,1)}_{n\mb} &= \frac{i\omega_\m\vec{\partial}_{\cal V}\bar h^{\{2\}}_{l\mb}}{\gamma^2r^2} +\O(1/r^3)
    \end{align}
\end{subequations}
for $\omega_\m\neq0$ modes, with all other tetrad components decaying as $1/r^3$ (except $G^{(1,1)}_{nm}=\bar G^{(1,1)}_{n\mb}$). Again, it should be understood that only the large-$r$ limit of $\gamma$ is to be used in Eq.~\eqref{eq:G11 Lorenz large r}. This $1/r^2$ falloff of the slow-evolution source appears here for the first time; previous analyses, which only considered $E^{(1)}_{\alpha\beta}$, always found more rapid falloff. 

We write the sum of Eqs.~\eqref{eq:G20 asymptotics Lorenz} and~\eqref{eq:G11 Lorenz large r}---the total source in Eq.~\eqref{tt EFE2R} outside the puncture region---as
\begin{equation}
    8\pi T^{(2)\rm eff}_{ab} = \frac{T^{\{2\}}_{ab}}{r^2} + \frac{T^{\{3\}}_{ab}}{r^3} +\O(1/r^4),
\end{equation}
absorbing the $8\pi$ for convenience. When applying $\hat\S_4$ to this source, referring to Eqs.~\eqref{eq:GHPS modes} and \eqref{eq:thornp modes}, we again note that $u$ derivatives do not introduce a power of $1/r$, while all other terms in $\hat\S_4$ do. At the level of modes, we can then read off
\begin{equation}\label{eq:S24 large r}
    S^{(2)}_{4,\l\m} = \begin{cases}
         \displaystyle\frac{\omega_\m^2\;T^{\{2\}}_{\mb\mb,\l\m}}{\gamma^2r^2} +\O(1/r^3), & \omega_\m \neq 0,\\[.7em]
         - \displaystyle\frac{\mu^\l_{\ 2}\;T^{\{2\}}_{nn,\l\m}}{4r^4} +\O(1/r^5), & \omega_\m = 0.
    \end{cases}
\end{equation}
We emphasize that the components of $T^{\{2\}}_{ab}$ entering Eq.~\eqref{eq:S24 large r} come entirely from the nonlinear source $G^{(2,0)}_{\alpha\beta}$; because of the particular components involved, the newfound $1/r^2$ terms in the slow-evolution source $G^{(1,1)}_{\alpha\beta}$ do not contribute to the leading behavior of the Teukolsky source.  Equation~\eqref{eq:S24 large r} is then equivalent to Eqs.~(3.29) and (3.31) of Ref.~\cite{Spiers:2026yqx}. 

Next consider the source for $\psi^{(2)}_{0L}$.  Since $\hat\S_0$ is given by Eq.~\eqref{eq:GHPS modes} with the interchanges $l^\alpha\leftrightarrow n^\alpha$ and $m^\alpha\leftrightarrow \mb^\alpha$ (and therefore $\thorn_\m\leftrightarrow\thorn_\m'$), we see it involves no $u$ derivatives. This implies $\omega_\m=0$ and $\omega_\m\neq0$ modes both decay as $S^{(2)}_{0,\l\m} = \O(1/r^4)$.

Now, the source in the compactified equation~\eqref{eq:teukolsky_compactified} is multiplied by an additional factor $\propto \sigma^{-\s-5}$ coming from Eqs.~\eqref{eq:scaled variables} and~\eqref{eq:-2Slm} (and below the latter, for $S^{(2)}_0$). Altogether, this leaves us with a source of the form 
\begin{equation}\label{eq:S2 near scri}
\tilde S = \begin{cases}
    j_{\infty}\sigma^{-(\s/2+2)}, & \omega_\m\neq0,\\
    k_{\infty}\sigma^{-(\s+1)} + \O(\sigma^{-\s}),  & \omega_\m = 0,
\end{cases}
\end{equation}
for some constants $j_\infty$ and $k_\infty$. The behavior of the field $\tilde\psi$ then depends strongly on whether we consider stationary ($\omega_\m=0$) or radiative ($\omega_\m\neq0$) modes. In Lorenz-gauge calculations, $\omega_\m=0$ modes are the ones that face intransigent divergences, which cannot be cured without access to the SF-MPM memory integrals~\cite{Cunningham:2024dog}, while $\omega_\m\neq0$ modes encounter mild logarithmic singularities associated with the perturbations to light cones; the latter do not require the SF-MPM expansion to determine boundary conditions. Only the radiative modes are needed for the computation of 2SF fluxes, but for completeness we consider both types here.


For $\omega_\m=0$, we can approximate Eq.~\eqref{eq:teukolsky_compactified} near $\sigma=0$ as
\beq
\sigma^2 \partial^2_\sigma \tilde\psi + 2\sigma(s+1)\partial_\sigma \tilde\psi - {}_s\lambda_\ell\tilde\psi  
= k_\infty\sigma^{-(\s+1)},\label{Teukolsky eq near scri}
\eeq
where we have used the fact that each derivative with respect to $\sigma$ lowers the power of $\sigma$ by one. The general solution to Eq.~\eqref{Teukolsky eq near scri} is
\begin{equation}
\tilde\psi = -\frac{k_\infty\sigma^{-(\s+1)}}{\l(\l+1)} + C_1\sigma^{\l-\s}+ C_2\sigma^{-(\l+\s+1)},\label{Teukolsky soln near scri OK}
\end{equation}
with arbitrary constants $C_n$. Here we see that the solution for $s=+2$ is singular at $\sigma=0$, but the solution for $s=-2$ is regular if we set $C_2=0$. In either case, this does not necessarily indicate a breakdown of the ``compactification + regularity'' method; if we introduce a new variable $\Psi\coloneqq \sigma^{\s+1} \tilde\psi$, then it has a smooth source. If we then assume $\Psi$ is regular and evaluate the field equation for $\Psi$ at $\sigma=0$, we immediately obtain a boundary condition of the form~\eqref{eq:sigma=0 eqn}, and $C_2$ is automatically eliminated. Hence, no external information is required. 

Since the solution for the original variable is singular, one might question whether regularity of $\Psi$ is the correct condition to impose. This is a sensible question because the physical variables, $\psi^{(2)}_{4,\l\m}$ and $\psi^{(2)}_{0,\l\m}$, behave as $\sigma^{\s+2}$, meaning that $\psi^{(2)}_{4,\l\m}$ ($\s=-2$) goes to a constant at \scri rather than falling off. However, this is a gauge artifact. One is not prevented from solving the field equation for $\Psi$ under the assumption of regularity and then transforming to a Bondi-Sachs gauge to obtain the well-behaved $\psi^{(2)}_{4,\l\m}$ at \scri. We contrast this with the situation for the Lorenz-gauge field equations in Appendix~\ref{sec:infrared Lorenz}.

Now turning to $\omega_\m\neq0$, we find we cannot easily write a useful small-$\sigma$ field equation analogous to Eq.~\eqref{Teukolsky eq near scri}. However, we can straightforwardly seek a small-$\sigma$ solution to the full field equation~\eqref{eq:teukolsky_compactified} with the coefficients~\eqref{eq:A[u]} and~\eqref{eq:B[u]} and the source~\eqref{eq:S2 near scri}. 
For $\s=+2$, the source diverges as $1/\sigma^3$, and one can quickly find a small-$\sigma$ solution,
\beq\label{eq:psi omega!=0 near scri s=2}
\tilde\psi = \frac{j_\infty}{4i\varpi\sigma^2} + \O(1/\sigma) \quad (\s=2).
\eeq
As in the $\omega_\m=0$ case, we can also define a rescaled variable $\Psi\coloneqq \sigma^2\tilde\psi$. Multiplying the field equation~\eqref{eq:teukolsky_compactified} by $\sigma^2$ then yields an equation for $\Psi$ with a regular source. Assuming a regular $\Psi$, we can evaluate at $\sigma=0$ to obtain a condition of the form~\eqref{eq:sigma=0 eqn}. That condition reads explicitly
\begin{equation}
    4i\varpi \Psi\bigr|_{\sigma=0} = j_\infty,
\end{equation}
recovering Eq.~\eqref{eq:psi omega!=0 near scri s=2}. Therefore, we conclude that for $\s=2$ and $\omega_\m\neq0$, the ``compactification + regularity'' approach again works, despite the singular behavior of the field variable $\tilde\psi$. Importantly, the resulting physical Weyl scalar $\psi^{(2)}_{0L}$ is regular, behaving as $\sigma^{\s+2} \tilde\psi=\O(\sigma^2)$.

However, we now finally encounter a case where this approach fails: $\omega_\m\neq0$ and $s=-2$. This is precisely the case of interest in the computation of energy fluxes at \scri, as the flux is directly determined by oscillatory modes of $\psi_4$. The source diverges as $1/\sigma$, and we can quickly find the small-$\sigma$ solution begins with a log divergence,
\beq\label{eq:Teukolsky log soln}
\tilde\psi = -\frac{j_\infty}{2i\varpi}\ln\sigma + a_0+\O(\sigma\ln\sigma),
\eeq
where $a_0$ is an arbitrary constant. One can easily verify that there is no possible rescaled variable $\Psi\coloneqq\sigma^k\tilde\psi$ for which the field equation has a regular limit to $\sigma=0$. We can hence conclude that ``compactification + regularity'' fails in this essential instance. 

The only way to obtain a regular field equation in this case is to work with a \emph{shifted} rather than rescaled variable, 
\begin{equation}\label{eq:Teukolsky infinity residual field}
\tilde\psi^{\cal R}_\infty\coloneqq \tilde\psi -\tilde\psi^{\cal P}_\infty,
\end{equation}
by subtracting a ``puncture at infinity'', $\tilde\psi^{\cal P}_\infty$. The field equation then reads
\begin{multline}
(1-\sigma) \sigma^2 \partial^2_\sigma \tilde\psi^{\cal R}_\infty + {\cal B}\,\partial_\sigma \tilde\psi^{\cal R}_\infty + {\cal A}\tilde\psi^{\cal R}_\infty \\
= \tilde S-\bigl[(1-\sigma) \sigma^2 \partial^2_\sigma \tilde\psi^{\cal P}_\infty + {\cal B}\,\partial_\sigma \tilde\psi^{\cal P}_\infty + {\cal A}\tilde\psi^{\cal P}_\infty\bigr],\label{eq:teukolsky_compactified_effective}
\end{multline}
which, unlike the original field equation, has a source that is regular at \scri. One can easily confirm that only using the leading term in Eq.~\eqref{eq:Teukolsky log soln} leads to a logarithmically divergent effective source in Eq.~\eqref{eq:teukolsky_compactified_effective}, but that is rectified by including the first subleading term in the puncture:
\begin{equation}\label{eq:Teukolsky infinity puncture}
    \tilde\psi^{\cal P}_\infty = -\frac{j_\infty}{2i\varpi}\ln\sigma + \frac{({}_{-2}\lambda_\l-4 i \varpi)j_\infty }{(2 i \varpi)^2}\sigma\ln\sigma.
\end{equation} 
At $\sigma=0$ the field equation then reduces to a boundary condition of the form~\eqref{eq:sigma=0 eqn}, relating $\tilde\psi^{\cal R}_\infty|_{\sigma=0}$ and $\partial_\sigma\tilde\psi^{\cal R}_\infty|_{\sigma=0}$. 

This effective-source approach is the method we used in our previous, Lorenz-gauge calculations~\cite{Miller:2023ers}. Instead of pursuing this method, in this paper we eliminate the singularity by transforming to a Bondi-Sachs gauge. We describe that approach next.

\subsection{Transformation to the Bondi-Sachs gauge} 
\label{sec:Bondi-Sachs}

In our previous, Lorenz-gauge calculations~\cite{Pound:2019lzj,Warburton:2021kwk}, second-order quantities at \scri were extracted by transforming from the asymptotically singular Lorenz gauge to the well-behaved Bondi-Sachs gauge \emph{after} solving the second-order EFE for the Lorenz-gauge $h^{(2)}_{\alpha\beta}$. This procedure was detailed in Ref.~\cite{Cunningham:2024dog}. More recently, in Refs.~\cite{Spiers:2023cip,Spiers:2026yqx}, two of us presented an alternative scheme: rather than carrying out an SF-MPM expansion to obtain Lorenz-gauge boundary conditions for $h^{(2)}_{\alpha\beta}$, solving the ``near zone'' field equations, and then transforming the result to Bondi-Sachs gauge, one can instead compute $h^{(1)}_{\alpha\beta}$ in the Lorenz gauge and then transform it to the Bondi-Sachs gauge \emph{before} proceeding to second order. This entirely removes the need for the SF-MPM expansion, and it leads to a second-order source that decays sufficiently well at large $r$ that no ``puncture at infinity'' is required. As we described in Ref.~\cite{Spiers:2026yqx}, working in the Bondi-Sachs gauge also allows us to control the choice of asymptotic (Bondi-Metzner-Sachs) frame and to naturally incorporate gravitational-wave memory into the field equations.

The prescription in Ref.~\cite{Spiers:2026yqx} does not require an exact transformation to Bondi-Sachs gauge in an open neighborhood of \scri. It only requires that $h^{(1)}_{\alpha\beta}$ asymptotically approaches the Bondi-Sachs gauge. Concretely, we must enforce the following asymptotic gauge conditions:
\begin{subequations}
\begin{align}
    h^{(1)}_{ll} &= \O(1/r^3),\\
    h^{(1)}_{ln} &= \O(1/r^2),\\
    h^{(1)}_{lm} &= \O(1/r^2),\\
    h^{(1)}_{m\mb} &= \O(1/r^2),
\end{align}
\end{subequations}
with other components allowed to fall off as $1/r$. For $\omega_\m\neq0$ modes, these conditions differ from the Lorenz-gauge condition~\eqref{eq:h1 Lorenz m!=0} only in that $h^{(1)}_{ll} = \O(1/r^2)$ and $h^{(1)}_{ln} = \O(1/r)$ in the Lorenz gauge. For $\omega_\m=0$, the Bondi-Sachs gauge conditions enforce
\begin{equation}\label{eq:h1 Bondi monopole}
    h^{\{1\}}_{ab} = 2M^{(1)} l_a l_b \quad (\l=0)
\end{equation}
for the monopole, in place of Eq.~\eqref{eq:h1 Lorenz m=0}. This change in the $\omega_m=0$ mode ensures that the perturbed light cones, as perturbed by the orbital energy, are surfaces of constant retarded time $u$ (at least asymptotically). In Ref.~\cite{Spiers:2026yqx} we also highlighted extensively that $\omega_\m=0$ modes appear for $\l\geq2$ in the Bondi-Sachs gauge, unlike in the Lorenz gauge, and that these encode the effect of gravitational memory.

With the above gauge conditions satisfied, the second-order source term~\eqref{eq:G20 asymptotics Lorenz} becomes~\cite{Spiers:2026yqx}
\begin{align}\label{eq:G20 Bondi-Sachs}
    G^{(2,0)}_{nn} &= \frac{\Omega^2\Bigl[2\Re \partial_{\phi_p}\bigl(h^{\{1\}}_{mm}\partial_{\phi_p} h^{\{1\}}_{\mb\mb}\bigr)-\bigl|\partial_{\phi_p} h^{\{1\}}_{mm}\bigr|^2\Bigr]}{r^2} \no\\
    &\quad + \O(1/r^3),
\end{align}
with all other components falling off at least as fast as $1/r^{3}$. Critically, there is no $1/r^2$ term in $G^{(2,0)}_{\mb\mb}$; the $1/r^2$ term in the Lorenz-gauge source~\eqref{eq:G20 asymptotics Lorenz} was due to coupling between the Lorenz-gauge monopole~\eqref{eq:h1 Lorenz m=0} and oscillatory modes, and it is eliminated with the change to the Bondi-Sachs monopole~\eqref{eq:h1 Bondi monopole}. With no $1/r^2$ term in $G^{(2,0)}_{\mb\mb}$, there is also no $1/r^2$ term in the Teukolsky source~\eqref{eq:S24 large r}. In fact, one can show $G^{(2,0)}_{\mb\mb}=\O(1/r^4)$ and $G^{(2,0)}_{n\mb}=\O(1/r^3)$~\cite{Spiers:2023cip}. Equation~\eqref{eq:GHPS modes} then implies
\begin{multline}\label{eq:S4[G20] Bondi large r}
    \hat\S_{4,\l\m}[G^{(2,0)}] = \frac{\omega_\m^2}{\gamma^{2}r^4}\,G^{(2,0)\{4\}}_{\mb\mb,\l\m} - \frac{\sqrt{{}_{-2}\lambda_\l}\, i\omega_\m}{\gamma r^4} G^{(2,0)\{3\}}_{n\mb,\l\m} \\
    -\frac{\mu^\l_{\ 2}}{4r^4} G^{(2,0)\{2\}}_{nn,\l\m} +\O(1/r^5),
\end{multline}
for both $\omega_\m=0$ and $\omega_\m\neq0$ modes. This is two orders more rapid decay than in the Lorenz gauge. The corresponding contribution to the source $\tilde S$ therefore scales linearly with $\sigma$ as $\sigma\to0$, which is more than sufficient for the ``compactification + regularity'' approach.

The second source, $G^{(1,1)}_{\alpha\beta}$, falls off as $1/r^2$, but only in the $\omega_\m=0$ piece, which reads (adapted from Ref.~\cite{Spiers:2026yqx})
\begin{equation}\label{eq:G11 Bondi-Sachs omega=0}
G^{(1,1)}_{nn,\l0} = - \frac{1}{4r^2}\frac{d}{d\tilde u}\bigl(8M^{(1)}_{\mathscr{B},\l0} -\mu^\l_{\ 2}\Re h^{\{1\}}_{mm,\l0}\bigr) +\O(1/r^3), 
\end{equation}
with all other components falling off more rapidly. Here $\tilde u = \e u$ is a slow retarded time, and $M^{(1)}_{\mathscr{B}}$ is the Bondi mass aspect, given by $M^{(1)}_{\mathscr{B}}=\frac{1}{2}h^{\{1\}}_{uu}=\frac{\gamma^2}{4}h^{\{1\}}_{nn}$ in the Bondi-Sachs gauge. As explained in Ref.~\cite{Spiers:2026yqx}, the quantity $h^{\{1\}}_{mm,\l0}$ appearing in Eq.~\eqref{eq:G11 Bondi-Sachs omega=0} represents the leading-order gravitational-wave memory, which encodes infinitely many slowly evolving degrees of freedom beyond the two-body parameters ${\cal J}_I$. This feature of the problem does not emerge in the Lorenz-gauge expression~\eqref{eq:G11 Lorenz large r} because $h^{\{1\}}_{mm,\l0}$ itself vanishes in the Lorenz gauge, which is a ``forgetful gauge'' in the terminology of Ref.~\cite{Spiers:2026yqx}. For simplicity, as we explain in Sec.~\ref{sec:nonlinear_source}, we do not consider the effects of $h^{\{1\}}_{mm,\l0}$ in this paper.

One can also check that the other components of $G^{(1,1)}_{\alpha\beta}$ mimic the falloff of $G^{(2,0)}_{\alpha\beta}$: $G^{(1,1)}_{n\bar m}=\O(1/r^3)$ and $G^{(1,1)}_{\mb\mb}=\O(1/r^4)$. Showing this is a nontrivial but straightforward application of the \textsc{PerturbationEquations} package, for example. Therefore, $\hat\S_4[G^{(1,1)}]$ falls off as $1/r^4$, in analogy with Eq.~\eqref{eq:S4[G20] Bondi large r}.  

So, we conclude that if $h^{(1)}_{\alpha\beta}$ is transformed to the Bondi-Sachs gauge, then the Teukolsky source is given by 
\begin{multline}\label{eq:source falloff Bondi-Sachs}
    S^{(2)}_{4,\l\m} = \frac{\omega_\m^2}{\gamma^{2}r^4}\,T^{\{4\}}_{\mb\mb,\l\m} - \frac{\sqrt{{}_{-2}\lambda_\l}\, i\omega_\m}{\gamma r^4} T^{\{3\}}_{n\mb,\l\m} \\
    -\frac{\mu^\l_{\ 2}}{4r^4} T^{\{2\}}_{nn,\l\m} +\O(1/r^5),
\end{multline}
in terms of the coefficients $T^{\{k\}}_{ab,\l\m}$ of $1/r^k$ in the source $-(G^{(2,0)}_{\alpha\beta}+G^{(1,1)}_{\alpha\beta})$. Again, this $1/r^4$ falloff of $S^{(2)}_{4,\l\m}$ is more than sufficiently rapid to enable the ``compactification + regularity'' method. 

Our argument here has extended the one in Ref.~\cite{Spiers:2026yqx}, both in the link to compactification and in considering the contribution of $G^{(1,1)}_{\alpha\beta}$. We describe the actual transformation of $h^{(1)}_{\alpha\beta}$ from the Lorenz gauge to the Bondi-Sachs gauge in Sec.~\ref{sec:nonlinear_source}.

\section{Construction of the second-order source}
\label{sec:source}

\subsection{Sources outside the worldtube: nonlinear source}
\label{sec:nonlinear_source}

An explicit mode-coupling formula for the spin-weighted spherical harmonic modes of the quadratic source term $\hat\S_{4,\l\m}[G^{(2,0)}]$ was derived in Ref.~\cite{Spiers:2023mor} and is available in the \textsc{PerturbationEquations} package~\cite{PerturbationEquations}. That formula takes the schematic form 
\begin{equation}
    \hat\S_{4,\l\m}[G^{(2,0)}] = \sum_{\l_1\m_1\l_2\m_2} \hat\S^{abcd}_{4,\l\m\,\l_1\m_1\l_2\m_2}h^{(1)}_{ab,\l_1\m_1}h^{(1)}_{cd,\l_2\m_2}, \label{eq:coupling}
\end{equation}
where the operator $\hat\S^{abcd}_{4,\l\m\,\l_1\m_1\l_2\m_2}$ includes up to four radial derivatives and acts bilinearly on the metric perturbation mode functions $h^{(1)}_{ab,\l_1\m_1}$ and $h^{(1)}_{cd,\l_2\m_2}$. We compute these input mode functions on a radial grid using the \textsc{h1Lorenz} code \cite{h1Lorenz,Akcay:2013wfa}. 

Equation~\eqref{eq:coupling} contains up to four radial derivatives of $h^{(1)}_{ab,\l\m}$. We obtain the mode functions and their first derivatives directly from the \textsc{h1Lorenz} code, and we use the homogeneous Lorenz-gauge field equations to express higher-order derivatives in terms of the mode functions and their first derivatives. We then feed the results into the C++ implementation of the mode coupling formula~\eqref{eq:coupling} in the \textsc{SecondOrderTeukolskySource}~\cite{TeukolskySource} code to obtain the source on a radial grid.

\subsubsection{Transformation to Bondi-Sachs gauge}

If we were to compute the nonlinear source using the Lorenz-gauge mode functions, we would immediately encounter the problem described in the previous section: the source we obtain would fall off at large radius as $\O(\frac{1}{r^2})$, as in Eq.~\eqref{eq:S24 large r}, which we recall is two orders too slow for a freely propagating solution. Instead, we address the slow-falloff issue and obtain two orders faster falloff in $1/r$ by introducing a first-order gauge transformation, $h^{(1)}_{\alpha\beta}\to h^{(1)}_{\alpha\beta} + {\cal L}_\xi g_{\alpha\beta}$, which puts $h^{(1)}_{\alpha\beta}$ in Bondi-Sachs gauge, following Ref.~\cite{Spiers:2026yqx}. We do this at the level of spherical-harmonic modes, and after applying the gauge transformation we use the resulting Bondi-Sachs metric perturbation mode functions as input into Eq.~\eqref{eq:coupling}. In this section and \cref{sec:slow-time-outside-worldtube-source} we impose the Carter tetrad ($\gamma=1$).

The spherical-harmonic modes of the gauge vector that transforms from Lorenz gauge to Bondi-Sachs gauge are given by
\begingroup\allowdisplaybreaks%
\begin{subequations}\label{eq:mneq0xi}%
    \begin{align}
    \xi_{l,\l\m}&=\frac{h_{ll,\l\m}^{\{2\}}}{\sqrt{2}r} ,\\
    \xi_{n,\l\m} &= 0, \\
    \xi_{m,\l\m} &= \frac{-\mu^\l_{\ 1}\;h_{ll,\l\m}^{\{2\} }}{2\sqrt{2}r}\;   
    \end{align}
\end{subequations}\endgroup%
in the $\m\neq 0$ sector\footnote{That is, the only non-Bondi-Sachs behavior in the $\m\neq0$ Lorenz gauge metric perturbation is $h_{ll,\l\m}^{\{2\} }$ and $h_{ln,\l\m}^{\{1\}}$ being non-zero; both are corrected by the gauge transformation~\eqref{eq:mneq0xi} as $h_{ln,\l\m}^{\{1\}}= +2i\omega_\m \frac{h_{ll,\l\m}^{\{2\}}}{\sqrt{2}}$ in the Lorenz gauge.} and 
\begingroup\allowdisplaybreaks%
\begin{subequations}\label{eq:m=0xi}
\begin{align}
    \xi_{l,00}&=-\frac{h_{ll,00}^{\{1\}}}{\sqrt{2}}\ln \frac{r}{M}+\frac{h_{ll,00}^{\{2\}}+2M h_{ll,00}^{\{1\}}}{\sqrt{2}r}\no\\*& \quad +\frac{M h_{ll,00}^{\{1\}}\ln \frac{r}{M}}{\sqrt{2}r}, \\
    \xi_{n,00}&=\frac{h_{m\mb,00}^{\{1\}}}{\sqrt{2}}-\frac{h_{ll,00}^{\{1\}}}{\sqrt{2}}\ln \frac{r}{M} \no\\*& \quad+\frac{h_{ll,00}^{\{2\}}+2M h_{ll,00}^{\{1\}}}{\sqrt{2}r}+\frac{M h_{ll,00}^{\{1\}}\ln \frac{r}{M} }{\sqrt{2}r},\\
    \xi_{m,00}&=0,
\end{align}
\end{subequations}\endgroup%
in the $\m=0$, $\l=0$ sector. For $\m=0$, $\l>0$, we set the gauge vector to zero;\footnote{However, we stress that this does not cause our metric perturbation to violate the Bondi-Sachs gauge conditions, even if it causes violation of a specific sector of the multiscale EFE in Bondi-Sachs gauge. The only non-Bondi-Sachs behavior in the $\m=0$ Lorenz-gauge metric perturbation is the nonvanishing $\l=0$ coefficients $h_{ll,00}^{\{1\} }$, $h_{ll,00}^{\{2\} }$ and $h_{m\mb,00}^{\{1\}}$, which are eliminated by the gauge transformation~\eqref{eq:m=0xi}.} this amounts to neglecting slowly evolving BMS transformations~\cite{Bondi:1962px,Sachs:1962wk} that appear in a complete treatment~\cite{Spiers:2026yqx}. Neglecting these specific transformations has the consequence of neglecting gravitational-wave memory in $h^{\{1\}}_{mm,\l0}$ and memory distortion in $h^{(2)}_{\alpha\beta}$, which in turn means we fail to satisfy a specific sector of the full multiscale EFE at second order. Future work will incorporate those effects, particularly the memory-distortion contribution to the energy flux. But for our initial implementation we satisfy ourselves with achieving consistency with previous Lorenz-gauge calculations, where those effects were likewise ignored. We expect the effect on the fluxes to be small in any case.

\begin{figure}
\centering
\includegraphics[width=\columnwidth]{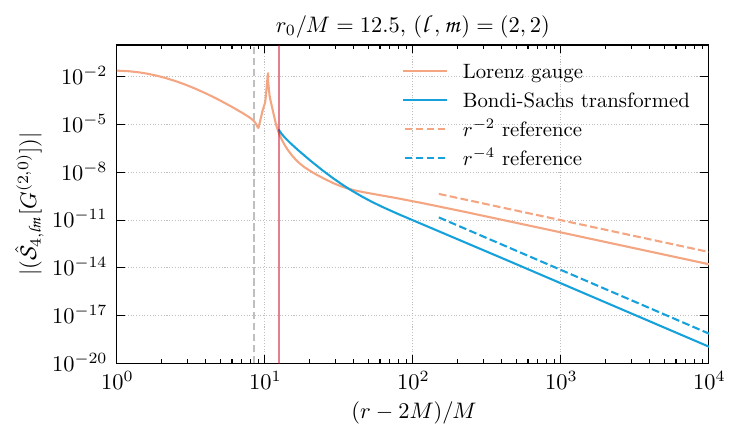}
\caption{\label{fig:Source22modeBS}The $\l=2, \m=2$ mode of the nonlinear source $\hat\S_{4,\l\m}\left[G^{(2,0)}\right]$ for $r_{0} / M = 12.5$ in the Carter tetrad and BL coordinates.
The source calculated purely from Lorenz-gauge perturbations is shown in orange, while the source calculated using metric perturbation data with a Bondi-Sachs transformation is shown in blue.
Each source has been calculated using $\l_{max}=50$ input modes of $h^{(1)}_{\l\m}$.
The Bondi-Sachs transition is implemented sharply at the worldtube boundary $r = \rplus = 14.5M$, indicated by the red line.
In the asymptotic large-$r$ regime, the falloff of the Bondi-Sachs transformed source is two orders faster, $r^{-4}$, than the original source term, which falls off as $r^{-2}$.}
\label{fig:sg20_source}
\end{figure}

The transformation to Bondi-Sachs gauge is only required near \scri. We stick with Lorenz gauge inside the worldtube around the particle and only apply the transformation to Bondi-Sachs gauge for points from the outer edge of the worldtube to \scri. This is demonstrated in \cref{fig:Source22modeBS}, which shows both the original Lorenz-gauge source and the source after transforming to Bondi-Sachs gauge.

Note that we deliberately choose to make a sharp transition from Lorenz gauge to Bondi-Sachs gauge at a single radial point. Compared to a smooth transition, for example, this has the advantage of improved computational efficiency: it allows us to avoid introducing additional detailed structure in the source that would need to be resolved. However, it comes at the cost of having to account for a corresponding jump in $\psi_{4L}^{(2)}$ that is induced by the gauge transformation at the transition point. This jump is obtained by substituting the gauge vector in Eqs.~\eqref{eq:mneq0xi} and \eqref{eq:m=0xi} into Eq.~\eqref{eq:psi4L2GaugeTransform}. The resulting expressions are somewhat unwieldy; rather than giving them here, we provide the two pieces, $\frac{1}{2} \hat\T_{4,\l\m}[{\cal L}^2_\xi g]$ and $\hat\T_{4,\l\m}[{\cal L}_\xi h^{(1)}]$, in an accompanying Mathematica notebook (the slow-evolution term, $\hat\T_{4,\l\m}[{\cal L}^{(1)}_\xi g]$, will be addressed in the next subsection).

\subsection{Sources outside the worldtube: slow-evolution source\label{sec:slow-time-outside-worldtube-source}}

We next consider the slow-evolution contribution to the source, $\hat\S_4[G^{(1,1)}_{ab}]$. An expression for this can be obtained by substituting $G^{(1,1)}_{ab}$ from Eq.~\eqref{G11} into the operator $\hat{\cal S}_4$ defined in Sec.~\ref{sec:second-order-teukolsky} and decomposing into spherical-harmonic modes as in Sec.~\ref{sec:sph_harm}. The resulting expression for the mode decomposition of the slow-evolution piece of the source is too long to give here; instead we give it as a Mathematica notebook in the supplemental material. Schematically, two types of term appear, one coming from the slow evolution of $h_{ab}$, and the other from the slow evolution of $\omega_\m$. We use Eq.~\eqref{Omegadot} to write the latter as an expression involving $F^{(0)}_\Omega$ multiplied by the modes of $h^{(1)}_{ab}$. For the former, we use the results of Ref.~\cite{Durkan:2022fvm} to obtain data for $\partial_{r_0} h^{(1)}_{ab,\l\m}$ and then relate this to the required slow-evolution data using
\begin{align}\label{eq:partialV}
    \vec{\partial}_{\cal V} \to F^{(0)}_\Omega \frac{\partial}{\partial\Omega}=F^{(0)}_\Omega \frac{\partial r_0}{\partial\Omega} \frac{\partial}{\partial r_0},
\end{align}
where $\frac{\partial r_0}{\partial\Omega}=-\frac{2M^{\frac{1}{3}}}{3\Omega^{\frac{5}{3}}}$ for the present case of quasi-circular orbits in Schwarzschild spacetime. Numerical data for $F^{(0)}_\Omega$ is obtained from its expression in terms of the energy flux, which in turn we extract from the numerical solutions for $h^{(1)}_{ab,\l\m}$.

\subsubsection{Transformation to Bondi-Sachs gauge}

As with the nonlinear source in the previous subsection, we must account for the transformation from Lorenz gauge to Bondi-Sachs gauge in the slow-evolution contribution to the source. The additional contribution from the gauge transformation is given by $G^{(1,1)}[\Lie^{(0)}_\xi g]$ and consists of two types of term: terms involving $\vec{\partial}_{\cal V} (\Lie^{(0)}_{\xi} g)_{\mu\nu}$ and terms involving $F_\Omega^{(0)} (\Lie^{(0)}_\xi g)_{\mu\nu}$. It turns out that the second of these always comes multiplied by a factor of $1-H^2$ so it vanishes on the $u$ slicing region where the gauge transformation. We are then left with only terms of the first type. For the $m\ne 0$ gauge vector given in Eq.\eqref{eq:mneq0xi} these terms are readily obtained from
\begingroup\allowdisplaybreaks%
\begin{subequations}%
\begin{align}
     \big(\partial_{r_0} \Lie_{\xi}^{(0)} g\big)_{ll}^{\l\m} &= \frac{ (M -  r)\partial_{r_0}h_{ll,\l\m}^{\{2\}} }{f^{1/2} r^3}, \\
   \big(\partial_{r_0} \Lie_{\xi}^{(0)} g\big)_{ln}^{\l\m} &= \frac{ \bigl(r-3M - 2i r^2 \omega_\m \bigr)\partial_{r_0}h_{ll,\l\m}^{\{2\}} }{2 f^{1/2} r^3}  \no\\*
   &\quad -\frac{2i(\partial_{r_0} \omega_\m)  h_{ll, \l\m}^{\{2\}}}{2 f^{1/2} r},  \\
   \big(\partial_{r_0} \Lie_{\xi}^{(0)} g\big)_{lm}^{\l\m} &= \frac{ \mu^{\l}_{\ 1} (-1 + f^{1/2})\partial_{r_0}h_{ll,\l\m}^{\{2\}}  }{2 r^2}, \\
   \big(\partial_{r_0} \Lie_{\xi}^{(0)} g\big)_{nn}^{\l\m} &= 0,  \\
   \big(\partial_{r_0} \Lie_{\xi}^{(0)} g\big)_{nm}^{\l\m} &=  \frac{ \mu^{\l}_{\ 1}  \bigl(2 M -  r + i r^2 \omega_\m \bigr)\partial_{r_0}h_{ll,\l\m}^{\{2\}}}{2 f^{1/2} r^3}\no\\* 
   &\quad + \frac{i \mu^{\l}_{\ 1} r^2(\partial_{r_0} \omega_\m )h_{ll,\l\m}^{\{2\}}}{2 f^{1/2} r^3}, \\
   \big(\partial_{r_0} \Lie_{\xi}^{(0)} g\big)_{mm}^{\l\m} &= \frac{ (\mu^{\l}_{\ 1})^2 \partial_{r_0}h_{ll,\l\m}^{\{2\}}  }{2 r^2}, \\
   \big(\partial_{r_0} \Lie_{\xi}^{(0)} g\big)_{m\mb}^{\l\m} &= \frac{ \bigl[2 f^{1/2} -  (\mu^{\l}_{\ 1})^2\bigr] \partial_{r_0}h_{ll,\l\m}^{\{2\}}}{2 r^2} ,
\end{align}
\end{subequations}%
\endgroup%
along with Eq.~\eqref{eq:partialV}.
In this work we focus only on computing fluxes for circular orbits, and these require only $m\ne0$ modes of $\psi_{4,\l\m}^{(2)}$. Since $G^{(1,1)}$ is a linear operator, we therefore do not require the $m=0$ modes of $\vec{\partial}_{\cal V} (\Lie^{(0)}_{\xi} g)_{\mu\nu}$.

As before, we must also account for the jump in $\psi_{4L}^{(2)}$ from the slow-time gauge transformation, given by the final term in \cref{eq:psi4L2GaugeTransform}. Using  \cref{eq:mneq0xi}, we find,
\begin{multline}
\hat \T_{4,\l\m}\bigl[ \Lie^{(1)}_{\xi}g\bigr] = \frac{\mu^\l_{\ 1} \mu^\l_{\ 2}}{8fr^4}\Bigl[2(M-i\omega_\m r^2)\partial_{r_0}h_{ll,\l\m}^{\{2\} } \\  \quad - i r^2 ( \partial_{r_0}\omega_\m)h_{ll ,\l\m}^{\{2\}} \Bigr],
\end{multline}
for $m\neq0$ in $u$-slicing.

\begin{figure}
\centering
\includegraphics[width=\columnwidth]{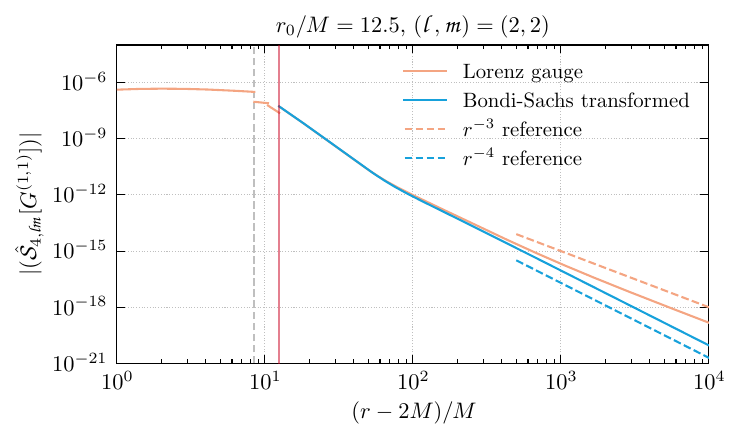}
\caption{The $\l = 2$, $\m = 2$ mode of the slow-time source $\hat{\cal S}_{4,\l\m}[G^{(1,1)}]$ for $r_{0}/M = 12.5$ in the Carter tetrad and BL coordinates. 
As in Fig.~\ref{fig:sg20_source}, we have constructed the source using Lorenz gauge metric perturbation data in orange and with a Bondi-Sachs transformation in blue at the same worldtube boundary.
Unlike $\hat{\cal S}_{4,\l\m}[G^{(2,0)}]$, the fall-off of the Bondi-Sachs transformed source is only improved by one order: from $r^{-3}$ to $r^{-4}$.}
\end{figure}







\subsection{Effective source inside the worldtube}\label{sec:Seff}

Inside the worldtube the effective source is given by $S_{4,\eff}^{(2)} = \hat{S}_4[T^{(2),\eff}_{\mu\nu}]$ with $T^{(2),\eff}_{\mu\nu}$ given by the right-hand side of Eq.~\eqref{tt EFE2R}. In practice, since we use integration by parts to shift $\hat{S}_4$ onto the Green's function (see Sec.~\ref{sec:mixed method}) we only need to compute the modes of $T^{(2),\eff}_{nn}$, $T^{(2),\eff}_{n\bar{m}}$ and $T^{(2),\eff}_{\bar{m}\bar{m}}$. We obtain these following closely the methods described in Ref.~\cite{Upton:2025bja}:
\begin{itemize}
    \item We split the first-order metric perturbation into puncture and residual fields $h^{(1)}_{\mu\nu} = h^{(1)\calP}_{\mu\nu} + h^{(1)\calR}_{\mu\nu}$ and use this to split the nonlinear term in the source into four pieces,
        \begin{align}
            G_{\mu\nu}^{(2,0)}&(h^{(1)},h^{(1)}) = G_{\mu\nu}^{(2,0)}(h^{(1)\calP},h^{(1)\calP})\nonumber \\
            & + G_{\mu\nu}^{(2,0)}(h^{(1)\calP},h^{(1)\calR}) + G_{\mu\nu}^{(2,0)}(h^{(1)\calR},h^{(1)\calP})\nonumber \\
            & + G_{\mu\nu}^{(2,0)}(h^{(1)\calR},h^{(1)\calR}).
        \end{align}
    \item We evaluate the spherical-harmonic modes of the second, third and fourth of these terms in exactly the same way as we did in Ref.~\cite{Upton:2025bja}. This is similar to what we did in Sec.~\ref{sec:nonlinear_source}, except that we now use the C++ implementation of the mode coupling formula for the second-order Ricci tensor in the \textsc{SecondOrderRicci} \cite{SecondOrderRicci} code.
    \item We evaluate the first term in the same way as described in Sec.~IV~B~2 of Ref.~\cite{Upton:2025bja}. In summary, the expression is evaluated as a coordinate series in rotated coordinates in which the particle is instantaneously at the north pole. We decompose into modes by performing a 2D numerical integration over the sphere and then use the Wigner-D matrix to transform to modes in the original unrotated coordinate system. We make two further tweaks to the calculation
        \begin{enumerate}
            \item We cancel the divergent contribution to the source from $h^{\SS}$ before decomposing into modes, by first obtaining a coordinate expression for $2 \breve G_{\mu\nu}^{(2,0)}[h^{(1)\calP},h^{(1)\calP}] - \brE^0_{\alpha \beta}[h^{\SS}]$ and then numerically integrating over the sphere. This significantly reduces numerical noise in the result.
            \item We add the mode decomposition $\Delta T^{(2)\eff}_{\mu\nu}$, which accounts for the conversion from gauge-damped Lorenz gauge equations to linearised Einstein equations (see Appendix \ref{sec:Lorenz conversion} for details).
        \end{enumerate}
    \item With the split of the second order puncture into its constituent pieces,
        \begin{equation}
            h^{(2)\calP}_{\mu\nu} = h^{\SS}_{\mu\nu} + h^{\SR}_{\mu\nu} + h^{\delta m}_{\mu\nu} + h^{\delta z}_{\mu\nu} + h^{\ms}_{\mu\nu}, \label{eq:h2P}
        \end{equation}
        we have already accounted for the contribution from $h^{\SS}_{\mu\nu}$ so we only need to compute the modes of $G^{(1,0)}_{\mu\nu}[h^{\SR}_{\mu\nu} + h^{\delta m}_{\mu\nu} + h^{\delta z}_{\mu\nu} + h^{\ms}_{\mu\nu}]$. These are obtained directly from the results of Ref.~\cite{Upton:2025bja} with the only required adjustments being the conversion from Lorenz-gauge to linearised Einstein operator (see Appendix \ref{sec:Lorenz conversion}) and the conversion from BLS modes to tetrad modes (this is given in Table II of Ref.~\cite{Spiers:2023mor}).
    \item The only remaining contribution to the source inside the worldtube is the slow-evolution piece, $G^{(1,1)}_{\mu\nu}[h^{(1)}]$. This is obtained as described in Ref.~\cite{Upton:2025bja}, adjusted to convert from Lorenz-gauge to linearised Einstein operator (see Appendix \ref{sec:Lorenz conversion}).
\end{itemize}

Each of the pieces of the puncture in Eq.~\eqref{eq:h2P} has a different physical origin:  $h^{\SS}_{\mu\nu}$ is sourced by quadratic combinations of the first-order puncture field, $h^{\SR}_{\mu\nu}$ by products of the first-order puncture and residual fields, $h^{\delta m}_{\mu\nu}$ gives the correction to the small object's mass monopole and solely features terms that are $\l=0$ on spheres centered on the small object, $h^{\delta z}_{\mu\nu}$ tracks corrections to the position of the object's worldline, and $h^{\ms}_{\mu\nu}$ describes the slow evolution of that worldline and is sourced by terms $\sim \vec{\partial}_{\cal V}$ acting on the first-order puncture.

We validated our calculations by verifying that each piece of our source satisfies the appropriate smoothness conditions, which follow from the fact that
\begin{align}
    \brE^{(0)}_{\mu\nu}[\barh^{\SS}] \simeq{}& - \brG^{(2,0)}_{\mu\nu}[h^{(1)\calP},h^{(1)\calP}], \label{eq:hSSEFE} \\
    \brE^{(0)}_{\mu\nu}[\barh^{\SR}] \simeq{}& - 2\brG^{(2,0)}_{\mu\nu}[h^{(1)\calP},h^{(1)\R}], \\
    \brE^{(0)}_{\mu\nu}[\barh^{\delta m}] \simeq{}& 0, \\
    \brE^{(0)}_{\mu\nu}[\barh^{\delta z}] \simeq{}& 0, \\
    \brE^{(0)}_{\mu\nu}[\barh^{\ms}] \simeq{}& - \brE^{(1)}_{\mu\nu}[\barh^{(1)\calP}],
\end{align}
where we use $\simeq$ to indicate this is true up to some order in distance away from the worldline (the specific order is dictated by the order of the expansion of the puncture fields). 
To satisfy the gauge conditions, the pieces must be combined into ``singular times singular'' terms $\sim m \times m$ and ``singular times regular'' terms $\sim m \times h^{(1)\calR}$:
\begin{align}
    Z^{(0)}_\mu[\barh^{\SS}] \simeq{}& 0, \\
    Z^{(0)}_{\mu}[\barh^{\SR} + \barh^{\delta m} + \barh^{\delta z} + \barh^{\ms}] \simeq{}& - Z^{(1)}_{\mu}[\barh^{(1)\calP}].
\end{align}
We use these gauge conditions as an additional check on our results.

\section{Spectral method}
\label{sec:spectral_method}
We solve the compactified radial equation in \eqref{eq:teukolsky_compactified} using a multidomain Chebyshev collocation method.
As described in the previous section, the worldtube construction naturally divides the radial domain into three regions.
The region extending from the worldtube to future null infinity is described on outgoing $u$ slices, the worldtube region on $t$ slices, and the region extending from the worldtube to the future horizon on ingoing $v$ slices.

We use the same compactification introduced in \eqref{eq:teukolsky_compactified}, which we restate here for convenience:
\beq
    \sigma=\frac{2M}{r},
    \qquad
    {\text \scri}:\ \sigma=0,
    \qquad
    \mathscr{H}^{+}:\ \sigma=1.
    \label{eq:spectral_compactification}
\eeq
For later use, we let $\rminus < \rp < \rplus$ denote the inner wall, particle, and outer wall.
Since $\sigma$ decreases as $r$ increases, we label the compact-coordinate walls according to their orientation in the $\sigma$ domain:
\begin{align}
    \sigmaplus &\coloneqq \sigma(r_{-}) = \frac{2M}{r_{-}},\qquad
    \sigmap \coloneqq \sigma(r_{p}) = \frac{2M}{r_{p}},\\
    \sigmaminus &\coloneqq \sigma(r_{+}) = \frac{2M}{r_{+}}.
\end{align}
Thus $0 < \sigmaminus < \sigmap < \sigmaplus < 1$.

\subsection{Multidomain setup and junction conditions}

The source construction also determines the relevant data required to join the regional solutions.
Three changes enter this matching: reconstruction of the total field from the residual field, the Bondi-Sachs gauge transformation, and the slow evolution of the orbital parameters between slices.
We shall describe each contribution before giving the junction conditions.

We use the rescaled field, $\tilde{\psi}$, and the source, ${\tilde{S}}$ of Eq.~\eqref{eq:scaled variables}, with $\s=-2$.
Each region obeys Eq.~\eqref{eq:teukolsky_compactified}, with the coefficients ${\cal A}_{[s]}$ and ${\cal B}_{[s]}$ evaluated in the appropriate slicing: $s = u, t, v$ corresponds to $H = 1, 0, -1$. 
Spin, mode and second-order labels remain suppressed, as in \secref{compactification}.

Inside the worldtube we solve for ${\tilde{\psi}}^{{\cal R} [t]}$.
The total Lorenz-gauge field is reconstructed as ${\tilde{\psi}}^{[t]} = {\tilde{\psi}}^{{\cal R} [t]} + {\tilde{\psi}}^{{\cal P} [t]}$, where the puncture of Eq.~\eqref{eq:worldtube_puncture_definition} is rescaled by the same factor in Eq.~\eqref{eq:master_rescaling}.
Here, and in the remainder of this section, ${\tilde{\psi}}^{{\cal P}}$ is evaluated in $t$-slicing.
No additional field normalisation is introduced.
To compare the exterior fields with the interior field, we restore their Fourier phases: $e^{i\omega_{\m}r_{*}}\tilde{\psi^{[u]}}$ at the outer worldtube boundary and $e^{-i\omega_{\m}r_{*}}\tilde{\psi^{[v]}}$ at the inner worldtube boundary.

At the outer worldtube boundary, the first-order Bondi-Sachs transformation induces the second-order change in the metric perturbation such that
\beq
    \Delta_{\xi} h^{(2)} := {\cal L}^{(0)}_{\xi}h^{(1)} + \frac{1}{2}{\cal L}^{(0)}_{\xi}{\cal L}^{(0)}_{\xi}g + {\cal L}^{(1)}_{\xi}g,
    \label{eq:second_order_bs_transformation}
\eeq
with all the quantities evaluated in $t$-slicing.
The corresponding change in the compactified master field is
\beq
    \Delta_{\xi}\tilde{\psi} = \frac{\sigma^{3}}{(1 - \sigma)^{2}}
    \frac{r^{2}\Delta}{2} \hat{\cal T}_{4}^{[t]}[\Delta_{\xi}h^{(2)}].
    \label{eq:master_field_bs_transformation}
\eeq
Here the $r^{2}\Delta/2$ is the conversion of the Carter-tetrad Weyl scalar to the spin-weight $s=-2$ master variable.
The final term of Eq.~\eqref{eq:second_order_bs_transformation} is the slow-evolution part of the gauge transformation, distinct from the slicing correction we describe below.

At first-order, changing from $t$ to $s$-slicing only changes the Fourier phase.
At second-order, however, the amplitudes and the orbital frequency also evolve between the two slices.
Expanding the mechanical variables and phase gives \cite{Miller:2016hjv, Miller:2023ers}
\begin{align}
    h^{(1) [s]}_{ab} &= e^{-i\omega_{\m}k} h^{(1)[t]},\\
    h^{(2) [s]}_{ab} &= e^{-i\omega_{\m}k} \bigg[h^{(2)[t]} + k\vec{\partial}_{\cal V} h^{(1)[t]} \nn\\
    &\qquad\qquad\qquad-\frac{i}{2}k^{2}\m F^{(0)}_{\Omega} h^{(1)[t]}\bigg].
\end{align}
All amplitudes on the right-hand side are evaluated at the same parameter values.
The term linear in $k$ appears due to the amplitude evolution, while the quadratic term comes from the frequency evolution of the phase \cite{Miller:2023ers}.

For $k = -r_{*}$ at the inner worldtube boundary and $k = r_{*}$ at the outer worldtube boundary, we define the jump corrections in $t$-slicing as
\begin{align}
    \delta^{-}_{\cal V} h^{(2)}
    &= r_{*} \vec{\partial}_{\cal V} h^{(1) [t]} + \frac{i}{2}r^{2}_{*}\m F^{(0)}_{\Omega} h^{(1)[t]},\\
    \delta^{+}_{\cal V} h^{(2)}
    &= r_{*} \vec{\partial}_{\cal V} h^{(1) {\rm BS} [t]} - \frac{i}{2}r^{2}_{*}\m F^{(0)}_{\Omega} h^{(1) {\rm BS} [t]}.
\end{align}
The inner correction is the $t$-minus-$v$ difference, whereas the outer correction is the $u$-minus-$t$ difference after restoring the phase.
The outer correction uses the Bondi-Sachs first-order metric perturbation since the infinity side-source is constructed in that gauge.

The corresponding matching terms for the reduced Teukolsky field are then
\beq
    \Delta^{\pm}_{\cal V}\tilde\psi = \frac{\sigma^{3}}{(1 - \sigma)^{2}}
    \frac{r^{2}\Delta}{2} \hat{\cal T}^{[t]}_{4}[\delta^{\pm}_{\cal V} h],
    \label{eq:slow_evolution_master_jump}
\eeq
evaluated at the respective walls.
The operator acts on the complete metric correction, including its factors of $r_{*}$. 
It cannot be interchanged with multiplication by $r_{*}$ or with $\vec\partial_{\cal V}$, since it contains radial derivatives and explicit factors of $\omega_{\m}$. 
A scalar slicing formula applied directly to $\tilde\psi^{(1)}$ therefore does not give the matching term for the reduced field $\psi^{(2)}_{4L}$. 
The slow-evolution curvature contribution in Eq.~\eqref{eq:psi42deconstructed} must be distinguished from that reduced field.

The total retarded-field relations at the worldtube boundaries are
\begin{align}
 \tilde\psi^{[t]}-e^{-i\omega_\m r^*}\tilde\psi^{[v]} &= \Delta_{\cal V}^{-}\tilde\psi,
 &&r=\rminus,\label{eq:inner_total_evolution}\\
 e^{i\omega_\m r^*}\tilde\psi^{[u]}-\tilde\psi^{[t]} &= \Delta_\xi\tilde\psi +\Delta_{\cal V}^{+}\tilde\psi,
 &&r=\rplus.\label{eq:outer_total_evolution}
\end{align}
Radial derivatives are taken at fixed mechanical parameters before evaluating the wall data. They act on both normalisation factors in Eq.~\eqref{eq:slow_evolution_master_jump} and on the full jump in the metric perturbation, with $\partial_r r^*=f^{-1}$.

We define the radial jump relation
\beq
    \jump{F}_{r} \coloneqq F(r^{+}) - F(r^{-}),
    \label{eq:jump_convention}
\eeq
where we note the limits $r^{\pm}$ should not be confused with the worldtube boundaries $r_{\pm}$.
The worldtube matching conditions are then given by
\begin{align}
 \bigl[\tilde\psi^{{\cal R}[t]}
       -e^{-i\omega_\m r^*}\tilde\psi^{[v]}\bigr]_{\rminus}
 &=J_{-},\label{eq:inner_wall_condition_field}\\
 \bigl[\partial_r\tilde\psi^{{\cal R}[t]}
       -\partial_r(e^{-i\omega_\m r^*}\tilde\psi^{[v]})\bigr]_{\rminus}
 &=J^{\prime}_{-},\label{eq:inner_wall_condition_derivative}\\
 \bigl[e^{i\omega_\m r^*}\tilde\psi^{[u]}
       -\tilde\psi^{{\cal R}[t]}\bigr]_{\rplus}
 &=J_{+},\label{eq:outer_wall_condition_field}\\
 \bigl[\partial_r(e^{i\omega_\m r^*}\tilde\psi^{[u]})
       -\partial_r\tilde\psi^{{\cal R}[t]}\bigr]_{\rplus}
 &=J^{\prime}_{+}.\label{eq:outer_wall_condition_derivative}
\end{align}
Each field in these expressions is evaluated from its own side of the wall.
These conditions match the exterior total field and its radial derivative to the total field reconstructed from the residual variable inside the worldtube.
At the horizon-side worldtube boundary there is no gauge transformation, but the $v$--$t$ slicing change contributes $\Delta_{\cal V}^{-}\tilde\psi$.
At the infinity-side, the matching contains the puncture, Bondi--Sachs gauge, and $t$--$u$ slicing contributions.
It follows that the data supplied to Eqns.~\eqref{eq:inner_wall_condition_field}-\eqref{eq:outer_wall_condition_derivative} are
\begin{align}
    J_{-} &= -\tilde{\psi}^{\cal P}(\rminus)
    + \Delta_{\cal V}^{-}\tilde\psi(\rminus),
    \label{eq:inner_wall_jump}
    \\
    J^{\prime}_{-} &= -\partial_{r}\tilde{\psi}^{\cal P}(\rminus)
    + \partial_{r}\Delta_{\cal V}^{-}\tilde\psi(\rminus), 
    \label{eq:inner_wall_jump_derivative}
    \\
    J_{+} &= \tilde{\psi}^{\cal P}(\rplus)
    + \Delta_{\xi}\tilde{\psi}(\rplus)
    + \Delta_{\cal V}^{+}\tilde\psi(\rplus),
    \label{eq:outer_wall_jump}
    \\
    J^{\prime}_{+} &= \partial_{r}\tilde{\psi}^{\cal P}(\rplus)
    + \partial_{r}\Delta_{\xi}\tilde{\psi}(\rplus)
    + \partial_{r}\Delta_{\cal V}^{+}\tilde\psi(\rplus), 
    \label{eq:outer_wall_jump_derivative}
\end{align}
Thus the inner worldtube boundary jump conditions contain the puncture contribution, while the outer worldtube boundary jump condition contains both the puncture data and the Bondi-Sachs gauge contribution.
In the compact coordinate the derivative jumps are obtained without a further sign convention such that
\beq
    \jump{\partial_{\sigma}\Psi}_{r_{a}} = \frac{dr}{d\sigma}\bigg|_{\sigma(r_{a})}
    \jump{\partial_{r}\Psi}_{r_{a}},
    \qquad
    \frac{dr}{d\sigma} = -\frac{2M}{\sigma^{2}}.
    \label{eq:derivative_jumps}
\eeq

The derivatives of the exterior fields include the restored phases. Explicitly,
\begin{align}
 \partial_\sigma\Psi_u
 &=
 e^{+i\omega_m r_*}
 \left(
  \partial_\sigma\widetilde{\psi}^{[u]}_{\ell m}
  +i\omega_m\frac{dr_*}{d\sigma}
    \widetilde{\psi}^{[u]}_{\ell m}
 \right),
 \label{eq:completed-u-phase-derivative}\\
 \partial_\sigma\Psi_v
 &=
 e^{-i\omega_m r_*}
 \left(
  \partial_\sigma\widetilde{\psi}^{[v]}_{\ell m}
  -i\omega_m\frac{dr_*}{d\sigma}
    \widetilde{\psi}^{[v]}_{\ell m}
 \right).
 \label{eq:completed-v-phase-derivative}
\end{align}
The particle location, $\sigma_{p}$, is also taken to be a domain boundary.
The compact interval is consequently divided initially as
\beq
    \sigma \in [0,\sigmaminus] \cup [\sigmaminus, \sigmap]
    \cup [\sigmap, \sigmaplus] \cup [\sigmaplus, 1].
    \label{eq:domain_boundaries}
\eeq
This division isolates the particle from every open spectral subdomain and preserves spectral convergence when the effective-source has only finite differentiability at $\sigma_{p}$.

\subsection{Multidomain Chebyshev collocation}
\label{sec:chebyshev_collocation}
We reserve the term domains for each of the regions in Eq.~\eqref{eq:domain_boundaries}.
Each of the four intervals is further subdivided into Chebyshev subdomains $D_{j} = [\sigma_{j},\sigma_{j+1}]$.
For an interval $[a, b]$, we place the domain boundaries at
\beq
    \sigma^{(a, b)}_{j} = \frac{a + b}{2} + \frac{b - a}{2\chi}
    {\rm arcsinh}\left[\left(\frac{2j}{N_{\rm sub}} - 1\right)\sinh\chi\right],
\eeq
where $\chi$ controls how strongly the domains cluster toward the interval's endpoints, $N_{\rm sub}$ is the number of subdomains in the interval, and $j = 0, \dots, N_{\rm sub}$.
We apply this distribution separately to each of the four intervals.
The local subdomain width scales as
$[1+\sinh^2\chi(2j/N_{\rm sub}-1)^2]^{-1/2}$, so the ratio of the widest (central) to the narrowest
(endpoint) subdomain grows like $\cosh\chi$, while $\chi \to 0$ recovers a uniform
partition. 
The endpoints of the intervals are precisely the points at which the solution is least regular or the coordinates least uniform: the worldtube boundaries $\sigmaminus$ and $\sigmaplus$, where the regional fields satisfy the jump conditions above; the particle location $\sigmap$, where the effective source has only finite differentiability; and $\sigma = 0$, where the compactification maps a subdomain of fixed width in $\sigma$ onto an unbounded range in $r$. 
Increasing $\chi$ therefore refines the resolution near these boundaries, at the cost of a coarser central region, so that $N_{\rm sub}$ must be raised alongside $\chi$. 
A baseline collocation configuration uses $\chi = 5$ with $N_{\rm sub} = 30$ in each of the two exterior intervals $[0,\sigmaminus]$ and $[\sigmaplus,1]$, and $N_{\rm sub} = 15$ in each half of the worldtube interval, giving 90 subdomains in total, with $32$ Chebyshev nodes per subdomain. 
The refined settings used for the flux results, the radial convergence tests, and the independent check of the infinity-side solution are described in Appendix~\ref{sec:error_endpoints_discretization}.

On $D_{j}$, we introduce the affine coordinate
\beq
    x_{j} = \frac{2\sigma - (\sigma_{j+1}+\sigma_{j})}{\sigma_{j+1}-\sigma_{j}}
    \label{eq:affine_coordinate}
\eeq
and write
\beq
    \tilde\psi^{[s]}(\sigma)
    = \sum^{N-1}_{n = 0}V^{(j)}_{n}T_{n}(x_{j}(\sigma)), \quad \sigma\in D_j,
    \label{eq:chebyshev_collocation}
\eeq
where $N$ is the number of Chebyshev nodes per subdomain and $T_n$ denotes a Chebyshev polynomial of the first kind. 
The equation is imposed at the Chebyshev-Gauss-Lobatto points 
\beq
    x_{k} = \cos\left(\frac{k\pi}{N-1}\right),
    \qquad k = 0, \ldots, N-1.
\eeq
Writing $h_{j} = \sigma_{j+1} - \sigma_{j}$ and $\sigma_{jk}=\sigma(x_{k})$, we express the collocation matrix $L^{(j)}$ as
\begin{multline}
    L^{(j)}_{kn} = (1 - \sigma_{jk})\sigma^{2}_{jk}\left(\frac{2}{h_{j}}\right)^{2} T^{\prime\prime}_{n}(x_{k}) \\
    + {\cal B}_{[s]}(\sigma_{jk})\left(\frac{2}{h_{j}}\right) T^{\prime}_{n}(x_{k})
    + {\cal A}_{[s]}(\sigma_{jk}) T_{n}(x_{k}),
\end{multline}
such that the spectral coefficients obey
\beq
    \sum^{N-1}_{n = 0}L^{(j)}_{kn}V^{(j)}_{n} = 
    \tilde S^{[s]}(\sigma_{jk}).
\eeq

At an ordinary subdomain boundary, the two adjacent expansions are matched in value and first derivative.
If $\tilde\psi_{(j)} = \sum_{n}V^{(j)}_{n}T_{n}(x_{j})$, the two conditions at $\sigma_{j+1}$ are
\begin{align}
    \tilde\psi_{(j)}(\sigma_{j+1}) - \tilde\psi_{(j + 1)}(\sigma_{j+1}) &= 0,\\
    \partial_{\sigma}\tilde\psi_{(j)}(\sigma_{j+1}) - \partial_{\sigma}\tilde\psi_{(j + 1)}(\sigma_{j+1}) &= 0.
\end{align}
At $\sigmaminus$, these conditions are replaced by the outer-wall conditions in Eqs.~\eqref{eq:outer_wall_condition_field}-\eqref{eq:outer_wall_condition_derivative}; at $\sigmaplus$, they are replaced by the inner-wall conditions in Eqs.~\eqref{eq:inner_wall_condition_field}-\eqref{eq:inner_wall_condition_derivative}.
In each case the radial derivative is converted using Eq.~\eqref{eq:derivative_jumps}.
At $\sigma = 0$ and $\sigma = 1$, the principal part of the differential operator, $(1 - \sigma)\sigma^{2}$ vanishes, and the endpoint collocation equations reduce to
\begin{align}
    {\cal B}_{[u]}(0)\partial_{\sigma}\tilde\psi^{[u]}(0)+{\cal A}_{[u]}(0)\tilde\psi^{[u]}(0) &= \tilde S^{[u]}(0),\\
    {\cal B}_{[v]}(1)\partial_{\sigma}\tilde\psi^{[v]}(1)+{\cal A}_{[v]}(1)\tilde\psi^{[v]}(1) &= \tilde S^{[v]}(1),
\end{align}
which are the regularity conditions.
No asymptotic extrapolation or separately fitted boundary expansion is therefore required.
The values of the rescaled field at the endpoint collocation points directly determine the amplitudes at \scri and $\mathscr{H}^+$.

If $N_{\rm sub}^{\rm tot}$ denotes the total number of subdomains across the three slicing domains, the expansion contains $N \times N_{\rm sub}^{\rm tot}$ coefficients.
The same number of equations is obtained by retaining the two endpoint regularity equations and replacing two bulk endpoint equations at each internal boundary by the corresponding pair of matching conditions.
The discretisation therefore preserves the order of the differential equation at every interface and produces a square global system.

The resulting complex linear system is solved after row equilibration, and its accuracy is assessed from the collocation residual, the decay of the Chebyshev coefficients, and the independent variation of both the number of subdomains and the polynomial order.

\section{Method mixing: spectral outside, variation of parameters inside}
\label{sec:mixed method}

The only previous second-order implementation was formulated in terms of variation of parameters on an unbounded radial domain~\cite{Miller:2023ers}. That approach could be used to solve our second-order Teukolsky equation as well, but as mentioned in the Introduction, it does not take maximal advantage of hyperboloidal slicing. Specifically, as we describe in this section, the variation-of-parameters method requires dealing with ill-behaved integrals and oscillatory integrands even in regions where the physical fields are well behaved and slowly varying. 

Conversely, the spectral construction can naturally treat the entire spacetime outside the worldtube with only moderate resolution because it only ever needs to resolve the physical, slowly varying fields there. This is particularly true once we impose the Bondi-Sachs gauge conditions to remove any delicate asymptotics. 

On the other hand, the relative merits of the two methods are largely reversed inside the worldtube. Our spectral method involves taking two extra derivatives of $T^{(2)\rm eff}_{ab}$, leading to stronger singularities that are more difficult to numerically resolve and to numerical derivatives of quantities with limited accuracy near the worldline. Variation of parameters more naturally deals with the Teukolsky source through integration by parts, moving derivatives off $T^{(2)\rm eff}_{ab}$ and onto the Green's function, thereby avoiding ever computing the numerically problematic source $\hat\S_4[T^{(2)\rm eff}]$. 

In this section we formulate a mixed method that exploits the advantages of both approaches: we use the spectral method in the two exterior regions and construct the interior worldtube solution by variation of parameters.

\subsection{Variation of parameters in the worldtube interior}

We first examine the problems that arise when using variation of parameters to solve the second-order Teukolsky equation over the entire domain. We then describe how to restrict it to the worldtube interior and reduce the singularity of the integrand through integration by parts. 


It will be convenient to write the $\s=-2$ Teukolsky equation~\eqref{eq:master eqn} as 
\beq\label{eq:master eqn v2}
    c(r)\partial^{2}_{r}\psi + b(r)\partial_{r}\psi + a(r)\psi = 8\pi d(r)\hat\S_{4,\l\m}[T^{(2)\rm eff}],
\eeq
where we introduce the shorthand $\psi\coloneqq {}_{-2}\psi_{\l\m}$ outside the worldtube and $\psi\coloneqq {}_{-2}\psi^{\cal R}_{\l\m}$ inside, and we note the particularly relevant coefficients
\begin{equation}\label{eq:c and d defs}
c = \Delta, \qquad d = -2r^6(\gamma/\gamma_{\rm K})^2.
\end{equation}
Here we work with $r$ rather than $\sigma$ as it is the more familiar coordinate in the variation-of-parameters approach, and we use non-calligraphic coefficients to distinguish the field equation from the one in compactified coordinates, Eq.~\eqref{eq:teukolsky_compactified}.

The retarded Green's function for Eq.~\eqref{eq:master eqn v2}, derived via variation of parameters, is~\cite{Pound:2021qin,Miller:2023ers}
\begin{multline}\label{eq:glm}
    g_{\l\m}(r,r') = \frac{1}{c(r')W(r')}\bigl[\psi^{\rm up}(r)\psi^{\rm in}(r')\Theta(r-r')\\
    +\psi^{\rm in}(r)\psi^{\rm up}(r')\Theta(r'-r)\bigr],
\end{multline}
where
\begin{equation}
    W = \psi^{\rm in} \partial_r\psi^{\rm up} - \psi^{\rm up} \partial_r\psi^{\rm in}
\end{equation}
is the Wronskian. Here the two homogeneous solutions, $\psi^{\rm up}$ and $\psi^{\rm in}$, correspond to pure outgoing waves at \scri and pure ingoing waves at the horizon, respectively. If we assume $u$ slicing toward \scri and $v$ slicing toward $\mathscr{H}^+$, the solutions behave as  
\begin{align}
    \psi^{\rm up} &\sim \begin{cases}
        r^{3}, & r\to\infty,\\
        A^{\rm up}_{\rm inc}e^{2i\omega r^*} + A^{\rm up}_{\rm ref}\,\psi^{\rm in},& r\to2M,
        \end{cases}\\
    \psi^{\rm in}&\sim\begin{cases}
        A^{\rm in}_{\rm inc}r^{-1}e^{-2i\omega r^*} + A^{\rm in}_{\rm ref}\,\psi^{\rm up},& r\to\infty,\\
        \Delta^2, & r\to2M,\\
    \end{cases}
\end{align}
with constant incidence and reflection coefficients $A^{\rm up/in}_{\rm inc/ref}$.

One might suppose that the physical solution to Eq.~\eqref{eq:master eqn v2} is
\begin{subequations}
    \begin{align}
        \psi &= \int_{2M}^\infty g_{\l\m}(r,r')d(r')S^{(2)}_{4,\l\m}(r')dr'\\
        &= \frac{\psi^{\rm up}}{\Delta^{-1}W}\int_{2M}^r \frac{e^{2i\omega_\m k'}\psi^{\rm in}(r')}{\Delta(r')c(r')}d(r')S^{(2)}_{4,\l\m}(r')dr'\no\\
        &\quad +\frac{\psi^{\rm in}}{\Delta^{-1}W}\int_{r}^\infty \frac{e^{2i\omega_\m k'}\psi^{\rm up}(r')}{\Delta(r') c(r')}d(r')S^{(2)}_{4,\l\m}(r')dr',\label{eq:ret integral}
    \end{align}
\end{subequations}
where $k'\coloneqq k(r'^*)$ is the height function at the integration point, and we used the facts that $\psi^{\rm in/up}_{[s]}=e^{-i\omega_\m k}\psi^{\rm in/up}_{[t]}$~\cite{Miller:2023ers} and that $\Delta^{-1}W$ is independent of $r$ in $t$ slicing~\cite{Pound:2021qin}.
However, with the integral written in the form~\eqref{eq:ret integral}, we immediately see problems arise. Near $r'=\infty$, the second integrand behaves as $\sim e^{2i\omega_\m k'}r'^{5}S^{(2)}_{4,\l\m}$. Even in the Bondi-Sachs gauge, this blows up linearly with $r'$, as we can see from Eq.~\eqref{eq:source falloff Bondi-Sachs}; in the Lorenz gauge, the integrand blows up as $r'^3$.

This divergence of the retarded integral is of a different nature than the infrared divergences related to gravitational memory in Lorenz-gauge calculations~\cite{Cunningham:2024dog}. This divergence is instead due to a well-known pathology of the Teukolsky equation: the retarded Green's function does not pick out the physical solution when the source does not decay extremely rapidly at large $r$~\cite{Poisson:1996ya} (nor, indeed, does it pick out \emph{any} solution, since the integral diverges).

One could construct a puncture at infinity, carried to sufficiently high order in $1/r$ to yield a convergent integral and the correct solution~\cite{Miller:2023ers}. But even then, this approach would fight against the goal of hyperboloidal slicing. The source and the physical solution on slices of constant $s$ should be slowly varying, with no oscillations, because the slice asymptotes to wavefronts toward \scri and $\mathscr{H}^+$; limiting the fields' variation in this way should then lead to rapid numerical convergence. The Green's-function method spoils this advantage by introducing rapidly oscillating factors $e^{\pm2i\omega_\m r^*}$ into the integrands. 

On the other hand, variation of parameters does not have any of these drawbacks if we restrict its use to inside the worldtube. We can write the general solution to Eq.~\eqref{eq:master eqn v2} for points $r\in(r_-,r_+)$ as
\begin{equation}\label{eq:vop v1}
    \psi^{\cal R} = b^{\rm up}\psi^{\rm up}  + b^{\rm in}\psi^{\rm in}+\psi^{\rm inh},
\end{equation}
where we introduced an explicit residual-field label as a reminder that this applies for points in the worldtube. Here $b^{\rm up}$ and $b^{\rm in}$ are constants, and the inhomogeneous solution $\psi^{\rm inh}$ is the integral~\eqref{eq:ret integral} with limits of integration restricted to $r_\pm$:
\begin{equation}\label{eq:psi inh v1}
    \psi^{\rm inh}(r) = C^{\rm in}(r)\psi^{\rm in}(r) + C^{\rm up}(r) \psi^{\rm up}(r),
\end{equation}
with
\begin{align}
    C^{\rm in}(r) &= 8\pi\int^{r_+}_{r}\phi^{\rm up}(r')\hat\S_{4,\l\m}[T^{(2)\rm eff}] dr',\label{eq:C1}\\
    C^{\rm up}(r) &= 8\pi\int^{r}_{r_-}\phi^{\rm in}(r')\hat\S_{4,\l\m}[T^{(2)\rm eff}] dr',\label{eq:C2}
\end{align}
and 
\begin{equation}\label{eq:phi up/in def}
\phi^{\rm in/up}(r)\coloneqq\frac{d(r) \psi^{\rm in/up}(r)}{c(r)W(r)}.    
\end{equation}
Unlike in Eq.~\eqref{eq:ret integral}, to keep expressions compact we have not moved $\Delta^{-1}W$ outside the integrals.

The general solution~\eqref{eq:vop v1} can now replace the spectral ansatz~\eqref{eq:chebyshev_collocation} inside the worldtube. In place of the Chebyshev coefficients, it contains two unknowns, $b^{\rm in}$ and $b^{\rm up}$. These unknowns can be determined through the junction conditions~\eqref{eq:inner_wall_jump}--\eqref{eq:outer_wall_jump_derivative}, as we detail in the next section. 

However, our motivation for using variation of parameters is to avoid the extra derivatives in $\hat\S_{4,\l\m}[T^{(2)\rm eff}]$. In the remainder of this section, we remove those derivatives, integrating by parts to move them off of $T^{(2)\rm eff}_{ab}$ and onto $\phi^{\rm in/up}$,

To most easily integrate by parts while remaining agnostic to the choice of tetrad, we define a one-dimensional radial version of Wald's adjoint~\cite{Wald:1978vm}. For any linear operator $\hat{\cal Q}$ involving radial derivatives and functions of $r$, its adjoint $\hat{\cal Q}^\dagger$ is defined by
\begin{equation}
    A \hat{\cal Q}B = B \hat{\cal Q}^\dagger A +\frac{\partial}{\partial r} \hat{\cal B}_{\cal Q}(A,B),
\end{equation}
for any two sufficiently differentiable $A$ and $B$, where $\hat{\cal B}_{\cal Q}(A,B)$ is bilinear in $A$ and $B$. For example, $\partial_r^\dagger = -\partial_r$, with $\hat{\cal B}_{\partial_r}(A,B)=AB$, and we always have $(\hat{\cal P}\hat{\cal Q})^\dagger = \hat{\cal Q}^\dagger\hat{\cal P}^\dagger$ for any linear radial operators $\hat{\cal P}$ and $\hat{\cal Q}$. We require that if $A$, $B$, and $\hat{\cal Q}$ are all boost weighted (as in our case), with weights $\b_A$, $\b_B$, and $\b_{\cal Q}$, then the sum of boost weights is zero (as is true in our case): $\b_A = -\b_B-\b_{\cal Q}$, ensuring that integrals are boost invariant.

If we now note that $\hat\S_{4,\l\m}[T]$, as defined in Eq.~\eqref{eq:GHPS modes}, is shorthand for $\hat\S^{ab}_{4,\l\m}T_{ab}$, then the above definition implies
\begin{equation}\label{eq:int phi S4}
    \int_{r_1}^{r_2}\phi \hat\S_{4,\l\m}[T]dr' = \hat{\cal B}_{{\cal S}_4}(\phi,T)\bigr|^{r_2}_{r_1} + \int_{r_1}^{r_2}T_{ab} \hat\S^\dagger_{4,\l\m}[\phi,T]dr'
\end{equation}
for any $\phi$ and $T_{ab}$. We can immediately read off
\begin{align}\label{eq:GHPSdagger}
    T_{ab}\,\hat\S^{ab\,\dagger}_{4,\l\m}[\phi] &= -\frac{\mu^{\l}_{\ 2}}{4r^2} T_{nn}\phi + \sqrt{{}_{-2}\lambda_\l} T_{n\mb}(\thorn'^\dagger_\m -3\rho')\frac{\phi}{\sqrt{2}r}\nonumber\\
    &\qquad -\frac{1}{2}T_{\mb\mb}(\thorn'^\dagger_\m-\rho')(\thorn'^\dagger_\m-5\rho')\phi
\end{align}
from Eq.~\eqref{eq:GHPS modes}. Specializing to $t$ slicing (meaning $H=0$), we can write Eq.~\eqref{eq:thornp modes} as
\begin{equation}
    \thorn'_\m  = 
    (-i\omega n^t + n^r\partial_r + \b\, \partial_r n^r),
\end{equation}
and a brief calculation determines
\begin{equation}\label{eq:thornpdagger}
    \thorn'^\dagger_\m = (-i\omega n^t - n^r\partial_r - \b\, \partial_r n^r).
\end{equation}
Note that $\b$ here is the boost weight of the object on which $\thorn'^\dagger_\m$ acts. The homogeneous fields $\psi^{\rm in/up}$ are solutions for the tetrad-independent master variable and therefore have boost weight zero; consequently, their Wronskian $W$ also has boost weight zero. 
Since $d\propto(\gamma/\gamma_{\rm K})^2$ has boost weight $+2$, the definition~\eqref{eq:phi up/in def} implies that $\phi^{\rm in/up}$ has boost weight $+2$. 
Each application of $\thorn'^\dagger_\m$ lowers the boost weight by one. Thus, in the double-adjoint term in Eq.~\eqref{eq:GHPSdagger}, the rightmost operator acts with $\b=2$, and the leftmost operator acts with $\b=1$.

When using Eq.~\eqref{eq:int phi S4} to rearrange the integrals~\eqref{eq:C1} and \eqref{eq:C2} in the solution~\eqref{eq:vop v1}, we observe two things about the boundary terms. First, boundary terms at the worldtube boundaries $r_\pm$ simply contribute constants multiplying $\psi^{\rm in/up}$ in Eq.~\eqref{eq:psi inh v1}, which can be absorbed into a redefinition of the coefficients $b^{\rm in/up}$ in Eq.~\eqref{eq:vop v1}. Second, boundary terms at $r$ that arise from a single radial derivative in $\hat\S_{4,\l\m}$ will automatically cancel in Eq.~\eqref{eq:psi inh v1} because they come with opposite sign from the upper limit of Eq.~\eqref{eq:C2} and the lower limit of Eq.~\eqref{eq:C1}; for example, the boundary terms at $r$ arising from an expression of the form
\begin{multline}\label{eq:1 deriv ints}
    \psi^{\rm in}(r)\int_r^{r_+}\phi^{\rm up}(r')\alpha(r')\partial_r\beta(r') dr' \\
    + \psi^{\rm up}(r)\int^r_{r_-}\phi^{\rm in}(r')\alpha(r')\partial_r\beta(r') dr'
\end{multline}
evaluate to
\begin{equation}\label{eq:1 deriv bdry terms}
    -\psi^{\rm in}\phi^{\rm up}\alpha\beta 
    + \psi^{\rm up}\phi^{\rm in}\alpha\beta = 0 
\end{equation}
by virtue of the definition~\eqref{eq:phi up/in def}, for any $\alpha$ and $\beta$. 

Therefore, the only boundary terms of interest are those at $r$ arising from second radial derivatives in $\hat\S_{4,\l\m}$. If we replace $\partial_r$ with $\partial_r^2$ in Eq.~\eqref{eq:1 deriv ints}, then integrating by parts twice yields the following boundary terms at $r$:
\begin{equation}
    \psi^{\rm in}\partial_r(\alpha\phi^{\rm up})\beta 
    - \psi^{\rm up}\partial_r(\alpha\phi^{\rm in})\beta = \frac{\alpha d}{c}\beta, 
\end{equation}
where the dependence on $\psi^{\rm in/up}$ and their derivatives have canceled with the Wronskian in Eq.~\eqref{eq:phi up/in def}. 
Referring to Eq.~\eqref{eq:GHPS modes}, we see that the only second derivative in $\hat\S_{4,\l\m}$ acts on $T_{\mb\mb}$, and we read off $\alpha=-\frac{1}{2}(n^r)^2$ and $\beta=8\pi T^{(2)\rm eff}_{\mb\mb}$. So, the only boundary contribution is a ``local'' term at $r$,
\beq
    \psi^{\rm loc} = -\frac{(n^{r})^2 d}{2c}8\pi T^{(2)\rm eff}_{\mb\mb}.
    \label{eq:local_term_kernel}
\eeq
Equations~\eqref{n def} and \eqref{eq:c and d defs} imply that this reduces to
\beq
    \psi^{\rm loc} = \frac{r^{2}\Delta}{4}8\pi T^{(2)\rm eff}_{\mb\mb}
    \label{eq:local_term_coordinate_form}
\eeq
in any tetrad.

Putting together these results, we use Eq.~\eqref{eq:int phi S4} to rewrite the general solution~\eqref{eq:vop v1} as
\begin{equation}\label{eq:vop v2}
    \psi^{\cal R} = c^{\rm up}\psi^{\rm up} + c^{\rm in}\psi^{\rm in} + \zeta^{\rm inh},
\end{equation}
where the coefficients $c^{\rm up/in}$ differ from $b^{\rm up/in}$ by boundary terms at $r_\pm$, and the inhomogeneous solution is
\begin{equation}\label{eq:zeta inh}
    \zeta^{\rm inh}(r) = D^{\rm in}(r)\psi^{\rm in}(r) + D^{\rm up}(r) \psi^{\rm up}(r) + \psi^{\rm loc}(r),
\end{equation}
with the modified coefficients
\begin{align}
    D^{\rm in}(r) &= 8\pi\int^{r_+}_{r}T^{(2)\rm eff}_{ab}\hat\S^{ab\,\dagger}_{4,\l\m}[\phi^{\rm up}]dr',\label{eq:Din}\\
    D^{\rm up}(r) &= 8\pi\int_{r_-}^{r}T^{(2)\rm eff}_{ab}\hat\S^{ab\,\dagger}_{4,\l\m}[\phi^{\rm in}]dr'.\label{eq:Dup}
\end{align}
Note that (i) $\zeta^{\rm inh}$ must include the local term to be an inhomogeneous solution, and (ii) it is a \emph{different} inhomogeneous solution than $\psi^{\rm inh}$ due to the absorption of boundary terms into $c^{\rm up/in}$.

We can make Eq.~\eqref{eq:vop v2} more concrete by evaluating the integrand with a particular choice of tetrad. Using Eq.~\eqref{eq:GHPSdagger} with Eqs.~\eqref{eq:thornpdagger} and \eqref{n def}, one finds that in the Carter tetrad ($\gamma=1$), the integrand evaluates to
\begin{align}
    &T_{ab}\,\hat\S^{ab\,\dagger}_{4,\l\m}[\phi]\no\\
    &\ =-\frac{\mu^\l_{\ 2}}{4r^2}T_{nn} \phi\no\\
    &\ \quad +\frac{\sqrt{{}_{-2}\lambda_\l}}{\sqrt f} T_{n\mb} \Biggl[\frac{f}{2r}\, \partial_r\phi +\biggl( \frac{5M-2r}{r^3} - \frac{i\omega_{\m}}{2r} \biggr) \phi \Biggr] \nn\\
    &\ \quad +T_{\mb\mb} \Biggl\{
    -\frac f4 \partial_r^2\phi +\left(\frac{i\omega_{\m}}{2} +\frac{3}{2r}-\frac{4M}{r^2}\right)
    \partial_r\phi \nn\\
    &\quad\qquad\qquad +\biggl[ \frac{\omega_{\m}^2}{4f}
    -\frac{i\omega_{\m}(3r-7M)}{2r^2f} \nn\\
    &\quad\qquad\qquad\qquad-\frac{5r^2-28Mr+36M^2}{2r^4f} \biggr]\phi \Biggr\}.
    \label{eq:vop_carter_expanded_source}
\end{align}

\subsection{Junction conditions at the worldtube boundaries}

The general solution~\eqref{eq:vop v2} inside the worldtube contains two unknowns, $c^{\rm up}$ and $c^{\rm in}$, which must be determined through the junction conditions at the worldtube boundary. Since the exterior solution is obtained for rescaled functions of $\sigma$, we first convert the functions in the worldtube using the rescaling in Eq.~\eqref{eq:master_rescaling}: $\tilde\psi^{\cal R} = (1-\sigma)^{-2}\sigma^3\psi^{\cal R}$, $\tilde\psi^{\rm up/in} = (1-\sigma)^{-2}\sigma^3\psi^{\rm up/in}$, and so on. The rescaled general solution in the worldtube is then
\begin{equation}\label{eq:vop v2 compactified}
    \tilde\psi^{\cal R} = c^{\rm up}\tilde\psi^{\rm up} + c^{\rm in}\tilde\psi^{\rm in} + \tilde\zeta^{\rm inh}.
\end{equation}

At the worldtube boundaries $\sigma_\pm$, the fields must satisfy the junction conditions~\eqref{eq:inner_wall_condition_field}--\eqref{eq:outer_wall_condition_derivative}, which we can write here as
\begin{align}
    c^{\rm up}\tilde\psi^{\rm up}_\pm + c^{\rm in}\tilde\psi^{\rm in}_\pm
    &= \tilde\psi^{[t]}_\pm - \tilde\zeta^{\rm inh}_\pm \pm J_\mp,\label{eq:vop BCs 1}\\
    c^{\rm up}\partial_\sigma\tilde\psi^{\rm up}_\pm + c^{\rm in}\partial_\sigma\tilde\psi^{\rm in}_\pm
    &= \partial_\sigma\tilde\psi^{[t]}_\pm \nonumber\\
    &\quad - \partial_\sigma\tilde\zeta^{\rm inh}_\pm
    \mp \frac{2M}{\sigma_\pm^2}J'_\mp.\label{eq:vop BCs 2}
\end{align}
where a $\pm$ subscript denotes evaluation at $\sigma_\pm$ (\emph{after} taking derivatives).

To combine this with our spectral collocation method outside the worldtube, we include $c^{\rm in}$ and $c^{\rm up}$ among our discrete set of unknowns---the other unknowns being the Chebyshev coefficients $V^{(j)}_{n}$ from Eq.~\eqref{eq:chebyshev_collocation} in the exterior regions. The field equations outside the worldtube require one boundary condition at each worldtube wall, and  $c^{\rm in}$ and $c^{\rm up}$ require two additional conditions. Equations~\eqref{eq:vop BCs 1} and~\eqref{eq:vop BCs 2} together supply these four necessary conditions.

We evaluate the quantities $\tilde\psi^{\rm in/up}$ and $\tilde\zeta^{\rm inh}$ entering Eqs.~\eqref{eq:vop BCs 1}-\eqref{eq:vop BCs 2} by representing the homogeneous solutions spectrally in terms of $\sigma$. Radial derivatives appearing in Eq.~\eqref{eq:vop_carter_expanded_source} are evaluated as
\begin{align}
    \partial_{r}\phi &= -\frac{\sigma^{2}}{2M}\partial_{\sigma}\phi, \\
    \partial^{2}_{r}\phi &= \frac{\sigma^{4}}{4M^{2}}\partial^{2}_{\sigma}\phi
    + \frac{\sigma^{3}}{2M^{2}}\partial_{\sigma}\phi.
    \label{eq:radial_derivative_expressions}
\end{align}
We evaluate the homogeneous solutions $\tilde\psi^{\rm in/up}$ and their first derivatives by integrating the homogeneous $t$-slicing equation as a first-order system across the worldtube. 
Their second derivatives are obtained directly from the homogeneous equation, while the first and second derivatives of $\phi^{\rm in/up}$ are constructed analytically from Eq.~\eqref{eq:phi up/in v2}, using Abel's identity for the derivatives of the Wronskian.

The local term~\eqref{eq:local_term_coordinate_form}, rescaled and expressed in the compactified coordinate, is given by
\beq
    \tilde\psi^{\rm loc}(\sigma) = \frac{4M^4}{\sigma(1-\sigma)} 8\pi T^{(2)\rm eff}_{\mb\mb}(\sigma).
    \label{eq:local_term_compactified}
\eeq
The integral terms $\tilde\zeta^{\rm inh}_{\pm}$ in Eqs.~\eqref{eq:vop BCs 1} and \eqref{eq:vop BCs 2} simplify since one or the other of the integrals~\eqref{eq:Din} or \eqref{eq:Dup} vanishes at the worldtube boundaries, reducing Eq.~\eqref{eq:zeta inh} to
\begin{align}
    \tilde\zeta^{\rm inh}_+ &= D^{\rm in}_+\tilde\psi^{\rm in}_+ + \tilde\psi^{\rm loc}_+ ,\\
    \tilde\zeta^{\rm inh}_- &= D^{\rm up}_-\tilde\psi^{\rm up}_- + \tilde\psi^{\rm loc}_- .
\end{align}
Similarly,
\begin{align}
    \partial_\sigma\tilde\zeta^{\rm inh}_+ &= D^{\rm in}_+\partial_\sigma\tilde\psi^{\rm in}_+ + \partial_\sigma\tilde\psi^{\rm loc}_+  \no\\
    &\quad +\frac{2M}{\sigma^2_+}\bigl(8\pi T^{(2)\rm eff}_{ab}\hat\S^{ab\,\dagger}_{4,\l\m}[\phi^{\rm up}]\bigr)_+ \tilde\psi^{\rm in}_+ \no\\
    &\quad -\frac{2M}{\sigma^2_+}\bigl(8\pi T^{(2)\rm eff}_{ab}\hat\S^{ab\,\dagger}_{4,\l\m}[\phi^{\rm in}]\bigr)_+\tilde\psi^{\rm up}_+,\\
    \partial_\sigma\tilde\zeta^{\rm inh}_- &= D^{\rm up}_-\partial_\sigma\tilde\psi^{\rm up}_- + \partial_\sigma\tilde\psi^{\rm loc}_-  \no\\
    &\quad +\frac{2M}{\sigma^2_-}\bigl(8\pi T^{(2)\rm eff}_{ab}\hat\S^{ab\,\dagger}_{4,\l\m}[\phi^{\rm up}]\bigr)_- \tilde\psi^{\rm in}_- \no\\
    &\quad -\frac{2M}{\sigma^2_-}\bigl(8\pi T^{(2)\rm eff}_{ab}\hat\S^{ab\,\dagger}_{4,\l\m}[\phi^{\rm in}]\bigr)_-\tilde\psi^{\rm up}_-.    
\end{align}

To evaluate the integrals $D^{\rm in/up}_\pm$, we change the integration variable to $\sigma$, using
\beq
\int^{r_+}_{r_-}dr' = 2M\int^{\sigma_+}_{\sigma_-}\frac{d\sigma'}{\sigma'^2}.
\eeq
In practice, each of the integrals is split at the particle's position $\sigmap$. The integrands are therefore interpolated on separate smooth intervals, and no derivative is taken through the position of the particle.

Finally, we stress that $\tilde\psi^{\rm in/up}$ are the homogeneous solutions to Eq.~\eqref{eq:teukolsky_compactified} in $t$ slicing, since we use $t$ slicing throughout the worldtube interior. The fields $\phi^{\rm in/up}$ defined in Eq.~\eqref{eq:phi up/in def} can be written in terms of these homogeneous solutions as
\begin{equation}\label{eq:phi up/in v2}
\phi^{\rm in/up} = -\frac{2M\sigma}{(1-\sigma)^2}\frac{d(\sigma) \tilde\psi^{\rm in/up}(\sigma)}{c(\sigma)\tilde W(\sigma)},    
\end{equation}
where $\tilde W \coloneqq \tilde\psi^{\rm in}\partial_\sigma\tilde\psi^{\rm up}-\tilde\psi^{\rm up}\partial_\sigma\tilde\psi^{\rm in}$, and $c$ and $d$ are given in Eq.~\eqref{eq:c and d defs} with $r=2M/\sigma$.

\section{Results}
\label{sec:results}
In this section we present the central numerical results of this work: the first self-force calculations of the second-order gravitational-wave energy flux obtained directly from the second-order Teukolsky equation. 
We compute the oscillatory $(\l,\m)=(2,1)$ and $(2,2)$ modes of the flux for quasicircular orbits at 27 orbital radii across the range $7M\leq \rp\leq20M$. 
The calculation uses the effective source inside the worldtube, the quadratic and slow-evolution sources outside it, and the junction conditions of Eqs.~\eqref{eq:inner_wall_jump}-\eqref{eq:outer_wall_jump_derivative}.
The global linear system is solved with the mixed scheme of \secref{mixed method}: multidomain spectral collocation in the two exterior regions, with the worldtube interior treated by variation of parameters in the integrated-by-parts form of Eq.~\eqref{eq:zeta inh}, such that the derivatives of ${\hat {\cal S}}_{4,\l\m}$ act on the smooth kernels $\phi^{\rm in / up}$ rather than on the effective-source components.

We highlight two features of these calculations at the outset. 
First, the numerical domain covers the entire exterior spacetime, with \scri{} and $\mathscr{H}^{+}$ included as the grid points $\sigma=0$ and $\sigma=1$, where the field equation reduces to the regularity conditions~\eqref{eq:sigma=0 eqn} and~\eqref{eq:sigma=1 eqn}.
The radiation amplitude is therefore read off from the endpoint values of the rescaled field, with no extraction at finite radius and no extrapolation. 
Second, because the first-order field has been transformed to the Bondi-Sachs gauge, the source remains regular all the way to $\sigma=0$, and no puncture at infinity of the kind constructed in \secref{compactification} is required.

\begin{figure*}
    \centering
    \includegraphics[width=\textwidth]{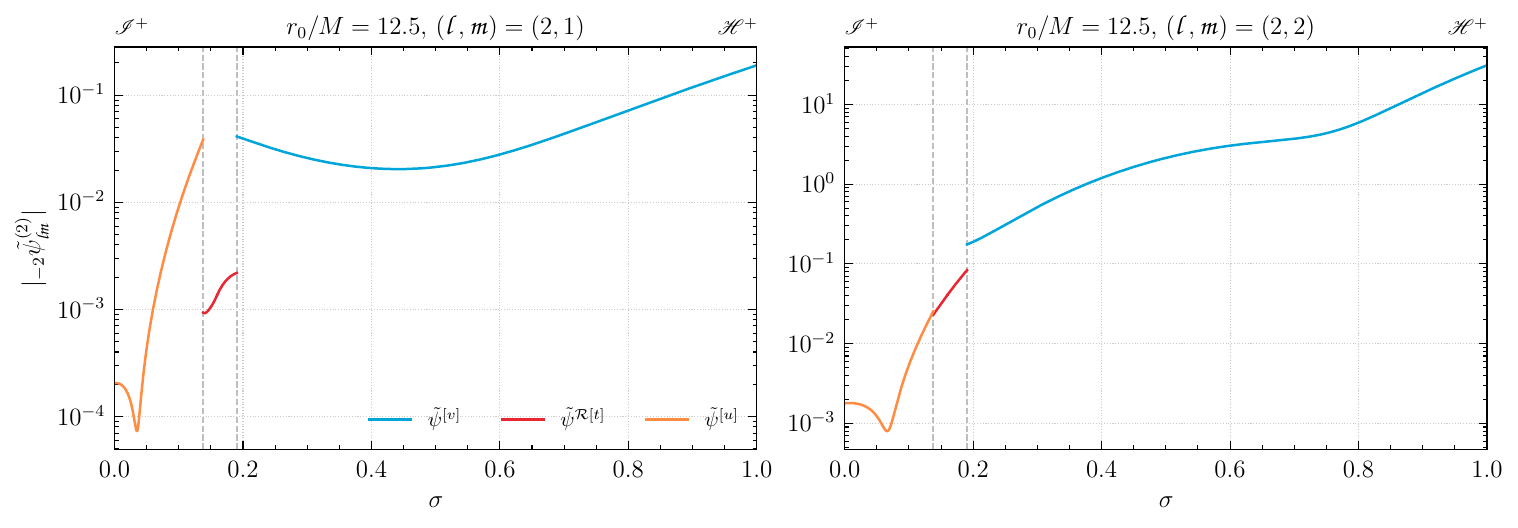}
    \caption{Magnitude of the rescaled second-order Teukolsky master variable,
    ${}_{-2}\tilde\psi^{(2)}_{\l\m}$, for an orbit at $r_0=12.5M$.
    The left and right panels show the $(\l,\m)=(2,1)$ and $(2,2)$ modes,
    respectively, over the compactified interval $\sigma=2M/r$,
    from future null infinity ($\sigma=0$) to the future horizon ($\sigma=1$).
    Blue and orange curves show the exterior fields in $v$ and $u$ slicing,
    respectively; red curves show the residual field in $t$ slicing inside
    the worldtube. The grey dashed lines mark the worldtube boundaries.
    The regional fields satisfy the junction conditions, with the offsets
    between them accounting for the puncture, gauge, and slow-evolution
    matching terms.}
    \label{fig:master_variable}
\end{figure*}

Figure~\ref{fig:master_variable} illustrates the radial solutions for an orbit at $r_{0}/M = 12.5$. 
The rescaled fields remain finite at both the horizon and future null infinity. 
Their complex boundary values at $\mathscr{I}^{+}$ provide the second-order amplitudes entering
the flux calculation below.

\subsection{Energy flux at \scri}
\label{sec:flux definition}

The gravitational-wave content at \scri{} is carried by the boundary value of the rescaled master variable of Eq.~\eqref{eq:master_rescaling}. 
For brevity, we write these boundary values as
\beq
    \tilde\psi^{(n)}_{\mathscr{I}^{+}}
    \coloneqq {}_{-2}\tilde\psi^{(n)[u]}_{\l\m}\bigr|_{\sigma=0},
    \qquad n=1,2,
    \label{eq:psi_scri_definition}
\eeq
leaving the mode labels implicit.
Both perturbative orders use the same rescaling and $u$ slicing. 
In the numerical units $M=1$, the flux coefficients for a single $\m>0$ mode are
\begin{align}
    {\cal F}^{(1)}_{\l\m}
    &=\frac{\bigl|\tilde\psi^{(1)}_{\mathscr{I}^{+}}\bigr|^2}{256\pi\omega_\m^2},
    \label{eq:flux1_result}\\
    {\cal F}^{(2)}_{\l\m}
    &=\frac{\Re\bigl[\overline{\tilde\psi^{(1)}_{\mathscr{I}^{+}}}\,\tilde\psi^{(2)}_{\mathscr{I}^{+}}\bigr]}
    {128\pi\omega_\m^2}.
    \label{eq:flux2_result}
\end{align}
We only compute modes with one sign of $\m$; the $-\m$ modes follow from the axial symmetry of the metric perturbation for a quasi-circular inspiral, so when comparing against a result summed over both signs of $\m$ we simply double the flux of $\m>0$.

Two distinct second-order flux quantities appear in our comparisons, and some care is required in distinguishing them. 
The first is ${\cal F}^{(2)}_{\l\m}$ itself. The amplitude $\tilde\psi^{(2)}_{\mathscr{I}^{+}}$ is sourced by every contribution assembled in \secref{source}: the quadratic source, the slow-evolution source $G^{(1,1)}_{\mu\nu}$ (which carries the leading inspiral rate $\dot{r}_{p}$), and the worldtube jumps described above. 
These are precisely the contributions contained in the tabulated Lorenz-gauge fluxes of Ref.~\cite{Warburton:2021kwk}, and Eq.~\eqref{eq:flux2_result} is therefore the quantity the two numerical calculations share; below we distinguish them with subscripts ``T'' and ``L'' for the Teukolsky and Lorenz-gauge results, respectively. 
The second quantity is the physical second-order flux, which contains one further term:
\begin{align}
    {\cal F}^{(2)\rm full}_{\l\m}
    &= {\cal F}^{(2)}_{\l\m}
    +{\cal F}^{(2){\cal V}}_{\l\m},
    \label{eq:full_flux}\\
    {\cal F}^{(2){\cal V}}_{\l\m}
    &=-\frac{\m F_\Omega^{(0)}}{2\pi\omega_\m^3}
    {\rm Im}\!\left[
    \overline{Z^{(1)}_{\l\m}}
    \partial_{\omega_\m}Z^{(1)}_{\l\m}
    \right],
    \label{eq:slow_time_flux}
\end{align}
where $Z^{(1)}_{\l\m}$ is the first-order Teukolsky amplitude in the convention of Refs.~\cite{Warburton:2021kwk,Warburton:2024xnr}. 
Here $\m F_\Omega^{(0)}=\vec\partial_{\cal V}\omega_\m$ is the leading evolution coefficient, with the explicit power of $\e$ removed. 
This asymptotic contribution arises when the slowly evolving curvature waveform is converted to strain. 
It is distinct from the slow-evolution source and matching terms, and is absent from the tabulated Lorenz-gauge coefficient. We therefore add it only for the PN comparison.

\subsection{Comparison with the Lorenz-gauge calculation}
\label{sec:Lorenz comparison}
The Lorenz-gauge calculation provides an independent solution of the second-order field equations, although the two calculations share first-order inputs. Directly computed reference fluxes are available at 22 of our radii. We use an interpolated comparator at the remaining  5 radii and distinguish these points below.
\begin{figure*}
    \centering
    \includegraphics[width=\textwidth]
    {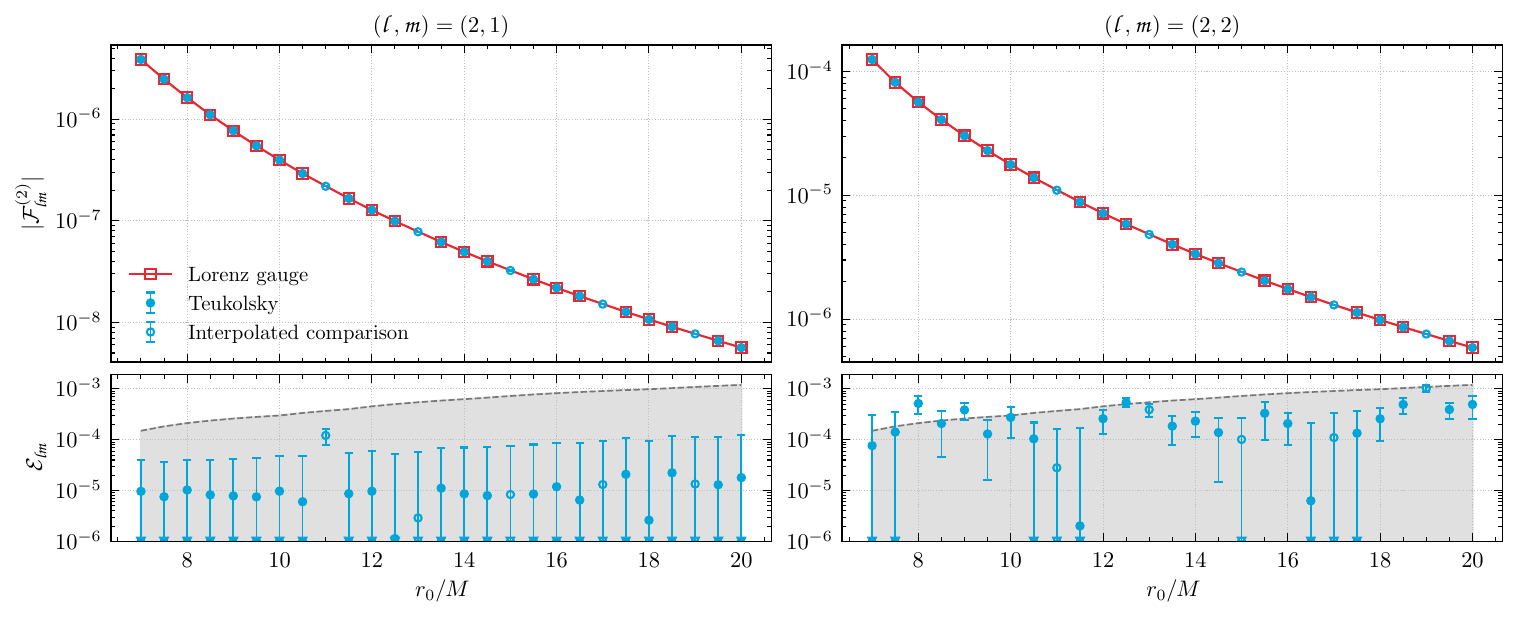}
    \caption{Comparison of the second-order energy flux ${\cal F}^{(2)}_{\l\m}$ computed from the second-order Teukolsky equation (points with error bars) with the Lorenz-gauge results of Ref.~\cite{Warburton:2021kwk}. Filled symbols are the 22 radii at which a Lorenz-gauge flux was computed; open symbols are the five radii at which the comparator is the interpolant~\eqref{eq:reference_interpolant_Lorenz} of those data, and their error bars additionally carry the interpolation uncertainty. The error bars show the numerical error of Eq.~\eqref{eq:error_quadrature}, while the gray bands show the calibrated numerical floor of the Lorenz-gauge calculation given in Eq.~\eqref{eq:Lorenz_floor}; the latter is not included in our error bars.}
    \label{fig:reference_flux_comparison}
\end{figure*}

One observes that the two calculations are indistinguishable on the scale of Fig.~\ref{fig:reference_flux_comparison}.
To resolve their difference, we therefore define the relative error
\beq
    {\cal E}_{\l\m}
    \coloneqq \left|
    \frac{{\cal F}^{(2)}_{\l\m,{\rm T}}}
      {{\cal F}^{(2)}_{\l\m,{\rm L}}}-1\right|.
    \label{eq:relative_flux_residual}
\eeq
The agreement is tighter for $(2,1)$: at the directly tabulated radii, the median relative difference is $8.65\times10^{-6}$, compared with $2.21\times10^{-4}$ for $(2,2)$. 
The corresponding root-mean-square differences are $1.11\times10^{-5}$ and $2.97\times10^{-4}$, respectively. Table~\ref{tab:flux_residuals} gives the individual values.

For $(2,2)$, the signed difference ${\cal F}^{(2)}_{22,{\rm T}}/{\cal F}^{(2)}_{22,{\rm L}}-1$ is negative at 20 of the 22 directly tabulated radii, with sign reversals only at $11.5M$ and $16.5M$. 
Since both coefficients are negative, the Teukolsky result is generally smaller in magnitude. 
At $20M$, the relative differences are $1.80\times10^{-5}$ for $(2,1)$ and $4.96\times10^{-4}$ for $(2,2)$.

The fixed-source radial tests give relative flux changes below $1.77\times10^{-11}$ for $(2,1)$ and $6.47\times10^{-9}$ for $(2,2)$, well below these discrepancies. 
The source and matching-data error calculations are less restrictive and show a stronger response in $(2,2)$, particularly for SS angular quadrature.  
These error estimates do not, however, uniquely identify the origin of the Teukolsky-Lorenz difference. 
The reference uncertainty must also be considered: at $20M$, the $(2,2)$ discrepancy lies below the approximate Lorenz-reference envelope of $1.2\times10^{-3}$. 
At this radius, the two results therefore agree within the estimated uncertainty of the Lorenz-gauge reference.

At radii without a tabulated Lorenz-gauge flux, we use the interpolant
\beq
    {\cal F}^{(2),{\rm int}}_{\l\m}(r_{0})
    = r_{0}^{-6}P_4(M/r_{0}),
    \label{eq:reference_interpolant_Lorenz}
\eeq
where $P_4$ is a polynomial of degree four fitted to $r_0^6{\cal F}^{(2)}_{\l\m,{\rm L}}$ at tabulated radii within $5M$ of the target. 
Multiplication by $r^{6}_{0}$ removes most of the steep radial decay, leaving a slowly varying function for the polynomial to approximate. 
Across the tabulated range, the scaled magnitudes vary by factors of only $\sim 1.3$ and $\sim 2.6$ for $(2,1)$ and $(2,2)$, respectively. 
This is a numerical rescaling, not an assumption that both modes have the same leading PN power, and no PN coefficients are imposed. 
We estimate the interpolation uncertainty by omitting nearby tabulated points in turn and comparing the fit with the omitted values. 
The five interpolated comparators are shown as open circles in Fig.~\ref{fig:reference_flux_comparison} and are excluded from the reference comparison statistics.

\subsection{Comparison with the 4.5PN expansion}
\label{sec:PN comparison}
We compare the physical flux~\eqref{eq:full_flux} with the 4.5PN expansion of Ref.~\cite{Blanchet:2023bwj} at all radii. 
Following that reference, we introduce $x=(M\Omega)^{2/3}=M/r_{0}$ and the symmetric mass ratio $\nu\coloneqq mM/(m+M)^2$. 
Converting from the expansion in $\e$ at fixed $M$ gives
\beq
 {\cal F}^{(2)}_{\nu,\l\m}
 = {\cal F}^{(2)}_{\e,\l\m}
 +2(2+\delta M_{\rm conv}){\cal F}^{(1)}_{\l\m}
 -\frac{2}{3}x(1+\delta M_{\rm conv})
 \partial_x{\cal F}^{(1)}_{\l\m},
 \label{eq:epsilon-nu-flux-conversion}
\eeq
where ${\cal F}^{(2)}_{\e,\l\m}={\cal F}^{(2)\rm full}_{\l\m}$ and $\delta M_{\rm conv}=-x/\sqrt{1-3x}$. 
The latter is a mass-convention term, distinct from the evolving parameters $\delta M_A$. 
Including both signs of $\m$ and dividing by the Newtonian flux factor gives the plotted coefficient
\beq
    \widehat{\cal F}^{(2)}_{\l\m}
    \coloneqq\frac{2{\cal F}^{(2)}_{\nu,\l\m}}{(32/5)x^5}.
    \label{eq:normalized-PN-coefficient}
\eeq
We evaluate $\partial_x{\cal F}^{(1)}_{\l\m}$ with a quintic spline through 544 first-order amplitudes. 
Replacing it with a cubic spline changes $\widehat{\cal F}^{(2)}_{\l\m}$ by at most $2.98\times10^{-8}$ for $(2,1)$ and $2.45\times10^{-6}$ for $(2,2)$. 
This sensitivity is separate from the second-order radial envelope. 
The flux coefficient is signed; a negative coefficient does not imply a negative total luminosity.
\begin{figure*}
    \centering
    \includegraphics[width=\textwidth]{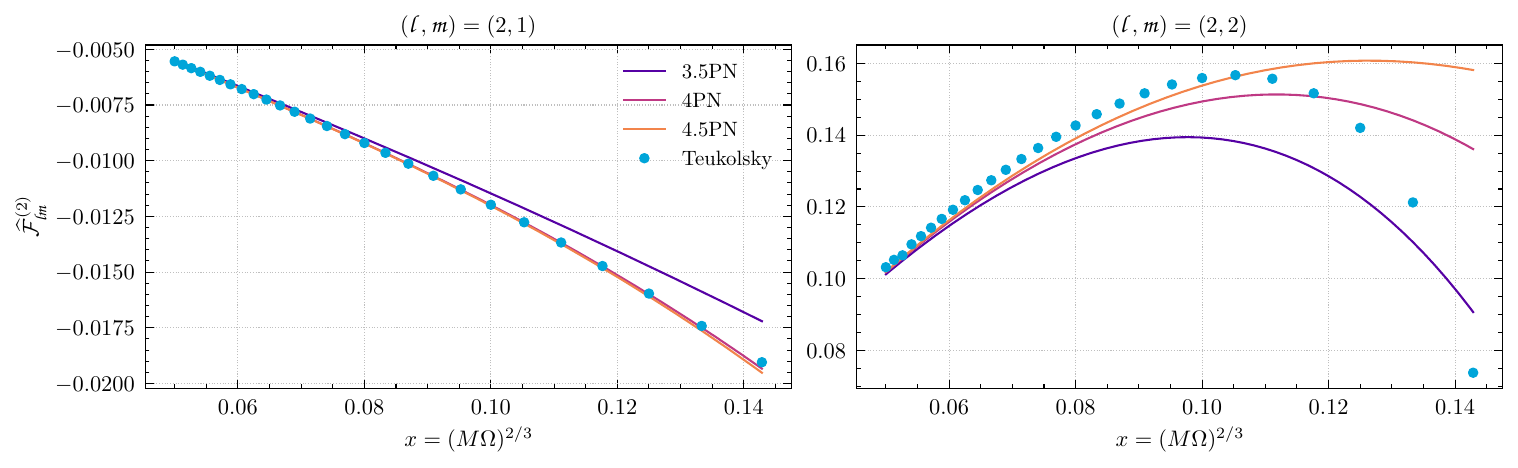}
    \caption{The Newtonian-normalized second-order flux coefficient compared with the 3.5PN, 4PN, and 4.5PN truncations. The Teukolsky points include the asymptotic slow-evolution contribution, the mass-ratio conversion~\eqref{eq:epsilon-nu-flux-conversion}, and both signs of $\m$.}
    \label{fig:4p5pn-mode-comparison}
\end{figure*}

\begin{figure*}
    \centering
    \includegraphics[width=\textwidth]{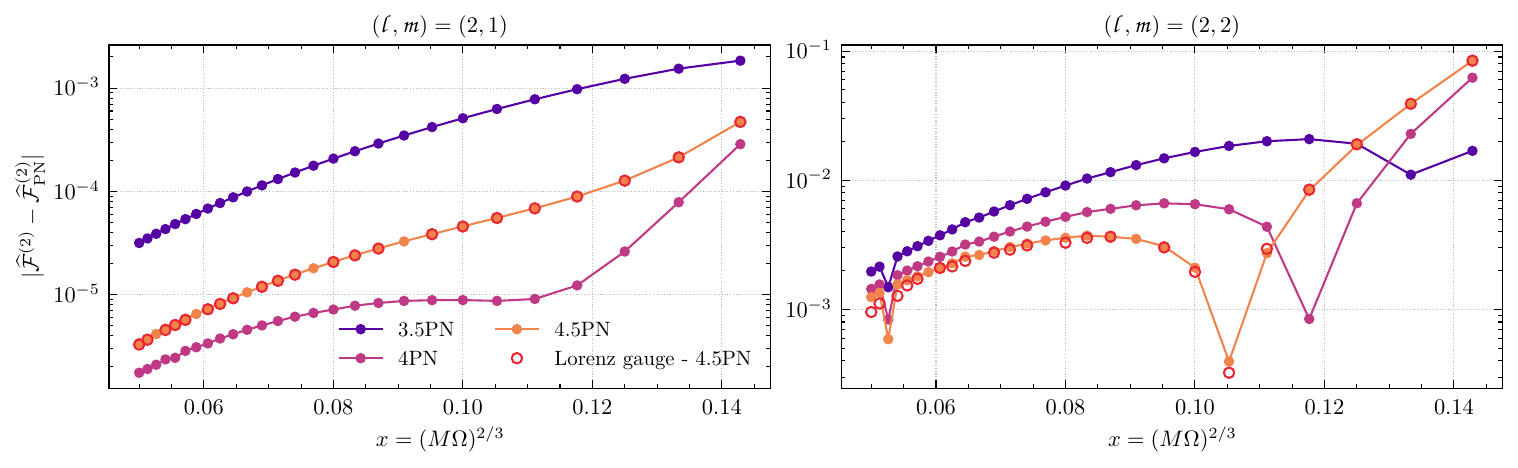}
    \caption{Absolute differences between the normalized numerical coefficient and the post-Newtonian truncations in Fig.~\ref{fig:4p5pn-mode-comparison}. The open circles give the corresponding Lorenz-gauge difference from 4.5PN at directly tabulated radii. These are remainders relative to a truncated analytic series, not estimates of numerical error.}
    \label{fig:4p5pn-residuals}
\end{figure*}

Over the 18 radii with $r_{0}\geq11.5M$, the root-mean-square differences from 4.5PN in the normalized coefficient are $1.33\times10^{-5}$ for $(2,1)$ and $2.57\times10^{-3}$ for $(2,2)$, with maxima $2.81\times10^{-5}$ and $3.73\times10^{-3}$, respectively. 
The mean absolute relative differences from the 4.5PN coefficients are $1.39\times10^{-3}$ and $1.91\times10^{-2}$.

Figures~\ref{fig:4p5pn-mode-comparison} and~\ref{fig:4p5pn-residuals} provide a quantitative check of the weak-field fluxes. 
Over the range $11.5M \leq r_{0} \leq 20M$, increasing the PN order from 3.5PN to 4.5PN reduces the root-mean-square residual by factors of $9.9$ and $2.3$ for $(2,1)$ and $(2,2)$, respectively. 
For $(2,2)$, each increase in PN order improves the agreement at every sampled radius in this range. 
For $(2,1)$, 4PN is closer than 4.5PN in both numerical calculations. 
This ordering is therefore not a feature specific to the Teukolsky result. 
More generally, at directly tabulated radii in this range, the Teukolsky--Lorenz difference is at most $3.0\%$ of the Teukolsky-4.5PN residual for $(2,1)$ and $24\%$ for $(2,2)$. 
The bulk of the PN residual is thus common to the two numerical calculations.

Together, these comparisons support the weak-field behaviour of the second-order flux and test the complete construction, including the asymptotic slow-evolution contribution and the mass-ratio conversion. 
The measured cubic-quintic spline sensitivity is below $7\times10^{-5}$ of the 4.5PN residual throughout this range in both modes, so the choice between these spline orders does not account for the residual. 
The PN comparison therefore supplies an analytic cross-check in addition to agreement between the two field-equation solvers. 
It does not, however, separate a PN truncation remainder from errors common to both numerical results, or establish coefficient-by-coefficient agreement through 4.5PN.

\section{Conclusion}
\label{sec:conclusion}

We have completed the first second-order gravitational self-force calculation based on the Teukolsky formalism and present numerical results for the radiated flux in the $\l=2$ mode for quasi-circular orbits around a Schwarzschild black hole.
Our results agree to $\lesssim 3\times10^{-3}$ (relative) with previous second-order results that were obtained by solving the second-order Einstein equation directly in the Lorenz gauge \cite{Warburton:2021kwk}.

Our scheme assembles several ingredients that have been developed in recent years that were absent from the previous approach.
We transformed the first-order field to the Bondi-Sachs gauge \cite{Spiers:2026yqx} before solving at second order, rather than after as in Refs.~\cite{Wardell:2021fyy,Cunningham:2024dog}, and worked with compactified hyperboloidal slicing, so that $\mathscr{I}^+$ and $\mathcal{H}^+$ are grid points that can be used in our numerical scheme.
This allows the radiation amplitude to be read off at $\mathscr{I}^+$ without any extrapolation from finite radii.
Outside of the particle's worldtube we use a multi-domain spectral collocation method \cite{PanossoMacedo:2022fdi}, and within we use variation of parameters in an integrated-by-parts form, which moves derivatives, that are challenging to compute numerically, off the effective source and onto the smooth homogeneous solutions. 
Our analysis of the infrared behavior of these equations is also a result in its own right:
we find that the second-order Teukolsky equation is free of the hereditary divergences that afflict the second-order Einstein equation in the Lorenz gauge, confirming the conclusion of Ref.~\cite{Spiers:2026yqx}.

There are several natural extensions of the present calculation.
The most immediate is to compute the flux for modes with $\l >2$ to compare with the Lorenz gauge calculation, and to compute the fluxes for larger orbital radii in order to make further comparisons with PN \cite{Warburton:2024xnr}.
Many of the ingredients deployed here are not specific to the Teukolsky formalism, and transforming to the Bondi-Sachs gauge at first order could equally be ported back to the Lorenz-gauge pipeline, removing its asymptotic puncture and its post-hoc gauge transformation \cite{Cunningham:2024dog}.
The Teukolsky and Lorenz-gauge calculations are largely independent, though they share the same first-order inputs.
This presents an opportunity to make detailed comparisons between the calculations that will enable more precise quantification of how numerical error accumulates in each.

Extension to Kerr is the principal motivation for our scheme.
The formalism carries over \cite{Spiers:2023cip}, as does the Bondi-Sachs analysis which underpins our treatment of $\mathscr{I}^+$ \cite{Spiers:2026yqx}.
The main challenge is that the source no longer separates due to the mixed radial and polar dependence.
There are two complementary ways forward.
One can work at the level of $m$-modes and solve the resulting elliptic equations, either for the metric perturbation \cite{Osburn:2026uct} or for the Teukolsky variables \cite{Cownden2026Capra,Vu:2026ypc,PanossoMacedo:2024pox,Osburn:2022bby}.
Alternatively, one can handle the spheroidal-to-spherical mode mixing semi-analytically to decompose the source onto a spheroidal harmonic basis \cite{Spiers:2024src}.
This would allow the spectral solver and variation of parameters method developed here to be ported over.
Whichever approach is taken, two additional ingredients are needed. The first is the first-order field, for which several options now exist, in the Lorenz gauge \cite{Dolan:2021ijg,Wardell:2024yoi}, or via metric reconstruction in a radiation gauge \cite{Green:2019nam,Hollands:2024iqp,vandeMeent:2017bcc,Toomani:2021jlo,Bourg:2024vre,Li:2026rkf}. The second is the second-order puncture in Kerr, which is actively in development \cite{Upton2026Capra}.

Pushing beyond circular orbits to eccentric and precessing systems brings new challenges, particularly due to the slow convergence of the Fourier sum over radial or polar harmonics that stems from the finite differentiability of the effective source at the particle's worldline.
Techniques exist to overcome this, but they are comparatively slow to converge \cite{Leather:2023dzj,Lu:2026jzh}.
This is particularly problematic because eccentric and precessing systems cover higher-dimensional parameter spaces and, at each point in the space, have a large set of harmonics to solve for.
Improving efficiency in future calculations will be crucial.

Looking beyond EMRIs to the intermediate mass ratios, the inspiral is no longer the whole signal, and the transition through the ISCO and the subsequent plunge must be modelled as well. A multiscale treatment of both regimes now exists \cite{Kuchler:2024esj,Honet:2025dho,Kuchler:2025hwx} but for each a new set of field equations must be solved. The ingredients provided in this work will also prove useful as these models are pushed to 1PA order.

The application of our methods and results extends beyond modeling binaries in general relativity to alternative theories of gravity. Modeling EMRIs in scalar-tensor theories of gravity has been shown to preserve the general relativity contributions of self-force calculations~\cite{Maselli:2021men,Spiers:2023cva} with the scalar contributions added on top. This modularity enables the streamlined integration of second-order self-force results in general relativity into scalar-tensor models, thereby improving their accuracy in the small-coupling limit~\cite{Barsanti:2026ulr}. There is a parallel investigation~\cite{Dyson:2025dlj} into the inclusion of environmental effects on EMRIs, such as scalar-boson clouds, using perturbation theory, which our results can be integrated into.

\begin{acknowledgments}
We thank Brad Cownden, Zach Nasipak, Nami Nishimura, Tommy Osburn, and Rodrigo Panosso Macedo for helpful discussions. 
BL gratefully acknowledges funding from the European Union’s Horizon Europe research and innovation programme under the Marie Sklodowska-Curie grant agreement No. 101209791.
AP and SDU acknowledge the support of a Royal Society University Research Fellowship and the ERC Consolidator/UKRI Frontier Research Grant GWModels (selected by the ERC and funded by UKRI [grant number EP/Y008251/1]). AP and AS acknowledge the support of a Royal Society University Research Fellowship Enhancement Award. AS acknowledges the partial support
from the STFC Consolidated Grant no. ST/V005596/1 and funding from the European Union's Horizon Europe research and innovation programme under the Marie Sklodowska-Curie grant agreement no. 101199153.
NW acknowledges support from a Royal Society – Research Ireland University Research Fellowship. 
This publication has emanated from research conducted with the financial support of Research Ireland under grant number 22/RS-URF-R/3825. 
This work made use of the Black Hole Perturbation Toolkit \cite{BHPToolkit}.
\end{acknowledgments}


\appendix


\section{Conversion of Lorenz-gauge sources}
\label{sec:Lorenz conversion}

As described in Sec.~\ref{sec:Seff}, to construct the source inside the worldtube we directly import the Lorenz-gauge effective source from Ref.~\cite{Upton:2025bja}. Doing so requires relating the field equation~\eqref{tt EFE2R} to the field equation used in our previous Lorenz-gauge calculations. We derive that relationship in this appendix.

In terms of the Lorenz-gauge linearized Einstein operator $E_{\mu\nu}[\bar h]$, the complete linearized Einstein tensor reads
\beq
G^{(1)}_{\mu\nu}[h] = -\frac{1}{2}E_{\mu\nu}[\bar h] + \overline{\nabla_{(\mu}Z_{\nu)}}[\bar h],
\eeq
where
\begin{align}
 E_{\alpha\beta}[\bar h] &\coloneqq  \Box \bar h_{\alpha\beta}+2R_\alpha{}^\mu{}_\beta{}^\nu \bar h_{\mu\nu},
\end{align}
and $\overline{\nabla_{(\mu}Z_{\nu)}}\coloneqq \left(g_\mu{}^\alpha g_\nu{}^\beta-\tfrac{1}{2}g_{\mu\nu}g^{\alpha\beta}\right)\nabla_{(\alpha}Z_{\beta)}$, with overlines denoting trace reversal. In the Lorenz-gauge calculations in Refs.~\cite{Miller:2020bft,Miller:2023ers,Pound:2019lzj,Warburton:2021kwk,Upton:2025bja}, we instead use a gauge-damped linearized operator,
\beq\label{eq:damped E}
\breve E_{\mu\nu}[\bar h] \coloneqq  E_{\mu\nu}[\bar h] - \frac{4M}{r^2}\n_{(\mu}\breve Z_{\nu)}.
\eeq
Here,  in Schwarzschild coordinates $(t,r,\theta,\phi)$, 
\beq
\n_\alpha\coloneqq -(1,1/f,0,0)=\sqrt{\frac2f}n_\alpha, 
\eeq
with $n_\alpha$ the Carter tetrad leg defined in Sec.~\ref{sec:regularity and tetrad}, and 
\beq
\breve Z_\alpha = (fZ_r,Z_r,Z_\theta,Z_\phi).
\eeq
The modified Lorenz-gauge operator~\eqref{eq:damped E} is inherited from Ref.~\cite{Barack:2005nr}\footnote{Note the damping term given in Ref.~\cite{Barack:2005nr} is erroneous and does not correspond to the mode-decomposed equations derived therein. The correct version, given in Ref.~\cite{Barack:2010tm}, is what appears in our Eq.~\eqref{eq:damped E}.}. In terms of this operator, we have
\beq
G^{(1)}_{\mu\nu}[h] = -\frac{1}{2}\breve E_{\mu\nu}[\bar h] -\frac{2M}{r^2}\n_{(\mu}\breve Z_{\nu)} + \overline{\nabla_{(\mu}Z_{\nu)}}[\bar h].\label{Einstein1Alt} 
\eeq

Now, the second-order Teukolsky equation~\eqref{Opsi} follows from the EFE~\eqref{tt EFE2} by virtue of the operator identity $\hat\S_4\hat\E=\hat\O\hat\T$. That identity can only be applied if $G^{(1,0)}_{\mu\nu}$ appears on the left-hand side of Eq.~\eqref{tt EFE2}, rather than $E^{(0)}_{\alpha\beta}$ or $\breve E^{(0)}_{\alpha\beta}$ (where we use a 0 to denote the replacement $\nabla_\mu\to\nabla^{(0)}_\mu$). For that reason, in the multiscale expansion in Sec.~\ref{sec:multiscale Teukolsky}, we expanded the full EFE $G_{\alpha\beta}=8\pi T_{\alpha\beta}$. However, such an approach differs from the one we took in the Lorenz gauge. There, we wrote the complete EFE as
\beq
\breve E_{\mu\nu}[\bar\hexact] = -16\pi T_{\mu\nu} + 2\breve{G}^{(2)}_{\mu\nu}[\hexact,\hexact] + {\cal O}(|\hexact|^3),
\eeq
where $\hexact_{\mu\nu}\coloneqq \gexact_{\mu\nu}-g_{\mu\nu}$ is the total perturbation, and $\breve{G}^{(2)}_{\mu\nu}$ is given by Eq.~(4) of Ref.~\cite{Pound:2021qin} with $\nabla_\nu\bar h^{\mu\nu}$ set to zero. We then performed a two-timescale expansion of this equation, leading to the hierarchy~\cite{Miller:2020bft}
\begin{align}
 \breve E^{(0)}_{\mu\nu}[\bar h^{(1)}] &= -16\pi T^{(1)}_{\mu\nu},\label{tt EFE1 Lorenz}\\
 \breve E^{(0)}_{\mu\nu}[\bar h^{(2)}] &= -16\pi T^{(2)}_{\mu\nu} + 2\breve G^{(2,0)}_{\mu\nu}[h^{(1)},h^{(1)}]-\breve E^{(1)}_{\mu\nu}[\bar h^{(1)}].\label{tt EFE2 Lorenz}
\end{align}
Again recall the notation from Sec.~\ref{sec:multiscale Teukolsky}: ``$(0)$'' on an operator indicates the operator with the replacement $\nabla_\alpha\to\nabla^{(0)}_\alpha$, and ``$(1)$'' indicates the term linear in ${\cal V}_I$. Explicitly, 
\beq\label{Ebreve}
\breve E^{(1)}_{\mu\nu}[\bar h] = E^{(1)}_{\mu\nu}[\bar h] - \frac{4M}{r^2}\n_{(\mu}\breve Z_{\nu)}^{(1)}[\bar h]
\eeq
with $\breve Z^{(1)}_\alpha = (fZ^{(1)}_r,Z^{(1)}_r,Z^{(1)}_\theta,Z^{(1)}_\phi)$ in Schwarzschild coordinates.

After moving the puncture to the right-hand side of Eq.~\eqref{tt EFE2 Lorenz}, we obtain an effective source
\begin{multline}\label{T2effbreve}
-16\pi \breve T^{(2)\rm eff}_{\mu\nu} \coloneqq  -16\pi T^{(2)}_{\mu\nu} +2\breve G^{(2,0)}_{\mu\nu}[h^{(1)},h^{(1)}]\\
-\breve E^{(1)}_{\mu\nu}[h^{(1)}]-\breve E^{(0)}_{\mu\nu}[h^{(2)\cal P}].
\end{multline}
This is the source used in our previous Lorenz-gauge calculations. In our second-order Teukolsky equation~\eqref{OpsiRv1}, we instead require the source $8\pi T^{(2)\rm eff}_{\mu\nu}$ appearing in Eq.~\eqref{tt EFE2R}. Hence, our effective source can be written as
\beq\label{eq:S4eff}
S^{(2)}_{4,\rm eff} = \hat\S_4\left[8\pi \breve T^{(2)\rm eff}_{\mu\nu}\right] +\hat \S_4\left[8\pi \Delta T^{(2)\rm eff}_{\mu\nu}\right],
\eeq
where $\Delta T^{(2)\rm eff}_{\mu\nu}\coloneqq T^{(2)\rm eff}_{\mu\nu}-\breve T^{(2)\rm eff}_{\mu\nu}$. Referring to Eqs.~\eqref{T2effbreve} and \eqref{Tneff def}, with Eqs.~\eqref{G11}, \eqref{tt EFE2R}, and \eqref{Ebreve}, we read off
\begin{align}\label{eq:DeltaT2eff}
8\pi\Delta T^{(2)\rm eff}_{\mu\nu} &= -\overline{\nabla^{(1)}_{(\mu}Z^{(0)}_{\nu)}}[\bar h^{(1)}] -\overline{\nabla^{(0)}_{(\mu}Z^{(1)}_{\nu)}}[\bar h^{(1)}] \nonumber\\ 
&\quad -\overline{\nabla^{(0)}_{(\mu}Z^{(0)}_{\nu)}}[\bar h^{(2)\cal P}] \nonumber\\
&\quad + \frac{2M}{r^2}\n_{(\mu}\left(\breve Z^{(1)}_{\nu)}[\bar h^{(1)}]  +\breve Z^{(0)}_{\nu)}[\bar h^{(2)\cal P}]\right).
\end{align}
The first term on the right vanishes if  $h^{(1)}_{\mu\nu}$ is in the Lorenz gauge, satisfying $Z^{(0)}_\mu[\bar h^{(1)}]=0$. All other terms would also identically vanish if the puncture identically satisfied the second-order Lorenz gauge condition,
\beq\label{Lorenz2}
Z^{(1)}_{\mu}[\bar h^{(1)}]+Z^{(0)}_{\mu}[\bar h^{(2)}] = 0.
\eeq
However, since the puncture does not satisfy that equation (except to some order in a local expansion around the particle), $\hat\S_4\bigl[\Delta T^{(2)\rm eff}_{\mu\nu}\bigr]$ is nonzero. It can alternatively be evaluated in the form
\beq\label{eq:DeltaT2effhR2}
8\pi\Delta T^{(2)\rm eff}_{\mu\nu} = -\overline{\nabla^{(0)}_{(\mu}Z^{(0)}_{\nu)}}[\bar h^{(2)\cal R}] + \frac{2M}{r^2}\n_{(\mu}\breve Z^{(0)}_{\nu)}[\bar h^{(2)\cal R}],
\eeq
which follows from Eq.~\eqref{Lorenz2}.


To include $\Delta T^{(2)\rm eff}_{\mu\nu}$ in the source calculations in Sec.~\ref{sec:Seff}, we decompose it in the same basis of Barack-Lousto-Sago harmonics used for $\breve T^{(2)\rm eff}_{\mu\nu}$:
\begin{equation}
    {8\pi}\Delta T^{(2)\rm eff}_{\mu\nu} = \sum_{\i\l\m} 8\pi\Delta T^{(2)\rm eff}_{\i\l\m}Y^{\i\l\m}_{\mu\nu}e^{-im\phi_p}.
\end{equation}
This is then converted to a tetrad and spin-weighted spherical harmonic basis.


\section{Infrared divergences in the Lorenz-gauge field equations}\label{sec:infrared Lorenz}

In Sec.~\ref{sec:compactification} we analyzed the compactified second-order Teukolsky equations, showing that they are free of the worst infrared problems encountered in Lorenz-gauge calculations. As first discussed in Ref.~\cite{Pound:2015wva} and more thoroughly in Ref.~\cite{Cunningham:2024dog}, Lorenz-gauge calculations encounter divergences that are cured by matching to an SF-MPM solution, which introduces memory integrals over the binary's past history. However, these analyses have always focused on the convergence or divergence of retarded integrals---integrals of sources against the retarded Green's function. Here we reanalyze the problem in terms of regularity of the compactified field equations. We show that this mode of analysis does not alter the conclusions of previous analyses, but it illuminates the contrasts between the Lorenz-gauge and Teukolsky equations.

We start from the field equations in the BLS harmonic basis~\cite{Barack:2005nr,Barack:2007tm}, specifically in the form presented in Sec.~IIIC of Ref.~\cite{Miller:2023ers}. Unlike in the Teukolsky case, where the field variable is a single complex scalar, here the field variable is originally a set of 10 metric-perturbation components. Using the gauge condition and the decoupling of even- and odd-parity perturbations, one reduces the field variable $\bm{\psi}$ to a tuple of up to five metric components, depending on the particular $\l\m$ mode considered. The field equations are then given by Eq.~(69) of Ref.~\cite{Miller:2023ers}. We consider the equations for $\l>0$ and for $\l=0$ separately, as the latter requires special treatment. In both cases, we write the field equations in the form
\beq
(1-\sigma) \sigma^2 \partial^2_\sigma \bm{\psi} + \bm{{\cal B}}\partial_\sigma \bm{\psi} + \bm{{\cal A}}\bm{\psi}  = \bm{S},\label{Lorenz compactified}
\eeq
using bold symbols to denote tuples and matrices. Like in the Teukolsky case in the body of the paper, the variable $\bm{\psi}$ will always be scaled with appropriate powers of $\sigma$ and $(1-\sigma)$ to ensure that it would approach a nonzero constant at each boundary if it represented a freely propagating wave there.

\subsection{$\l>0$ modes}

For $\l>0$ modes, $\bm{\psi}$ is a tuple of BLS coefficients: 
\begin{align}
\bm{\psi}=\begin{cases}
    (\bar h^{(2)}_1,\bar h^{(2)}_3,\bar h^{(2)}_5,\bar h^{(2)}_6,\bar h^{(2)}_7) & \l\geq2, \l+\m \text{ even}\\
    (\bar h^{(2)}_1,\bar h^{(2)}_3,\bar h^{(2)}_5,\bar h^{(2)}_6) & \l=1,\m=1\\
    (\bar h^{(2)}_9,\bar h^{(2)}_{10}) & \l\geq2, \l+\m \text{ odd}\\
    \bar h^{(2)}_9 & \l=1,\m=0,
    \end{cases}
\end{align}
where $\bar h_{\alpha\beta}$ is expanded as in BLS tensor spherical harmonics~\cite{Barack:2005nr, Barack:2007tm}, and we suppress $\l\m$ mode labels. The remaining BLS coefficients ($\bar h_2,\bar h_4,\bar h_8$) are obtained from those in $\bm{\psi}$ via the gauge condition, meaning we can focus on $\bm{\psi}$ alone. For $\omega_\m=0$ modes the number of variables can be further reduced using the gauge condition, but we skip that reduction here for simplicity.

Following the analysis in Sec.~\ref{sec:compactification}, we adopt  $u$ or $v$ slicing toward each boundary, meaning $H=\pm 1$. In terms of the compact coordinate $\sigma$, we can then straightforwardly write Eq.~(69) of Ref.~\cite{Miller:2023ers} in the form~\eqref{Lorenz compactified}, with the coefficients
\begin{align}
\bm{{\cal A}} &= -[\l(\l+1)+\sigma]\mathbf{1}_{d\times d}+\bm{{\cal N}_h},\\
\bm{{\cal B}} &= -\left(2i\varpi H +3 \sigma ^2-2 \sigma \right)\mathbf{1}_{d\times d} - \sigma^2(1-\sigma)\bm{{\cal N}_{\partial h}},
\end{align}
recalling $\varpi\coloneqq 2M\omega_\m$. Here $d$ is the dimension of the tuple $\bm{\psi}$ and of the source $\bm{S}$, and we have introduced the matrices $\bm{{\cal N}_h}$ and $\bm{{\cal N}_{\partial h}}$, which are related to the matrices $\bm{{\cal M}_h}$ and $\bm{{\cal M}_{\partial h}}$ in Appendix~A of Ref.~\cite{Miller:2023ers} by $\bm{{\cal N}_h} = r^2f\bm{{\cal M}_h}$ and $\bm{{\cal N}_{\partial h}} = \frac{r^2}{2M}\bm{{\cal M}_{\partial h}}$. $\bm{{\cal N}_h}$ goes smoothly to a nonzero constant matrix at both boundaries, while $\bm{{\cal N}_{\partial h}}$ is identically constant.

The source $\bm{S}$ is related to the source $\bm{J}$ in Sec.~IIIC of Ref.~\cite{Miller:2023ers} by
\begin{equation}
\bm{S} = \frac{4M^2}{\sigma^2}(1-\sigma)\bm{J}.
\end{equation}
Given the scalings from Sec.~IIID of that reference, we can infer the limits
\begin{equation}
    \bm{S} = \frac{\bm{j_\infty}}{\sigma} + \O(\sigma^0)
\end{equation}
near \scri and
\begin{equation}
    \bm{S} = \bm{j_H} + \O(1-\sigma)
\end{equation}
near the horizon, where $\bm{j_\infty}$ and $\bm{j_H}$ are constant vectors. These scalings follow straightforwardly from those of $G^{(2,0)}_{\alpha\beta}$ and $E^{(1)}_{\alpha\beta}$ in Eqs.~\eqref{eq:G20 asymptotics Lorenz} and Eq.~\eqref{eq:E1 asymptotics Lorenz} and from the smoothness of $G^{(2,0)}_{\alpha\beta}$ and $E^{(1)}_{\alpha\beta}$ at the horizon.

\subsubsection{Behavior at the horizon}

First consider the behavior at the horizon, which is quite simple. Assuming sufficient regularity of $\bm{\psi}$ at $\sigma=1$, we find the field equation at the horizon reduces to
\begin{multline}
\left\{\bm{{\cal N}_h}-[\l(\l+1)+1]\mathbf{1}_{d\times d}\right\}\bm{\psi}|_{\sigma=1}\\
+(2i\varpi-1)\partial_\sigma\bm{\psi}|_{\sigma=1} = \bm{j_H}.
\end{multline}
One can immediately solve this equation for $\partial_\sigma\bm{\psi}|_{\sigma=1}$ in terms of $\bm{\psi}|_{\sigma=1}$ and $\bm{j_H}$, regardless of the value of $\varpi$ and~$\l$. Using the explicit matrices in Eqs.~(A12)--(A14) of Ref.~\cite{Miller:2023ers}, one can also straightforwardly check that the matrix multiplying $\bm{\psi}|_{\sigma=1}$ is invertible for all $\l$, such that one can alternatively solve for $\bm{\psi}|_{\sigma=1}$ in terms of $\partial_\sigma\bm{\psi}|_{\sigma=1}$ and $\bm{j_H}$.

\subsubsection{$\omega_\m\neq0$ modes at \scri}

Next consider the behavior at \scri. If the source were continuous at $\sigma=0$ and we assumed $\bm{\psi}$ were sufficiently regular there, we would be able to take the limit of Eq.~\eqref{Lorenz compactified}, yielding
\begin{multline}
\left[\bm{{\cal N}_h}-\ell(\ell+1)\mathbf{1}_{d\times d}\right]\bm{\psi}|_{\sigma=0} \\
-2i\varpi\partial_\sigma \bm{\psi}|_{\sigma=0} = \bm{S}|_{\sigma=0}.
\end{multline}
However, this is prevented by the $1/\sigma$ divergence of the source. In this case, we quickly find the solution to Eq.~\eqref{Lorenz compactified} must contain a logarithmic divergence, with precisely the same form as the Teukolsky solution~\eqref{eq:Teukolsky log soln}:
\beq\label{eq:Lorenz log soln}
\bm{\psi} = -\frac{\bm{j_\infty}}{2i\varpi}\ln\sigma + \bm{a}_0+\O(\sigma\ln\sigma),
\eeq
where $\bm{a}_0$ is a constant. 

As we explained in the body of the paper, the presence of a leading-order logarithm means the divergence cannot be dealt with through any rescaling, but it can be dealt with using a puncture scheme, introducing a puncture and residual field 
\begin{equation}\label{eq:psiR omega!=0}
    \bm{\psi}^{\cal R}_\infty = \bm{\psi} - \bm{\psi}^{\cal P}_\infty.
\end{equation}
As in Eq.~\eqref{eq:Teukolsky infinity puncture}, the puncture must contain the first subleading logarithmic term  in order to yield an effective source that is continuous at $\sigma=0$. Including that term, the puncture reads
\begin{equation}
    \psi^{\cal P}_\infty = -\frac{\bm{j_\infty}}{2i\varpi}\ln\sigma - \frac{[\bm{{\cal N}_h}|_{\sigma=0}-\l(\l+1)]\bm{j_\infty}}{(2i\varpi)^2}\sigma\ln\sigma.
\end{equation}
The ``compactification + regularity'' approach then applies straightforwardly to the field equation for $\bm{\psi}^{\cal R}_\infty$.

\subsubsection{Generic $\omega_\m=0$ modes at \scri}

However, the problems at \scri worsen for $\omega_\m=0$ modes. Unlike in the Teukolsky case, the source $\bm{S}$ blows up as $1/\sigma$ for both $\omega_\m\neq0$ and $\omega_\m=0$ modes~\cite{Miller:2020bft,Miller:2023ers,Cunningham:2024dog}. The solution~\eqref{eq:Lorenz log soln} clearly does not exist for $\omega_\m=0$, and we now must seek an even more singular solution: 
\begin{equation}
    \bm{\psi} = \frac{\bm{a}_{-1}}{\sigma} + \O(\sigma^0). 
\end{equation}
Substituting this ansatz into Eq.~\eqref{Lorenz compactified}, we find
\begin{equation}\label{eq:a-1 omega=0}
    \frac{1}{\sigma}\bigl[\bm{{\cal N}_h}|_{\sigma=0} -\l(\l+1)\mathbf{1}_{d\times d}\bigr]\bm{a}_{-1}  = \frac{\bm{j}_\infty}{\sigma},
\end{equation}
noting that the $\partial_\sigma^2\bm{\psi}$ and $\partial_\sigma\bm{\psi}$ terms in Eq.~\eqref{Lorenz compactified} cancel at order $1/\sigma$. Therefore, we find a solution, albeit a singular one, so long as the matrix 
\beq
\bm{{\cal C}}_{\l\m}\coloneqq\bigl[\bm{{\cal N}_h}|_{\sigma=0} -\l(\l+1)\mathbf{1}_{d\times d}\bigr]
\eeq
is invertible. 

When $\bm{{\cal C}}_{\l\m}$ is invertible, we can define a new variable, $\bm{\Psi}\coloneqq \sigma \bm{\psi}$, that is regular at $\sigma=0$, and for which the field equation reduces to a regular boundary condition at $\sigma=0$: 
\begin{equation}\label{eq:Lorenz bdry condition omega=0}
    \bm{{\cal C}}_{\l\m}\bm{\Psi}|_{\sigma=0} = \lim_{\sigma\to0}(\sigma \bm{S}).
\end{equation}
This is the same scenario discussed around Eq.~\eqref{Teukolsky soln near scri OK} for $\omega_\m=0$ modes of the Weyl scalar $\psi^{(2)}_{4L}$. Although the solution for $\bm{\psi}$ is singular, the ``compactification + regularity'' method can still be applied for $\bm{\Psi}$ to obtain the unique solution. In this case, the field $\bm{\psi}$ blowing up as $1/\sigma$ corresponds to the physical metric components $h^{(2)}_{ab}$ going to a constant at infinity, seemingly in violation of asymptotic flatness. As discussed regarding $\psi^{(2)}_{4L}$, this singularity is a gauge artifact that can be removed through a gauge transformation after computing $h^{(2)}_{\alpha\beta}$; our analysis here shows that it does not obstruct that actual computation of $h^{(2)}_{\alpha\beta}$.

\subsubsection{$\l=2$, $\omega_\m=0$ mode at \scri}

Yet the situation worsens even further when the matrix $\bm{{\cal C}}_{\l\m}$ in Eq.~\eqref{eq:Lorenz bdry condition omega=0} is \emph{not} invertible. One can quickly verify, from the matrices in Eqs.~(A12)--(A14) of Ref.~\cite{Miller:2023ers}, that this is the case for exactly one mode: $\l=2,\m=0$, for which the $5\times 5$ matrix $\bm{{\cal C}}_{20}$ is rank 4. For this mode, we require an even more singular solution,
\begin{equation}\label{eq:psi20Lorenz near scri}
    \bm{\psi} = \frac{\bm{a}_{-1}^{\rm log}\ln\sigma+\bm{a}_{-1}}{\sigma} +\bm{a}^{\rm log}_0\ln\sigma +\bm{a}_0  + \O(\sigma\ln\sigma), 
\end{equation}
where we have displayed the necessary number of orders in $\sigma$ required to eliminate divergences from the field equations. 
Substituting this ansatz into Eq.~\eqref{Lorenz compactified}, we find that eliminating $1/\sigma$ terms in the field equations requires
\begin{equation}\label{eq:minus1 terms}
    \bm{{\cal C}}_{20}\,\bm{a}^{\rm log}_{-1} \ln\sigma
    - \bm{a}^{\rm log}_{-1} 
    + \bm{{\cal C}}_{20}\bm{a}_{-1}
    = \bm{j}_\infty,
\end{equation}
and eliminating the $\sigma^0\ln\sigma$ terms requires
\begin{equation}\label{eq:log terms}
    \bm{{\cal C}}_{20}\,\bm{a}_0^{\rm log} + \bigl[\bm{{\cal N}_{\partial h}}+\partial_\sigma\bm{{\cal N}_h}\bigr]\Bigr|_{\sigma=0}\bm{a}_{-1}^{\rm log} = \bm{0}.
\end{equation}

Consider the $\ln\sigma$ term in Eq.~\eqref{eq:minus1 terms}, which implies
\begin{equation}
    \bm{{\cal C}}_{20}\,\bm{a}^{\rm log}_{-1} =0.
\end{equation}
Since $\bm{{\cal C}}_{20}$ has rank 4, this equation determines four out of the five elements of $\bm{a}^{\rm log}_{-1}$. The remaining part of Eq.~\eqref{eq:minus1 terms},
\begin{equation}\label{eq:aminus1 equation}
     \bm{a}^{\rm log}_{-1} + \bm{{\cal C}}_{20}\bm{a}_{-1}
    = \bm{j}_\infty,
\end{equation}
can then be solved for the remaining element of $\bm{a}^{\rm log}_{-1}$ and for four of the five elements of $\bm{a}_{-1}$. Next, Eq.~\eqref{eq:log terms} can be solved for four of the five elements of $\bm{a}^{\rm log}_{0}$.

This leaves us with two unknowns, one in each of $\bm{a}_{-1}$ and $\bm{a}^{\rm log}_0$. However, we have not yet enforced the Lorenz gauge condition. By substituting our solution into the gauge condition~(68) from Ref.~\cite{Miller:2023ers}, we find that the gauge condition does fix the remaining unknown in $\bm{a}^{\rm log}_0$. On the other hand, we find that the unknown in $\bm{a}_{-1}$ remains freely specifiable.

We see from this analysis that, unlike all the cases examined above, the solution for $\l=2,\m=0$ \emph{cannot} be determined from the field equations alone: the field equations leave a free constant, corresponding to the amplitude of a homogeneous solution that diverges as $1/\sigma$. Even if we were to define a puncture
\begin{equation}\label{eq:psiP l=2,m=0}
\bm{\psi}^{\cal P} = \frac{\bm{a}_{-1}^{\rm log}\ln\sigma + \bm{a}_{-1}}{\sigma} + \bm{a}_{0}^{\log}\ln\sigma,
\end{equation}
the choice of puncture field would contain an arbitrary specification of one element of $\bm{a}_{-1}$. If we were to then impose regularity on the residual field $\psi^{\cal R}$, we would actually arrive at the \emph{wrong} total solution $\psi^{\cal R}+\psi^{\cal P}$---unless we happen to have guessed exactly the correct value of the free constant. 
This represents a genuine breakdown of the ``compactification + regularity'' approach. We return to the correct choice of free constant in Appendix~\ref{sec:infrared divergences comparison} below.

\subsection{Monopole}

The field equations for the monopole are not written explicitly in Ref.~\cite{Miller:2023ers}. Instead we take them from Eqs.~(D1) and~(D2) in Ref.~\cite{Wardell:2015ada}. These equations are for the BLS coefficients $\bar h^{(2)}_{1}$ and $\bar h^{(2)}_3$. We can put them in the form~\eqref{Lorenz compactified} by adopting the variable
\begin{equation}\label{eq:psi rescaling monopole}
    \bm\psi = \begin{pmatrix}
        (1-\sigma)^{-2}\bar h^{(2)}_1\\
        \bar h^{(2)}_3
    \end{pmatrix}.
\end{equation}
This overall scaling $(1-\sigma)^{-2}$ in the first element of $\bm{\psi}$ reflects the fact that for a \emph{static} perturbation (in which $h^{(n)}_{tr}=0$), $\bar h^{(n)}_1=\O[(1-\sigma)^2]$ is required for horizon-regularity of the metric components in ingoing Eddington-Finkelstein coordinates~\cite{Miller:2020bft}. With this rescaling, the coefficients in Eq.~\eqref{Lorenz compactified} are given by
\begin{equation}
\bm{{\cal A}} = \begin{pmatrix}
 -2 \sigma -1 & 1 \\
 4 \sigma^2-3 \sigma +1 & \sigma -1 \\
\end{pmatrix}
\end{equation}
and
\begin{equation}
    \bm{{\cal B}} = \begin{pmatrix}
 (1-4 \sigma ) \sigma  & \sigma  \\
 (1-2 \sigma) (1-\sigma)\sigma & (1-\sigma ) \sigma 
\end{pmatrix},
\end{equation}
while the source is related to the BLS modes of $T^{(2)\rm eff}_{\alpha\beta}$ by
\begin{equation}
    \bm{S} = -16\pi \frac{8M^3}{\sigma^3}\sqrt{2}\begin{pmatrix}
         (1-\sigma)^{-2}T^{(2)\rm eff}_1\\
         T^{(2)\rm eff}_3
    \end{pmatrix}.
\end{equation}

Toward the boundaries, $T^{(2)\rm eff}_{\i}$ goes to a constant at $\sigma=1$ and goes like $\sigma^2$ at $\sigma=0$. The source hence has the behavior
\begin{equation}
    \bm{S} = \begin{pmatrix}
        (1-\sigma)^{-2}j^H_1 + \O[(1-\sigma)^{-1}]\\
       j^H_3 + \O(1-\sigma)
    \end{pmatrix},
\end{equation}
and
\begin{equation}
    \bm{S} = \frac{\bm{j}_\infty}{\sigma} +\O(\sigma^0)
\end{equation}
for some constants $\bm{j}^H$ and $\bm{j}_\infty$ (different than those given the same symbols in preceding sections, recalling that we have suppressed $\l\m$ indices).

Near the horizon, there is no regular solution, unlike in the $\l>0$ case. Instead, we require the behavior
\begin{equation}\label{eq:00 Lorenz NH}
    \bm{\psi} = \begin{pmatrix}
        (1-\sigma)^{-1}b_{1,-1}\\
        b^{\rm log}_{3,0}\log(1-\sigma)
    \end{pmatrix} + \O[(1-\sigma)^0].
\end{equation}
Substituting this ansatz into the field equation~\eqref{Lorenz compactified} quickly determines $b_{1,-1}=-j_1^H=b^{\rm log}_{3,0}$. Note that this means $\bar h^{(2)}_{1}=\O(1-\sigma)$ due to the rescaling in Eq.~\eqref{eq:psi rescaling monopole}, meaning there is no actual divergence in $\bar h^{(2)}_1$. But there \emph{is} a logarithmic divergence in $\bar h^{(2)}_3$. This is the only instance where a singularity necessarily appears at the horizon in the Lorenz gauge. Moreover, like in the case of the $(2,0)$ mode at $\sigma=0$, there are homogeneous solutions less singular than the dominant singular behavior in Eq.~\eqref{eq:00 Lorenz NH}, which cannot be determined directly from the field equations. Again, the correct boundary conditions, which would fix the coefficient of this homogeneous solution, must be determined in some other way.

Near \scri, like in Eq.~\eqref{eq:psi20Lorenz near scri}, we find 
\begin{equation}\label{eq:psi00Lorenz near scri}
    \bm{\psi} = \frac{\bm{a}_{-1}^{\rm log}\ln\sigma+\bm{a}_{-1}}{\sigma} +\bm{a}^{\rm log}_0\ln\sigma +\bm{a}_0  + \O(\sigma\ln\sigma).
\end{equation}
Also just as for Eq.~\eqref{eq:psi20Lorenz near scri}, we find that the coefficient $\bm{a}^{\rm log}_{-1}$ is fully determined by the field equations, while $\bm{a}_{-1}$ is freely specifiable. Hence, like we found for the $(2,0)$ mode and for the monopole at the horizon, the ``compactification + regularity'' approach completely breaks down here.

\subsection{Comparisons with previous analyses}\label{sec:infrared divergences comparison}

Our analysis here has highlighted the limitations of the ``compactification + regularity'' approach, more so than our analysis of the Teukolsky equation in the body of the paper did. In the Teukolsky case, compactification + regularity could always be salvaged through rescaling variables or, at worst, by utilizing a puncture that could be determined directly from the field equations. Our analysis in this Appendix found two cases, specifically the $(\l,\m)=(2,0)$ and $(0,0)$ modes, in which there is no salvaging of the method: additional boundary conditions are fundamentally required. 

In our previous analyses in Refs.~\cite{Miller:2023ers,Cunningham:2024dog}, we arrived at the same conclusion, for the same two modes, by examining when the retarded integral of the source fails to converge. We noted in Sec.~\ref{sec:mixed method} that such an analysis can be misleading in the Teukolsky case, due to the known pathologies of the Teukolsky Green's function. But the better-behaved Green's function of the Lorenz-gauge Einstein operator appears to avoid such issues, providing a satisfying harmony between the Green's-function analysis and the compactified regularity analysis.

As described in Ref.~\cite{Cunningham:2024dog}, the SF-MPM approach allows one to determine the correct boundary conditions that cannot be determined directly from Lorenz-gauge multiscale field equations. Concretely, the physically correct small-$\sigma$ solutions are given by Eqs.~(146) and~(147) of Ref.~\cite{Cunningham:2024dog}. Equation~(147) of that reference, in particular, shows the uniquely determined term corresponding to $\bm{a}_{-1}/\sigma$ in Eqs.~\eqref{eq:psi20Lorenz near scri} and~\eqref{eq:psi00Lorenz near scri}. This term involves a memory integral over the system's entire past history, which no direct analysis of the compactified multiscale Lorenz-gauge field equations could possibly determine.

However, it is worth highlighting that the special cases of $(\l,\m)=(2,0)$ and $(0,0)$ are the only true failures of the ``compactification + regularity'' approach. All the other cases only require shifts or rescalings that can be determined directly from the field equations, and not even those are required at the horizon. 

In Ref.~\cite{Spiers:2026yqx} we already indicated why the problematic feature of the $\l=2,\omega_\m=0$ mode does not affect the second-order Teukolsky equation: $\psi_4$ asymptotes to $-\frac{1}{2}\partial_u^2 h_{\mb\mb}$ at large $r$, meaning that a slowly evolving, $\omega_\m=0$ mode of $h^{(n)}_{\alpha\beta}$ only enters $\psi_4$ at order $n+2$, effectively eliminating problematic memory contributions at first and second order. This is doubly true for the monopole, since $\psi_4$ contains no $\l=0$ modes.



\section{Consistency of Bondi-Sachs and Lorenz-gauge calculations}
\label{sec:Lorenz-Bondi consistency}

Our calculations in the body of the paper work with a source term constructed from a first-order metric perturbation in the Bondi-Sachs gauge. This source differs dramatically from what it would be if the $h^{(1)}_{\alpha\beta}$ were in the Lorenz gauge. One might reasonably wonder if the resulting waveform and flux should then differ from the results in the Lorenz gauge. Geometrically, one might think that since a $u=\text{constant}$ slice represents two different geometrical surfaces in the two gauges, the flux extracted on that slice should differ between the two. In this appendix we show that the two fluxes should, in fact, be the same. 

We start from the field equation~\eqref{Opsi}, reproduced here for convenience:
\begin{equation}\label{eq:Bondi-Sachs Teukolsky}
    \hat\O_4\bigl[\psi^{(2)}_{4L}\bigr] = \hat\S_4\Bigl\{8\pi T^{(2)} - G^{(2,0)}[h^{(1)},h^{(1)}]
    -G^{(1,1)}[h^{(1)}]\Bigr\},
\end{equation}
with $h^{(1)}_{\alpha\beta}$ related to the Lorenz-gauge perturbation by $h^{(1)}_{\alpha\beta} = h^{(1)\rm Lor}_{\alpha\beta}+{\cal L}_\xi^{(0)}g_{\alpha\beta}$. As explained in Sec.~\ref{sec:compactification}, if we were to leave the first-order perturbation in the Lorenz gauge, we would instead need to solve for a residual field:
\begin{multline}\label{eq:Lorenz Teukolsky}
    \hat\O_4\bigl[\psi^{(2)\cal R}_{4L,\infty}\bigr] 
    = \hat\S_4\Bigl\{8\pi T^{(2)} - G^{(2,0)}[h^{(1)}_{\rm Lor},h^{(1)}_{\rm Lor}]\\
    -G^{(1,1)}[h^{(1)}_{\rm Lor}]\Bigr\}
    - \hat\O[\psi^{(2){\cal P}}_{4L,\infty}].
\end{multline}
The puncture $\psi^{(2){\cal P}}_{4L,\infty}$ can be constructed from Eq.~\eqref{eq:Teukolsky infinity puncture}, noting that terms of higher order in $\sigma$ can also be freely added without changing the value of the full field $\psi^{(2)\cal R}_{4L,\infty} + \psi^{(2)\cal P}_{4L,\infty}$; this amounts to simply trading terms between the puncture and residual field~\cite{Miller:2023ers}. Thus, for conceptual simplicity, we can take $\psi^{(2)\cal P}_{4L,\infty}$ to precisely correspond to the puncture at infinity $h^{(2){\cal P}}_{\infty,\alpha\beta}$ used in previous Lorenz-gauge calculations, meaning
\begin{equation}\label{eq:psiPinf=ThPinf}
    \psi^{(2){\cal P}}_{4L,\infty} = \hat\T_4[h^{(2){\cal P}}_{\infty}].
\end{equation}

Now, the difference between the retarded solutions to Eqs.~\eqref{eq:Bondi-Sachs Teukolsky} and \eqref{eq:Lorenz Teukolsky} satisfies
\begin{align}
    \hat\O_4\bigl[\psi^{(2)}_{4L}-\psi^{(2)\cal R}_{4L,\infty}\bigr] &= -\hat\S_4\Bigl\{2G^{(2,0)}\bigl[h^{(1)},{\cal L}^{(0)}_\xi g\bigr] \nonumber\\
    &\qquad\qquad + G^{(2,0)}\bigl[{\cal L}^{(0)}_\xi g,{\cal L}^{(0)}_\xi g\bigr] \nonumber\\
    &\qquad\qquad + G^{(1,1)}\bigl[{\cal L}^{(0)}_\xi g\bigr]\Bigr\}\nonumber\\
   &\quad + \hat\O_4\bigl[\psi^{(2)\cal P}_{4L,\infty}\bigr],
\end{align}
where the stress-energy source terms canceled and we used $h^{(1)\rm Lor}_{\alpha\beta}=h^{(1)}_{\alpha\beta}-{\cal L}^{(0)}_{\xi}g_{\alpha\beta}$. Now, on the right-hand side we use the identity~\cite{Pound:2015fma}
\begin{multline}
    2G^{(2,0)}_{\alpha\beta}\bigl[h^{(1)},{\cal L}^{(0)}g\bigr] + G^{(2,0)}_{\alpha\beta}\bigl[{\cal L}^{(0)}g,{\cal L}^{(0)}g\bigr] \\
    = - G^{(1,0)}_{\alpha\beta}\bigl[\Delta_\xi^{(0)}h^{(2)}\bigr]
\end{multline}
with
\begin{equation}
    \Delta_\xi^{(0)}h^{(2)}_{\alpha\beta} = {\cal L}_\xi^{(0)}h^{(1)}_{\alpha\beta} + \frac{1}{2}{\cal L}_\xi^{(0)}{\cal L}_\xi^{(0)}g_{\alpha\beta},
\end{equation}
and 
\begin{equation}
    G^{(1,1)}_{\alpha\beta}\bigl[{\cal L}^{(0)}_\xi g\bigr] = - G^{(1,0)}_{\alpha\beta}\bigl[{\cal L}^{(1)}_\xi g\bigr], 
\end{equation}
which follows from $G^{(1)}_{\alpha\beta}[{\cal L}_\xi g]=0$. Together, these identities imply 
\begin{align}
    \hat\O_4[\psi^{(2)}_{4L}-\psi^{(2)\cal R}_{4L,\infty}] &= \hat\S_4\bigl\{ G^{(1,0)}\bigl[\Delta_\xi^{(0)}h^{(2)}_{\alpha\beta} + {\cal L}^{(1)}_\xi g \bigr]\bigr\} \no\\
    &\quad + \hat\O_4\bigl[\psi^{(2)\cal P}_{4L,\infty}\bigr].
\end{align}

Next we use Eq.~\eqref{eq:psiPinf=ThPinf} to rewrite this as
\begin{equation}
    \hat\O_4\bigl[\psi^{(2)}_{4L}-\psi^{(2)\cal R}_{4L,\infty}\bigr] =  \hat\O_4\hat\T_4\bigl[\Delta_\xi^{(0)}h^{(2)}_{\alpha\beta} + {\cal L}^{(1)}_\xi g + h^{(2){\cal P}}_\infty\bigr]
\end{equation}
or more simply as
\begin{equation}\label{eq:psi4LBS-psi4LR}
    \hat\O_4\bigl[\psi^{(2)}_{4L}-\psi^{(2)\cal R}_{4L,\infty}\bigr] = \hat\O_4\bigl[\Delta_\xi \psi^{(2)}_{4L} + \psi^{(2){\cal P}}_{4L,\infty}\bigr].
\end{equation}
with 
\begin{equation}
\Delta_\xi \psi^{(2)}_{4L} \coloneqq \hat\T_4\Bigl[{\cal L}_\xi^{(0)}h^{(1)} + \frac{1}{2}{\cal L}_\xi^{(0)}{\cal L}_\xi^{(0)}g+ {\cal L}^{(1)}_\xi g\Bigr].
\end{equation}
We can also write this in the mode-decomposed, compactified form of Eq.~\eqref{eq:teukolsky_compactified}:
\begin{equation}\label{eq:psi4LBS-psi4LR compactified}
    \tilde\O\bigl[\tilde\psi-\tilde\psi^{\cal R}_\infty\bigr] = \tilde\O\bigl[\Delta_\xi \tilde\psi + \tilde\psi^{{\cal P}}_\infty\bigr],
\end{equation}
where $\tilde\O\coloneqq (1-\sigma)\sigma^2\partial^2_\sigma + {\cal B}\partial_\sigma + {\cal A}$ with $s=-2$.

The operator $\tilde\O$ is not invertible, but if it were, we could immediately strip it off both sides of Eq.~\eqref{eq:psi4LBS-psi4LR compactified} to obtain 
\begin{equation}\label{eq:psi-psiR = Dpsi + psiP}
    \tilde\psi - \tilde\psi^{\cal R}_\infty = \Delta_\xi\tilde\psi + \tilde\psi^{{\cal P}}_\infty.
\end{equation}
This will be the essential equality in our proof, as we describe at the end of this section. We establish the equality as follows. 

We first note that $\tilde\O\tilde\psi = \O(\sigma^0) = \tilde\O\tilde\psi^{\cal R}_\infty$, as explained in Sec.~\ref{sec:compactification}. Therefore, the right-hand side of Eq.~\eqref{eq:psi4LBS-psi4LR compactified} must be continuous at $\sigma=0$. This can also be deduced from the following two facts: again as explained in Sec.~\ref{sec:compactification}, $\tilde\O[\tilde\psi^{{\cal P}}_\infty] = \tilde S_{\rm Lor} +\O(\sigma^0)$, where $\tilde S_{\rm Lor}$ is $\tilde S$ constructed from first-order Lorenz gauge metric perturbations; and $\tilde\O[\Delta_\xi\psi]$ must be equal to $- \tilde S_{\rm Lor} + \O(\sigma^0)$ in order for the source in Bondi-Sachs gauge to be continuous at $\sigma=0$. Therefore, the $\tilde S_{\rm Lor}$ terms cancel on the right-hand side of Eq.~\eqref{eq:psi4LBS-psi4LR compactified}, leaving something continuous. The ``compactification + regularity'' method then applies, with Eq.~\eqref{eq:psi4LBS-psi4LR compactified} reducing to a boundary condition of the form~\eqref{eq:sigma=0 eqn} at  $\sigma=0$:
\begin{multline}\label{eq:psi-psiR=Dpsi+psiP at sigma=0}
    -\bigl(2i\varpi\partial_\sigma+{}_{-2}\lambda_\l-4i\varpi \bigr)(\tilde\psi-\tilde\psi^{\cal R}_\infty)\bigr|_{\sigma=0} \\
    = -\bigl(2i\varpi\partial_\sigma+{}_{-2}\lambda_\l-4i\varpi \bigr)(\Delta_\xi\tilde\psi+\tilde\psi^{\cal P}_\infty)\bigr|_{\sigma=0}.
\end{multline}

Next, we assume $\Delta_\xi\tilde\psi$ and $\tilde\psi^{\cal P}_\infty$ both vanish for all $\sigma>\sigma_0$, for some $\sigma_0<1$. They can either smoothly transition to zero at $\sigma\leq\sigma_0$ or sharply change to zero there (as is the case in our numerical implementations). Since the field equations and boundary conditions for $\tilde\psi$ and $\tilde\psi^{\cal R}_\infty$ are identical for all $\sigma>\sigma_0$, they satisfy the following junction conditions at $\sigma_0$: 
\begin{align}
    \tilde\psi|_{\sigma_0^-} &= \tilde\psi|_{\sigma_0^+} + \Delta_\xi\tilde\psi|_{\sigma_0},\\
    \tilde\psi^{\cal R}_\infty|_{\sigma_0^-} &= \tilde\psi|_{\sigma_0^+} - \tilde\psi^{\cal P}_\infty|_{\sigma_0},
\end{align}
where $|_{\sigma^\pm_0}$ indicates evaluation at $\sigma_0$ from above or below $\sigma=\sigma_0$. These two junction conditions together imply
\begin{equation}\label{eq:psi-psiR=Dpsi+psiP at sigma0}
    (\tilde\psi-\tilde\psi_\infty^{\cal R})|_{\sigma_0^-} =  (\Delta_\xi\tilde\psi+ \tilde\psi^{\cal P}_\infty)|_{\sigma_0}.
\end{equation}

If we discretize Eq.~\eqref{eq:psi4LBS-psi4LR compactified}---using a finite difference representation or the collocation method we employ in our numerics, for example---then the combined set of equations~\eqref{eq:psi4LBS-psi4LR compactified}, \eqref{eq:psi-psiR=Dpsi+psiP at sigma=0}, and \eqref{eq:psi-psiR=Dpsi+psiP at sigma0} can be written as a matrix equation:
\begin{equation}\label{eq:discretized Opsi-OpsiR}
    \hat\O_{ij}(\tilde\psi_{ j} - \tilde\psi^{\cal R}_{\infty, j}) = \hat\O_{ ij}(\Delta_\xi\tilde\psi_{ j} + \psi^{\cal P}_{\infty, j}). 
\end{equation}
At interior points $\sigma_{ j}$, $\tilde\O_{ ij}$ represents the discretization of Eq.~\eqref{eq:psi4LBS-psi4LR compactified}; at the boundary points it represents Eq.~\eqref{eq:psi-psiR=Dpsi+psiP at sigma=0} or Eq.~\eqref{eq:psi-psiR=Dpsi+psiP at sigma0}. Now, unlike $\tilde\O$, $\tilde\O_{ ij}$ is invertible (which is the essence of why the ``compactification + regularity'' method works). This invertibility implies we can peel the operator off Eq.~\eqref{eq:discretized Opsi-OpsiR}, leaving us with
\begin{equation}
\tilde\psi_{ j} - \tilde\psi^{\cal R}_{\infty, j} = \Delta_\xi\tilde\psi_{ j} + \tilde\psi^{\cal P}_{\infty, j}.
\end{equation}
Taking the continuum limit, we obtain Eq.~\eqref{eq:psi-psiR = Dpsi + psiP}.

Finally, we can check analytically that 
\begin{equation}\label{eq:Dpsi=-psiP}
\Delta_\xi\tilde\psi = -\tilde\psi^{\cal P}_\infty + \O(\sigma^2\ln\sigma).    
\end{equation}
To understand why this is true, note that, as explained above, $\tilde\O[\Delta_\xi\tilde\psi]=-\tilde S_{\rm Lor}+\O(\sigma^0)$ and $\tilde\O[\tilde\psi^{\cal P}_\infty] = \tilde S_{\rm Lor}+\O(\sigma^0)$. These equations are enough to imply that $\Delta_\xi\tilde\psi$ and $\tilde\psi^{\cal P}_\infty$ both take the form~\eqref{eq:Teukolsky infinity puncture}, \emph{up to regular terms}; more explicitly, they both have the form
\begin{multline}
       -\frac{j_\infty}{2i\varpi}\ln\Bigl(\frac{2M}{r}\Bigr) + \alpha_0 \\
       + \frac{({}_{-2}\lambda_\l-4 i \varpi)j_\infty }{(2i\varpi)^2}\frac{2M}{r}\ln\Bigl(\frac{2M}{r}\Bigr) + \O(r^{-2}\ln r),
\end{multline}
where $\alpha_0$ is an arbitrary constant. Here we have deliberately replaced $\sigma$ with $2M/r$ to make clear the nature of the ambiguity: the constant $\alpha_0$ is degenerate with the choice of length scale in the logarithm. Because of this, one could easily end up with $\Delta_\xi\tilde\psi = -\tilde\psi^{\cal P}_\infty + \alpha_0 + \O(\sigma^2\ln\sigma)$ instead of Eq.~\eqref{eq:Dpsi=-psiP}. In practice, when deriving $\Delta_\xi\tilde\psi$ and $\tilde\psi^{\cal P}_\infty$, we use $M$ rather than $2M$ as the length scale in the logarithm and then set the constant terms to zero. In other words, in terms of $\sigma$ we use
\begin{multline}
       -\frac{j_\infty}{2i\varpi}\ln(\sigma/2) 
       + \frac{({}_{-2}\lambda_\l-4 i \varpi)j_\infty }{(2i\varpi)^2}\sigma\ln(\sigma/2) \\+ \O(\sigma^{2}\ln \sigma).
\end{multline}

Combined with Eq.~\eqref{eq:psi-psiR = Dpsi + psiP}, Eq.~\eqref{eq:Dpsi=-psiP} implies
\begin{equation}
    \tilde\psi = \tilde\psi^{\cal R}_\infty + \O(\sigma^2\ln\sigma).
\end{equation}
Therefore, the waveforms that $\tilde\psi$ and $\tilde\psi^{\cal R}_\infty$ represent at $\sigma=0$ are identical, as are the fluxes computed from them. 

There is then one final subtlety: in the Lorenz gauge, the total field is $\tilde\psi^{\cal R}_\infty+\tilde\psi^{\cal P}_\infty$, not $\tilde\psi^{\cal R}_\infty$ alone. As described in Ref.~\cite{Cunningham:2024dog}, to extract the waveform from a Lorenz-gauge calculation, we ultimately transform to a Bondi-Sachs gauge; the essential difference is that this is done at the end of the second-order Lorenz-gauge calculation, as opposed to at the beginning. This transformation removes the logarithmic singularity from $\tilde\psi^{\cal P}_\infty$ and leaves $\tilde\psi^{\cal R}_\infty$ as the only contribution to the physical waveform (up to a constant 2PA phase shift described in Appendix~A.5 of Ref.~\cite{Cunningham:2024dog}). This is relevant because there is again an arbitrary choice of length scale in the logarithmic gauge transformation. However, as in our other calculations, we consistently use $M$ as the length scale. Hence, both calculations yield the same waveform, $\tilde\psi = \tilde\psi^{\cal R}_\infty$, and the same fluxes.

\section{Numerical errors}
\label{sec:numerical errors}

We estimate numerical errors by varying the source construction, matching data, and radial resolution. 
Table~\ref{tab:retained_error_components} summarises the source and junction condition responses; the individual source and matching-data tests are described in the following subsections. 
We combine these contributions in the error bars of Fig.~\ref{fig:reference_flux_comparison}, with the qualifications given in Appendix~\ref{sec:error_estimates}.
The Lorenz-reference uncertainty is treated separately.

\begin{table*}
\caption{Source and matching-data sensitivity estimates, relative to the Lorenz-gauge flux. The statistics cover the 27-radius sample and include interpolated responses where specified in the text. The ``at'' columns give the radii of the maxima. These source-response tests are distinct from the radial convergence tests of Table~\ref{tab:error_component_summary}.}
\label{tab:retained_error_components}
\begin{ruledtabular}
\begin{tabular}{lrrrrrr}
 & \multicolumn{3}{c}{$(\l,\m)=(2,1)$} & \multicolumn{3}{c}{$(\l,\m)=(2,2)$}\\
Contribution & median & maximum & at $r_0/M$ & median & maximum & at $r_0/M$\\
\midrule
$\l^S_{\max}$ truncation & $7.64\times10^{-6}$ & $1.31\times10^{-5}$ & 7.0 & $4.75\times10^{-5}$ & $5.52\times10^{-5}$ & 12.5\\
$h^{\SS}_{\mu\nu}$ angular quadrature & $1.11\times10^{-6}$ & $1.72\times10^{-5}$ & 7.0 & $4.66\times10^{-5}$ & $2.13\times10^{-4}$ & 20.0\\
Source sampling & $8.07\times10^{-7}$ & $2.78\times10^{-6}$ & 11.5 & $2.32\times10^{-6}$ & $1.39\times10^{-4}$ & 10.0\\
Inner matching data $J_-$ & $2.42\times10^{-5}$ & $5.42\times10^{-5}$ & 20.0 & $3.17\times10^{-5}$ & $5.66\times10^{-5}$ & 19.5\\
Outer matching data $J_+$ & $5.29\times10^{-5}$ & $9.07\times10^{-5}$ & 20.0 & $9.52\times10^{-5}$ & $2.28\times10^{-4}$ & 7.0\\
Slow-evolution matching-data fits & $2.28\times10^{-14}$ & $2.12\times10^{-7}$ & 11.5 & $1.99\times10^{-13}$ & $1.88\times10^{-5}$ & 11.5\\
\end{tabular}
\end{ruledtabular}
\end{table*}

\subsection{Angular truncation}
\label{sec:error angular}

The angular cutoff limits the first-order modes entering the quadratic source, not the second-order mode being computed. 
An output mode with $\l=2$ still receives contributions from products with arbitrarily large $\l_1$ and $\l_2$. 
We test this truncation separately inside and outside the worldtube. 
The interior test supplies an error-bar contribution; the exterior test determines how many source modes we retain in the reported calculation.

Inside the worldtube, we increase the puncture cutoff $\l^S_{\max}$ from 30 to 40 in the puncture-residual couplings. 
We keep the residual-residual cutoff at 40 and leave the SS angular quadrature unchanged. 
We rebuild the source, repeat the flux calculation, and use the absolute flux difference as $\delta_{\l^S_{\max}}{\cal F}_{\l\m}$ in Eq.~\eqref{eq:error_quadrature}. 
This comparison is performed at $r_0/M=7$, $10$, $11.5$, $12.5$, and $19.5$; the fractional response is interpolated at the other radii. 
The resulting estimates appear in the first row of Table~\ref{tab:retained_error_components}. They test the puncture-residual truncation, not the SS quadrature or the residual-residual cutoff.

Outside the worldtube, we start with the quadratic source summed through $\l_1,\l_2\leq50$. 
We then add the higher-mode contributions one cutoff at a time: increasing the cutoff from $L-1$ to $L$ adds all couplings with $\max(\l_1,\l_2)=L$. 
Let $a_L$ be the corresponding change in the flux of the output mode. 
All computed additions are retained in the final source; no fitted tail is added.

We choose the final cutoff separately at each radius and require two conditions in both output modes. 
The sum of the magnitudes of the last five flux changes must be less than $10^{-6}$ of the flux magnitude obtained with the source with the cutoff of $50$ modes. 
We also estimate the omitted tail by assuming geometric decay. 
If $q$ is the largest of the last five ratios $|a_j/a_{j-1}|$, this estimate is
\beq
 I_{\rm tail}=\frac{|a_L|q}{1-q},
 \qquad q<1.
 \label{eq:angular_tail_indicator}
\eeq
We require $q<1$ and $I_{\rm tail}$ to satisfy the same $10^{-6}$ criterion. 
The use of absolute changes prevents cancellation between successive additions from satisfying the test artificially. 
These conditions select the exterior cutoff; $I_{\rm tail}$ is neither a correction to the flux nor a separate contribution to the error bars.

The added source modes are computed on finite radial intervals outside the worldtube. 
At the ends nearest the horizon and \scri, their magnitude must be below $10^{-10}$ of their peak magnitude. 
The source beyond these intervals, the worldtube source, and the matching data are held fixed. 
The exterior test therefore constrains the additions only on the intervals recomputed, and its tail estimate assumes continued decay beyond the largest calculated mode. 
It is not a bound on the full angular-truncation error.

\subsection{SS angular quadrature}
\label{sec:error_SS}

We estimate the SS angular-quadrature error by changing the polar integration used to decompose
\beq
 X_{\mu\nu}\coloneqq
 \brE^{(0)}_{\mu\nu}[\barh^{\SS}]
 +\brG^{(2,0)}_{\mu\nu}[h^{(1)\calP},h^{(1)\calP}].
 \label{eq:SS-4D-combination}
\eeq
The two terms are combined before the angular integrals are evaluated. 
In the rotated coordinates $(\alpha,\beta)$, the particle lies at $\alpha=0$. 
The standard rule uses 768 Gauss-Legendre integration points over $\alpha\in[10^{-3},\pi]$ and 128 equally spaced azimuthal points. 
We repeat the polar integral with 1056 points distributed over smaller intervals extending down to $\alpha=10^{-6}$, keeping the azimuthal grid fixed. 
This increases the polar resolution near the particle and reduces the excluded polar cap.

We propagate the change in the BLS coefficients into the flux and take the absolute difference,
\beq
 \delta_{\SS}{\cal F}_{\l\m}
 =\left|{\cal F}^{(2)}_{\l\m}[X^{\rm ref}]
       -{\cal F}^{(2)}_{\l\m}[X]\right|,
 \label{eq:ss-error-definition}
\eeq
where the superscript $\mathrm{ref}$ labels the finer polar rule. 
This is the $\delta_{\SS}{\cal F}_{\l\m}$ contribution to Eq.~\eqref{eq:error_quadrature}. 
The comparison is performed at 22 radii, with fractional responses interpolated at the remaining 5.

The relative median and maximum responses are $1.11\times10^{-6}$ and $1.72\times10^{-5}$ for $(2,1)$, and $4.66\times10^{-5}$ and $2.13\times10^{-4}$ for $(2,2)$. 
The latter maximum occurs at $20M$. 
These values are listed in Table~\ref{tab:retained_error_components}.

The estimate has two limitations. 
The azimuthal grid is not varied, so this test does not measure its convergence. 
The quoted values are therefore approximate sensitivities to the polar-rule change, not bounds on the complete error in the SS piece.

\subsection{Source sampling and matching data}
\label{sec:error_sampling_jumps}

We estimate the source-sampling error by reducing the radial data supplied to the solver while keeping the junction data fixed. 
On each of the two exterior grids and the two one-sided worldtube grids, we retain every other source point and both endpoints, then rebuild the cubic interpolant. 
We repeat the flux calculation with this coarsened source and use
\beq
    \delta_{\tilde S}{\cal F}_{\l\m}
    =\left|{\cal F}^{(2)}_{\l\m}[\tilde S_{2h}]
       -{\cal F}^{(2)}_{\l\m}[\tilde S_h]\right|.
    \label{eq:source_grid_error}
\eeq
Here $\tilde S_h$ and $\tilde S_{2h}$ denote the interpolants of the full and coarsened source grids. 
Their absolute flux difference supplies the source-sampling contribution to Eq.~\eqref{eq:error_quadrature}. 
It tests how the sampled source is represented between grid points, but cannot detect structure absent from the full grid.

This is distinct from changing the Gauss quadrature of the worldtube integrals, which is included in the radial test of Appendix~\ref{sec:error_endpoints_discretization}. 
In the integrated-by-parts construction, the worldtube integrals act on the effective-source tetrad components, with derivatives transferred to the kernels through Eq.~\eqref{eq:GHPSdagger} and the local term~\eqref{eq:local_term_coordinate_form} retained. 
The SS angular quadrature determines those source components; the radial quadrature determines how their worldtube integrals are evaluated.

We test the matching data separately. 
The junction conditions contain the puncture, Bondi-Sachs gauge change, and slow-evolution terms. 
For the fitted puncture derivatives at each wall, we vary the polynomial degree and the number of nearby points included in the one-sided fit. 
The largest absolute deviation from the adopted derivative sets the perturbation size. We apply this perturbation to the real and imaginary parts of the derivative jump in separate flux calculations. 
The two flux changes are combined in quadrature, giving one contribution $\delta_{J_-}{\cal F}_{\l\m}$ for the inner wall and one $\delta_{J_+}{\cal F}_{\l\m}$ for the outer wall. 
These terms measure the sensitivity to the derivative fits, not an uncertainty in every part of the junction conditions.

For the slow-evolution matching terms, we also vary the polynomial degree and the number of fitted points, taking the largest absolute flux change as $\delta_{\cal V}{\cal F}_{\l\m}$. 
The matching-data tests are performed at $r_0/M=7$, 10, 11.5, 12.5, 14.5, 18.5, 19.5, and 20, with fractional responses interpolated between them. 
The three matching contributions enter Eq.~\eqref{eq:error_quadrature} separately. 
All physical jump terms remain present in these tests; only their numerical evaluation is varied.

\subsection{Endpoint treatment and radial convergence}
\label{sec:error_endpoints_discretization}
We estimate the radial error by repeating the exterior solves and worldtube integrals at different numerical resolutions.
The source samples, their cubic interpolants, the first-order amplitudes, and the junction conditions are held fixed.
Unlike the source sampling test in Appendix \secref{error_sampling_jumps}, this comparison does not change the source supplied to the solver.
All calculations use Chebyshev collocation in both exterior regions and the integrated-by-parts worldtube construction of \secref{mixed method}.
We impose the regularity conditions \eqref{eq:sigma=0 eqn} and \eqref{eq:sigma=1 eqn} at exactly $\sigma = 0$ and $\sigma = 1$, retaining the source value at each endpoint.
These conditions and the endpoint locations are unchanged in the convergence tests.
The flux is extracted exactly at \scri itself, so no finite-radius extrapolation enters the calculation.

At each of the 27 radii, we perform three calculations in Table \ref{tab:error_component_summary} for both modes.
Calculation $\mathrm{c}$ uses coarser exterior resolution.
Calculation $\mathrm{r}$ increases the exterior collocation resolution and infinity-side arithmetic precision, while leaving the worldtube quadrature unchanged.
Calculation $\mathrm{q}$ keeps the exterior solutions of $\mathrm{r}$ and increases the number of Gauss points in each worldtube integration interval from $6$ to $10$.
These intervals end at the source grid points and at the particle, hence the last comparison tests the integration without changing the source interpolation.

\begin{table*}
\caption{Settings for the radial convergence tests. Calculation $\mathrm{q}$ supplies the reported flux; calculations $\mathrm{c}$ and $\mathrm{r}$ determine its radial error estimate~\eqref{eq:numerical_error_budget}. All three use identical source and junction data. The coarse and fine infinity-side grids are defined in Sec.~\ref{sec:chebyshev_collocation}. On the horizon side, each calculation uses 45 subdomains with the distribution controlled by $\chi$ in the same section.}
\label{tab:error_component_summary}
\begin{ruledtabular}
\begin{tabular}{lccc}
 & $\mathrm{c}$ & $\mathrm{r}$ & $\mathrm{q}$\\
\hline
Infinity-side precision (decimal digits) & 45 & 65 & 65\\
Infinity-side nodes per subdomain & 16 & 24 & 24\\
Infinity-side grid & coarse & fine & fine\\
Horizon nodes per subdomain & 60 & 100 & 100\\
Horizon map parameter $\chi$ & 7 & 5 & 5\\
Worldtube Gauss points per interval & 6 & 6 & 10\\
\end{tabular}
\end{ruledtabular}
\end{table*}

We report the flux from calculation $\mathrm{q}$.
At each radius and for each mode, we take the larger of its absolute differences from $\mathrm{c}$ and $\mathrm{r}$ as the radial error estimate:
\beq
    \delta_{\rm rad}{\cal F}_{\l\m}
    =\max_{a\in\{\mathrm{c},\mathrm{r}\}}
    |{\cal F}^{(2)}_{\l\m,a} - {\cal F}^{(2)}_{\l\m,\mathrm{q}}|.
    \label{eq:numerical_error_budget}
\eeq
The $\mathrm{r}-\mathrm{q}$ difference isolates the worldtube quadrature.
The $\mathrm{c}-\mathrm{q}$ difference also includes the change in the exterior solution.
We use the maximum, rather than adding the differences, because both comparisons include the quadrature change.

This absolute error is the radial contribution to the combined estimate \eqref{eq:error_quadrature}.
The relative error listed in Table \ref{tab:flux_residuals} is
\beq
    \epsilon^{\cal F}_{\l\m}
    =\frac{\delta_{\rm rad}{\cal F}_{\l\m}}
    {|{\cal F}^{(2)}_{\l\m,\mathrm{q}}|}.
    \label{eq:order_change}
\eeq
It measures sensitivity to radial resolution, not the total flux error.
Source and junction-condition data uncertainties are estimated separately in the preceding subsections; the Lorenz-gauge and PN comparisons are not used to set this estimate.

Across the 27 radii, the largest values of $\epsilon^{\cal F}_{\l\m}$ are $1.77\times10^{-11}$ for $(2,1)$ and $6.47\times10^{-9}$ for $(2,2)$.
The worldtube-quadrature change alone is below $10^{-15}$ in relative flux both modes.
The radial contribution is therefore much smaller than the dominant source and matching-data contributions in Table~\ref{tab:retained_error_components}, and much smaller than the Teukolsky-Lorenz differences discussed in \secref{Lorenz comparison}.

We also compare the complex amplitudes before projecting them into the flux through Eq.~\eqref{eq:flux2_result}:
\beq
    \epsilon^{\psi}_{\l\m} = \max_{a\in\{\mathrm{c},\mathrm{r}\}} \frac{|\tilde{\psi}^{(2)}_{\mathscr{I}^{+}, a} - \tilde{\psi}^{(2)}_{\mathscr{I}^{+}, \mathrm{q}}|}{|\tilde{\psi}^{(2)}_{\mathscr{I}^{+}, \mathrm{q}}|}
    \label{eq:amplitude_error}
\eeq
This checks changes in both magnitude and phase.
The maximum values are $1.74\times10^{-11}$ and $4.00\times10^{-9}$ for $(2,1)$ and $(2,2)$, respectively.
Since the resulting flux change is already included in $\delta_{\rm rad}{\cal F}_{\l\m}$, the amplitude comparison adds no further contribution to the error bars.

Finally, we check the infinity-side solution with an independent integration that advances in short steps using local Taylor series of the radial equation.
With the horizon solution and worldtube quadrature fixed, the relative flux differences from collocation are at most $4.67\times10^{-15}$ for $(2,1)$ and $1.31\times10^{-13}$ for $(2,2)$ across all 27 radii.
This is a consistency check, not the error estimate: all reported fluxes and the three calculations defining $\delta_{\rm rad}{\cal F}_{\l\m}$ use collocation.

\begin{table*}
\caption{Teukolsky-Lorenz relative differences and fixed-source radial flux envelopes. A dagger denotes an interpolated comparator, excluded from the direct-reference statistics. The radial envelope is normalized by the selected Teukolsky flux and excludes source, matching-data, and reference uncertainties.}
\label{tab:flux_residuals}
\begin{ruledtabular}
\begin{tabular}{r@{\qquad}rr@{\qquad}rr}
$r_0/M$ & ${\cal E}_{21}$ & $\epsilon^{\cal F}_{21}$ & ${\cal E}_{22}$ & $\epsilon^{\cal F}_{22}$\\
\hline
7.0 & $9.67\times10^{-6}$ & $1.37\times10^{-11}$ & $7.67\times10^{-5}$ & $2.83\times10^{-9}$\\
7.5 & $7.56\times10^{-6}$ & $7.88\times10^{-12}$ & $1.42\times10^{-4}$ & $2.32\times10^{-9}$\\
8.0 & $1.03\times10^{-5}$ & $1.77\times10^{-11}$ & $5.18\times10^{-4}$ & $5.69\times10^{-9}$\\
8.5 & $8.25\times10^{-6}$ & $1.45\times10^{-11}$ & $2.09\times10^{-4}$ & $5.28\times10^{-9}$\\
9.0 & $7.86\times10^{-6}$ & $5.89\times10^{-12}$ & $3.86\times10^{-4}$ & $2.50\times10^{-9}$\\
9.5 & $7.53\times10^{-6}$ & $4.88\times10^{-12}$ & $1.30\times10^{-4}$ & $2.45\times10^{-9}$\\
10.0 & $9.73\times10^{-6}$ & $1.17\times10^{-11}$ & $2.76\times10^{-4}$ & $6.47\times10^{-9}$\\
10.5 & $6.04\times10^{-6}$ & $3.16\times10^{-12}$ & $1.04\times10^{-4}$ & $2.03\times10^{-9}$\\
11.0\rlap{$^{\dagger}$} & $1.22\times10^{-4}$ & $5.37\times10^{-12}$ & $2.82\times10^{-5}$ & $3.44\times10^{-9}$\\
11.5 & $8.69\times10^{-6}$ & $6.30\times10^{-13}$ & $2.02\times10^{-6}$ & $5.25\times10^{-10}$\\
12.0 & $9.71\times10^{-6}$ & $2.21\times10^{-12}$ & $2.60\times10^{-4}$ & $1.81\times10^{-9}$\\
12.5 & $1.14\times10^{-6}$ & $3.76\times10^{-12}$ & $5.43\times10^{-4}$ & $4.28\times10^{-9}$\\
13.0\rlap{$^{\dagger}$} & $2.90\times10^{-6}$ & $1.79\times10^{-12}$ & $3.92\times10^{-4}$ & $1.96\times10^{-9}$\\
13.5 & $1.12\times10^{-5}$ & $1.06\times10^{-13}$ & $1.86\times10^{-4}$ & $2.14\times10^{-10}$\\
14.0 & $8.61\times10^{-6}$ & $2.74\times10^{-12}$ & $2.33\times10^{-4}$ & $3.83\times10^{-9}$\\
14.5 & $7.97\times10^{-6}$ & $1.42\times10^{-12}$ & $1.39\times10^{-4}$ & $2.25\times10^{-9}$\\
15.0\rlap{$^{\dagger}$} & $8.38\times10^{-6}$ & $6.20\times10^{-13}$ & $1.02\times10^{-4}$ & $1.22\times10^{-9}$\\
15.5 & $8.56\times10^{-6}$ & $9.36\times10^{-13}$ & $3.32\times10^{-4}$ & $1.73\times10^{-9}$\\
16.0 & $1.19\times10^{-5}$ & $5.61\times10^{-13}$ & $2.09\times10^{-4}$ & $1.38\times10^{-9}$\\
16.5 & $6.51\times10^{-6}$ & $1.97\times10^{-13}$ & $6.31\times10^{-6}$ & $7.47\times10^{-10}$\\
17.0\rlap{$^{\dagger}$} & $1.32\times10^{-5}$ & $1.66\times10^{-14}$ & $1.11\times10^{-4}$ & $9.77\times10^{-10}$\\
17.5 & $2.09\times10^{-5}$ & $2.09\times10^{-12}$ & $1.35\times10^{-4}$ & $4.74\times10^{-9}$\\
18.0 & $2.62\times10^{-6}$ & $1.34\times10^{-12}$ & $2.59\times10^{-4}$ & $2.96\times10^{-9}$\\
18.5 & $2.24\times10^{-5}$ & $4.89\times10^{-13}$ & $4.95\times10^{-4}$ & $5.42\times10^{-11}$\\
19.0\rlap{$^{\dagger}$} & $1.35\times10^{-5}$ & $1.69\times10^{-12}$ & $1.04\times10^{-3}$ & $2.88\times10^{-9}$\\
19.5 & $1.30\times10^{-5}$ & $1.42\times10^{-12}$ & $3.94\times10^{-4}$ & $1.27\times10^{-9}$\\
20.0 & $1.80\times10^{-5}$ & $1.34\times10^{-12}$ & $4.96\times10^{-4}$ & $5.13\times10^{-10}$\\
\end{tabular}
\end{ruledtabular}
\end{table*}

\subsection{Lorenz-gauge reference uncertainty}
\label{sec:Lorenz comparator errors}

The gray region in Fig.~\ref{fig:reference_flux_comparison} represents uncertainty in the Lorenz-gauge reference, separately from the Teukolsky error bars.
We construct it from a published error estimate and additional fixed-source tests of the Lorenz calculation.

The published estimate is the shaded $(2,2)$ region in Fig.~7 of Ref.~\cite{Warburton:2021kwk}. The resulting curve, $\epsilon^{\rm ref}_{\rm L}(r_0)$, rises from $1.5\times10^{-4}$ at $7M$ to $1.2\times10^{-3}$ at $20M$. 
We also use it for $(2,1)$ because the two modes share a computational scheme.

The additional tests are performed at $12.5M$ and $19.5M$, holding the second-order source fixed. 
We change the Green-function quadrature, tighten the tolerance used to compute the homogeneous solutions, and vary the order and radial range of the asymptotic fit. 
For the quadrature and tolerance tests, we use the absolute flux change. 
For the extraction test, we use half the range of fluxes obtained from the alternative fits. Adding these three contributions and dividing by the flux magnitude gives relative estimates of $9.99\times10^{-7}$ and $1.482\times10^{-5}$ for $(2,1)$, and $2.04\times10^{-6}$ and $4.164\times10^{-6}$ for $(2,2)$, at the two radii respectively. 
These tests do not include uncertainty in constructing the Lorenz source.

For each mode, we retain the larger of the two fixed-source estimates as $\epsilon^{\rm down}_{{\rm L},\l\m}$ throughout the plotted range. 
The final reference envelope is
\beq
 \epsilon^{\rm env}_{{\rm L},\l\m}(r_0)
 =\max\!\left[\epsilon^{\rm ref}_{\rm L}(r_0),
              \epsilon^{\rm down}_{{\rm L},\l\m}\right].
 \label{eq:Lorenz_floor}
\eeq
We take the maximum as a conservative estimate. 
The published, radius-dependent term is larger throughout our sample, so it sets the gray envelope. 
Neither the fixed-source tests nor this choice changes the tabulated Lorenz fluxes.

At radii without a directly tabulated reference flux, we interpolate the envelope in $r_{0}$. 
The uncertainty in interpolating the flux itself, described in \secref{Lorenz comparison}, is a separate contribution and is not included in the gray region. 
Neither reference uncertainty enters the Teukolsky error bars.

\subsection{Error estimates}
\label{sec:error_estimates}
The error bars in Fig.~\ref{fig:reference_flux_comparison} combine six source and junction-data contributions with the radial contribution. 
All terms are absolute flux changes. The relative source and matching estimates in Table~\ref{tab:retained_error_components} are multiplied by the magnitude of the Lorenz-reference flux; the relative radial estimate~\eqref{eq:order_change} is multiplied by the magnitude of the reported Teukolsky flux. 
We then combine them in quadrature:
\begin{multline}
    (\delta{\cal F})^2
    =(\delta_{\l^S_{\max}}{\cal F})^2+(\delta_{\SS}{\cal F})^2
    +(\delta_{\tilde S}{\cal F})^2\\
    +(\delta_{J_-}{\cal F})^2+(\delta_{J_+}{\cal F})^2
    +(\delta_{\cal V}{\cal F})^2+(\delta_{\rm rad}{\cal F})^2,
    \label{eq:error_quadrature}
\end{multline}
where mode labels are suppressed and $\delta_{\cal V}{\cal F}$ is the slow-evolution matching-data contribution. 
Each contribution is included once. 
In particular, the amplitude comparison and independent radial checks in Appendix~\ref{sec:error_endpoints_discretization} are not additional terms.

The bars summarize the measured sensitivities, not a statistical confidence interval or a complete error bound. 
Some source responses are interpolated between tested radii, and the source-response tests have not been repeated with the final radial settings. 
Correlations and source errors not exposed by these variations are not included.

The Lorenz-reference envelope and the uncertainty of interpolated comparators are kept separate from our bars. 
The PN residuals also contain the remainder of the truncated analytic series and the first-order derivative sensitivity in the mass-ratio conversion. 
They are therefore comparisons with an approximate analytic result, not numerical error estimates.

\bibliography{bib}

\end{document}